\documentclass[fleqn,usenatbib]{mnras}
\usepackage{newtxtext,newtxmath}
\usepackage[T1]{fontenc}
\usepackage{ae,aecompl}

\usepackage{graphicx} 
\usepackage{amsmath,amssymb,times}
\usepackage{deluxetable}
\usepackage{longtable}
\usepackage{epstopdf}
\usepackage[usenames,dvipsnames]{color}
\usepackage{float}
\usepackage[export]{adjustbox}
\usepackage{ulem}
\usepackage[export]{adjustbox}
\usepackage{hyperref}
\hypersetup{
  colorlinks= true, 
  urlcolor  = Blue, 
  filecolor=black,
  runcolor=blue,
  linkcolor = red, 
  citecolor = blue 
} 
\newcommand{\nad}{Na\,{D}}

\newcommand{\RN}[1]{%
  \textup{\uppercase\expandafter{\romannumeral#1}}%
}

\makeatletter
\def\@to{to}
\makeatother

\title[The properties of cold gas flows]{The prevalence and properties of cold gas inflows and outflows around galaxies in the local Universe}

\author[Roberts-Borsani \& Saintonge]{G.~W. Roberts-Borsani$^{1}$\thanks{E-mail: guidorb@star.ucl.ac.uk} and A. Saintonge$^{1}$
\\
$^{1}$Department of Physics and Astronomy, University College London, Gower Street, London WC1E 6BT, UK}

\date{Accepted XXX. Received YYY; in original form ZZZ}
\pubyear{2018}

\begin{document}
\label{firstpage}
\pagerange{\pageref{firstpage}--\pageref{lastpage}}
\maketitle

\begin{abstract}
We perform a stacking analysis of the neutral \nad\,$\lambda\lambda$5889,5895\,\AA\ ISM doublet using the SDSS DR7 spectroscopic data set to probe the prevalence and characteristics of cold (T\,$\lesssim$\,10$^{4}$\,K) galactic-scale gas flows in local (0.025$\leqslant z\leqslant$0.1) inactive and AGN-host galaxies across the SFR-M$_{*}$ plane. We find low-velocity outflows to be prevalent in regions of high SFRs and stellar masses (10 $\lesssim$log M$_{*}$/M$_{\odot}$ $\lesssim$ 11.5), however we do not find any detections in the low mass (log M$_{*}$/M$_{\odot}$ $\lesssim$ 10) regime. We also find tentative detections of inflowing gas in high mass galaxies across the star-forming population. We derive mass outflow rates in the range of 0.14-1.74\,M$_{\odot}$yr$^{-1}$ and upper limits on inflow rates <1\,M$_{\odot}$yr$^{-1}$, allowing us to place constraints on the mass loading factor ($\eta$=$\dot{M}_{\text{out}}$/SFR) for use in simulations of the local Universe. We discuss the fate of the outflows by comparing the force provided by the starburst to the critical force needed to push the outflow outward, and find the vast majority of the outflows unlikely to escape the host system. Finally, as outflow detection rates and central velocities do not vary strongly with the presence of a (weak) active supermassive black hole, we determine that star formation appears to be the primary driver of outflows at $z\sim$0.
\end{abstract}

\begin{keywords}
galaxies: evolution -- galaxies: starburst -- ISM: jets and outflows -- ISM: inflows
\end{keywords}

\section{Introduction}

The Lambda-Cold Dark Matter ($\Lambda$CDM) model has been extremely successful in reproducing observations of the large scale structure of the Universe (e.g., \citealt{planck15}). However, the assumption that the mean growth of DM halos dictates the cosmological accretion rate of cool gas onto galaxies causes the framework to fail by overpredicting the star formation (SF) activity in low-mass halos at early times, and in high-mass halos at later times (\citealt{bell03,li2009}).
The introduction of feedback mechanisms, such as outflows, goes a long way towards reconciling these important discrepancies \citep{somerville08,bouche10} by expelling gas or preventing accretion, and are now regularly invoked in hydrodynamical simulations (e.g., \citealt{oppenheimer10,vdv11}). 
The need to accommodate the cycling of gas in and out of galaxies (known as the "baryon cycle") also brought out new observational frameworks - known as "bathtub" or "equilibrium" models \citep{bouche10,dave12,lilly13} - which place emphasis on the accreted cold gas, the efficiency of star formation, and the role of metal enriched outflows ejected into the circumgalactic (CGM) and intergalactic medium (IGM). These models have now taken center stage in galaxy evolution \citep{saintonge13} and it is the interplay between inflows and outflows which dictates the position of a galaxy relative to the "main sequence" \citep{noeske07} on the SFR-M$_{*}$ plane, highlighting a necessity for a thorough theoretical and observational understanding of these processes.

Outflows can arise due to large amounts of energy and momentum given off by stellar winds, supernovae, or an active galactic nucleus (AGN). They have been observed and found to be ubiquitous at all epochs (see review by \citealt{veilleux05}), although most observations have typically focused on a variety of more extreme objects such as mergers, (U)LIRGs and QSO hosts (e.g., \citealt{cicone14,rupke05a,rupke05b,rupke17}). Much less is known on the more normal objects at each epoch.
Over the past decade, several pioneering studies have helped to observationally constrain the prevalence of low-$z$ ($z\lesssim$1) outflows in samples of less extreme star-forming galaxies (e.g., \citealt{weiner09,chen10,martin12,rubin14}). 
In one of the first major systematic searches, \citet{weiner09} used a sample of 1,406 DEEP2 galaxy spectra at $z\sim$1.4 to search for cool, low-ionization outflowing gas and found detections in more than half of their sample. They found evidence that their detection rate had a weak positive dependence on stellar mass and SFR.
\citet{martin12} followed up on this with an investigation of 200 deep Keck/LRIS spectra of highly star-forming galaxies with log M$_{*}$/M$_{\odot}>$ 9.4 at 0.4<$z$<1.4. They also found a high detection rate of $\sim$50\%, however unlike \citet{weiner09}, they did not find any dependence of outflow properties with stellar mass or SFR. Despite the success of the aforementioned studies in demonstrating the ubiquity of outflows at $z\sim$1, the selected samples were of a starburst nature and a higher redshift than the present day Universe. \citet{rubin14} improved on this with their sample of 105 galaxies derived from the GOODS fields and Extended Groth Strip, at a median redshift $z\sim$0.5. These objects spanned a larger range of stellar mass (log M$_{*}$/M$_{\odot}$ $\gtrsim$ 9.6) and SFRs (SFR $\gtrsim$ 2 M$_{\odot}$\,yr$^{-1}$) than the previous studies. The detection rate for their sample remained high ($\sim$66\%), despite sampling lower SFRs. Arguably the most representative study for normal star-forming galaxies of the local Universe, however, was that of \citet{chen10}. The authors selected a large sample of massive (log M$_{*}$/M$_{\odot}>$ 10.4) star-forming galaxies from the Sloan Digital Sky Survey (SDSS; \citealt{york2000}) at redshifts 0.05\,$\leqslant$$z$$\leqslant$0.18, and by means of stacking found strong and clear dependencies of neutral gas outflow properties with galaxy viewing angle, stellar mass, and star formation surface density ($\Sigma_{\text{SFR}}$).

Each of these studies have found winds to be prevalent in the low-$z$ Universe and, due to them arising in galaxies with high SFRs, have suggested they are generally a consequence of high levels of star formation or $\Sigma_{\text{SFR}}$ - in fact, a critical $\Sigma_{\text{SFR}}$ threshold of 0.1\,M$_{\odot}\,$yr$^{-1}$\,kpc$^{2}$ is regularly suggested in order to launch a galactic wind \citep{heckman02}. However, the detection rates of winds have also been found to vary strongly as a function of galaxy disk inclination: working under the assumption that outflows have a biconical structure which exits perpendicular to the disk, one would expect to have fewer detections in absorption at high inclinations (where one views the disk edge-on) and more at low inclination (viewing the disk face-on). Indeed this appears to be the case, with the majority of detections in absorption arising from low inclinations \citep{martin12,rubin14,chen10}. For starburst galaxies with inclinations less than $i\sim$60$^{\circ}$, \citet{heckman2000} found a $\sim$70\% probability of detecting outflows in absorption.

Deep observations of local galaxies also revealed the presence of a diffuse, secondary layer of extraplanar gas (known as a "lagging halo") which extends kiloparsecs out of the disk. The extraplanar gas has been observed in the atomic (e.g., \citealt{fraternali02,matthews03,oosterloo07,zschaechner15}) and ionised (e.g., \citealt{rossa03,rossa032,heald07,kamphuis08}) gas phases, in both external galaxies and the Milky Way \citep{marasco11}. Accompanying the extraplanar gas are often signatures of accretion (e.g., \citealt{fraternali02,fraternali08,zschaechner15}), and dynamical modeling of the gas suggests outflows or accretion alone cannot account for the observed kinematics and gas masses \citep{fraternali06}. As such, the emerging picture appears to be a cyclic scenario, where gas gets blown out from the disk by stellar winds and supernovae (this "blowout" phase has been observed by e.g., \citealt{boomsma08}, who report holes of H\,I gas in the disk of NGC\,6946 with high rates of star formation) and eventually condenses and mixes into colder gas which gets re-accreted and fuels star formation. This scenario is known as the "galactic fountain" \citep{shapiro76} and plays a crucial role in regulating the gas contents and SFRs of local galaxies.


Although star formation certainly appears to play an important role in launching winds, the dominant energy source for outflows in the present day Universe is not always obvious. Several recent studies have aimed to address this by comparing the detection rates of outflows in local galaxies displaying signatures of star formation and AGN. For instance, \citet{sarzi16} used a sample of 456 objects for which both optical and radio data were available and found that none of the 23 objects displaying signatures of neutral gas outflows showed radio emission or optical line ratios indicative of an AGN. \citet{concas17} conducted a similar study with SDSS-selected galaxies and found outflows traced by the same neutral gas to be present in both star-forming galaxies and AGN hosts. These results appear to suggest that weak, optically-selected AGN do not have a major influence on the detection rates of neutral gas outflows.
Studies of more extreme AGN/QSOs and starbursts, however, generally portray a more distinct picture: many such objects exhibit very powerful outflows (e.g., \citealt{walter02,cannon05,feruglio10,combes13,cicone14,garciaburillo15,fiore17}) which appear significantly enhanced by the presence of an AGN. For example, using a sample of 19 strong Seyferts, LINERs and "pure" starburst galaxies, \citet{cicone14} found strong molecular outflows in all galaxy types, but with significantly boosted outflow velocities and mass loss rates in the AGN hosts. The latter quantity was also found to increase with the AGN luminosity.



These results inevitably lead to the crucial question of whether outflows ultimately halt the SF processes in galaxies (coined "negative feedback") or not. This can happen via the removal of gas necessary to fuel star formation, prevention of accretion, or a combination of both. Spectroscopic studies have found cases of AGN-driven outflows expelling mass at a rate many times that of the host galaxy's SFR, thereby clearly able to remove significant fractions of gas and eventually quench the host (e.g., \citealt{cicone14,sturm11}). Additionally, some Integral Field Unit (IFU) studies have also shown a spatial coincidence with outflowing material and an absence of SF (e.g., \citealt{cresci15,carniani16}), although it is often unclear whether there is causality in this.
Further complicating this picture are instances where star formation has seen itself reignited due to the turbulence created by the presence of an outflow (coined "positive feedback", e.g., \citealt{maiolino17,gallagher18}). Simulations of the Milky Way even suggest weak outflows form a necessary ingredient to stimulate and \textit{sustain} accretion - and therefore star formation - by transferring gas from a surrounding hot corona to the disk (e.g., \citealt{marinacci10,marasco12}).

A quantity often used to describe how efficiently outflows can remove mass is the \textit{mass loading factor}, $\eta$, defined as the mass outflow rate divided by the SFR of the host. This value is used in simulations to dictate the strength of outflows (e.g., \citealt{oppenheimer10, muratov15, vogelsberger13}), yet important discrepancies exist with results found in observations, with the latter often finding order-of-magnitude lower values (e.g., \citealt{weiner09, martin12, rubin14, chisholm17}) for normal galaxies. This demonstrates the need to understand whether observations are missing large fractions of ejected mass traced by different gas phases, or simulations are invoking outflows that are larger than those seen in the present day Universe. 

\begin{figure}
 \includegraphics[width=\columnwidth]{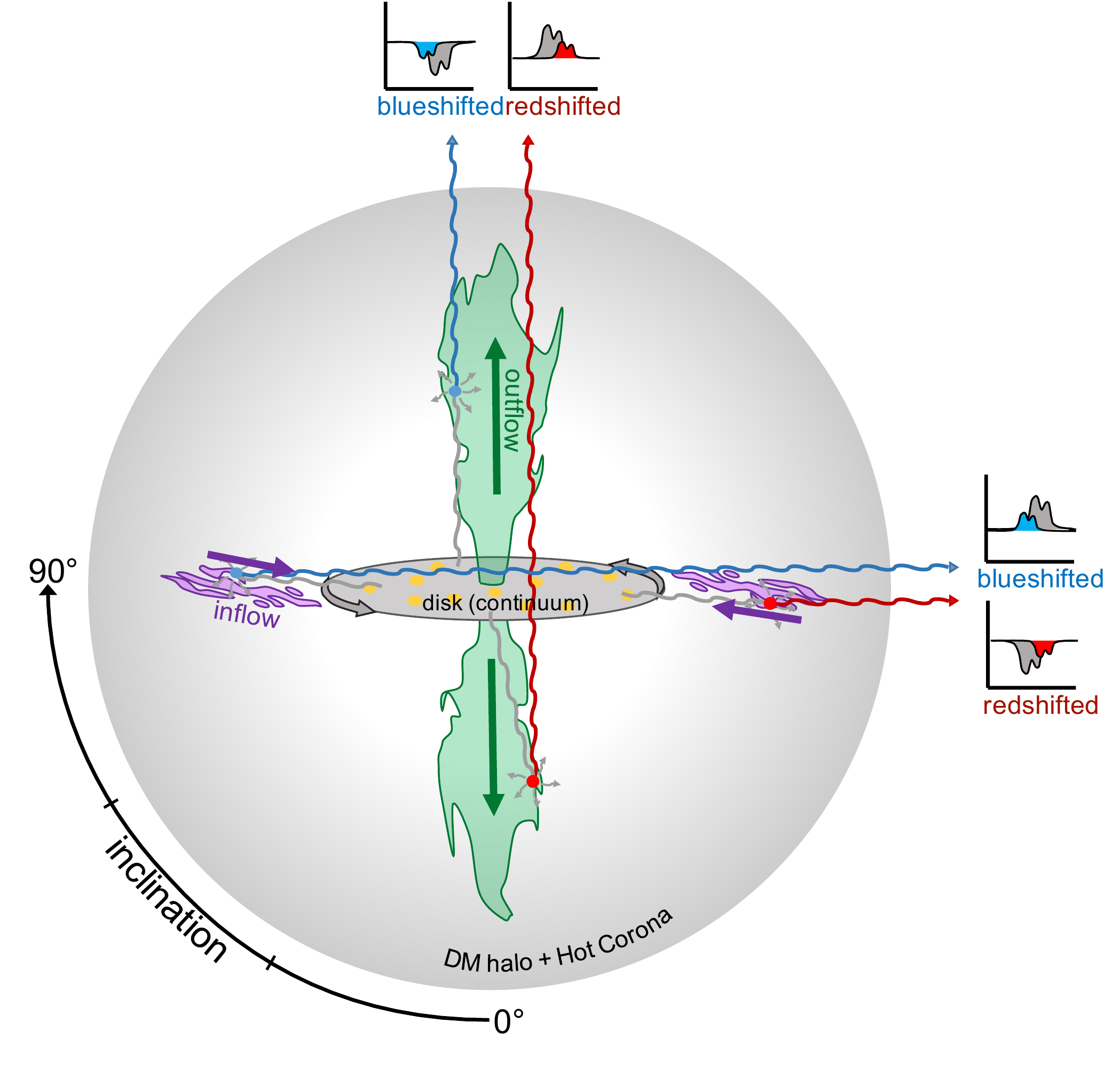}
 \caption{A schematic of the different types of Doppler shifts one can observe in the \nad\ transition. Foreground gas is dominated by absorption of the background continuum, with profiles either blueshifted (outflows) or redshifted (inflows). Background gas is seen in emission of re-emitted photons. Gray lines represent continuum photons from the galaxy disk, blue lines represent blueshifted signatures, and red lines represent redshifted signatures. Small blue or red circles represent absorption, from which re-emission of the photon occurs. Included in the schematic is the role of the viewing angle in detecting outflowing or potentially inflowing gas.}
 \label{fig:naddoppler}
\end{figure}

Outflows are typically observed via Doppler shifts of ISM gas, characterized by broad components in emission spectra or blueshifted signatures in absorption. The latter method is known as the "down the barrel" technique, where gas in front of a galaxy is illuminated by the background continuum. Since the gas is moving toward the observer along the line of sight (or "barrel"), it appears blueshifted with respect to the systemic (galaxy) component. Equally, redshifted absorption is suggestive of gas moving towards the galaxy, in the shape of inflowing gas. Although such signatures may arise from anywhere along the sight line to the galaxy, such that the technique offers no information as to whether the gas reaches (in the case of an inflow) or fully escapes (in the case of an outflow) the galaxy, red/blueshifted absorption has typically been interpreted as an unambiguous signature of in/outflowing gas relative to the galaxy. A schematic of this method is shown in Figure \ref{fig:naddoppler}. 

However, the use of different tracers in rest-frame UV or optical wavelengths (e.g., Fe\,\RN{2}\,$\lambda\lambda$2586,2600, Mg\,\RN{2}\,$\lambda\lambda$2796,2803, Na\,\RN{1}\,D\,$\lambda\lambda$5890,5896, [OIII]\,$\lambda$5007) and (in many cases) the use of a biased sample (e.g., containing objects selected \textit{a priori} to have high SFRs, $\Sigma_{\text{SFR}}$s or stellar masses), have limited the extent to which general conclusions can be made. An alternative is to use a single tracer and stacking approach over a large and representative sample to create much higher signal-to-noise (S/N) composite spectra, allowing for flow detections over the \textit{general} galaxy population. This is especially useful to probe regions of parameter space that single spectra cannot (e.g., very low M$_{*}$ or low SFR galaxies) and derive accurate measurements of flow properties. Recently, similar studies  were undertaken using stacked data from the SDSS and other surveys to constrain the properties of and links between neutral and ionized outflows in the general population of galaxies from the low-$z$ Universe (e.g., \citealt{chen10,cicone16,concas17,sugahara17}). These studies have allowed for strong constraints on the evolution of outflow properties as a function of key global galaxy parameters (e.g., SFR, M$_{*}$, $\Sigma_{\text{SFR}}$, $z$, inclination, excitation mechanism). However, still lacking are the crucial constraints on the mass loading factor, and potential inflowing gas.

In this study, we aim to use the SDSS Data Release 7 (DR7; \citealt{abazajian09}) data set and a stacking technique in order to sample large ranges of global galaxy properties with which to infer detection rates, properties, and mass flow rates of inflows and outflows. With this, we can place strong constraints on the mass loading factor in the local Universe. We focus on the resonant Na\,I absorption doublet at 5889.95\,\AA\ and 5895.92\,\AA\ (also referred to as \nad), which traces cool (T\,$\lesssim$\,10$^{4}$\,K), metal-enriched gas. We present our observational data set and selection criteria in Section \ref{sec:datasets}, stacking and fitting procedures in Section \ref{sec:analysis}, and present our results in Section \ref{sec:results}, including details on covering fractions, equivalent widths, mass inflow/outflow rates, central velocities and the mass loading factor. Section \ref{sec:discussion} discusses the implications of our detections and results by offering a comparison to recent simulation results, the role of SF vs AGN feedback, a dissection of the sources of inflow, and a brief discussion on the fate of the outflows. Finally, we summarize our main conclusions in Section \ref{sec:conclusions}.
Throughout this paper we adopt a $\Lambda$CDM cosmology with \textit{H}$_{0}$ = 70 km/s/Mpc, $\Omega_{m}
=$ 0.3, and $\Omega_{\wedge} =$ 0.7, and assume a Chabrier IMF.

\section{Sample Definition \& Measurements}
\label{sec:datasets}
\begin{figure*}
 \includegraphics[width=\textwidth]{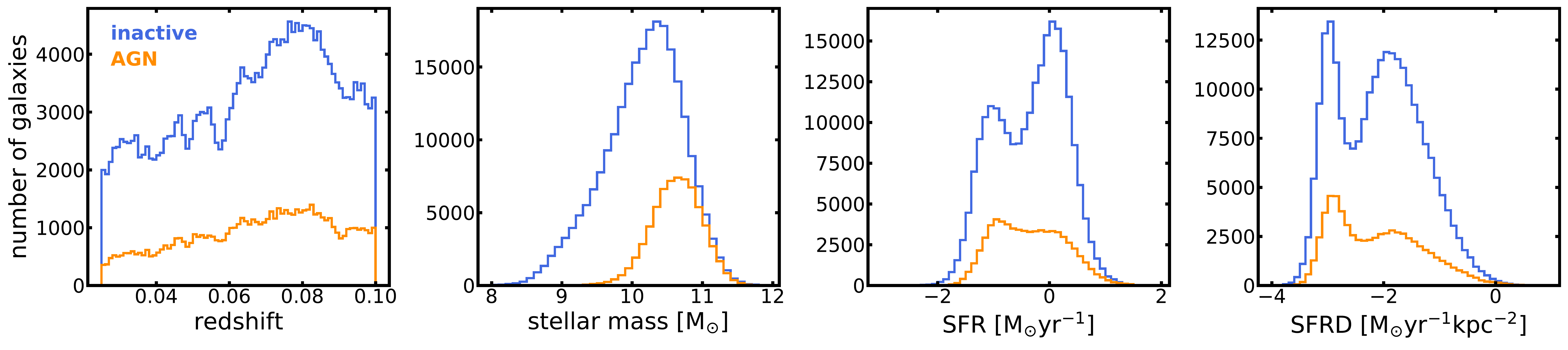}
 \caption{The distributions of redshift, stellar mass, SFR, and $\Sigma_{\text{SFR}}$ of our main samples. The blue histograms represent our inactive galaxies sample and the orange histograms represent our AGN sample.}
 \label{fig:distributions}
\end{figure*}

We make use of the full SDSS DR7 catalogs. After requiring all objects to satisfy an SDSS "type" of "GALAXY", we select all objects with a redshift of 0.025\,$\leqslant$$z$$\leqslant$0.1. This 
redshift range allows for a robust derivation of the galaxy morphology whilst the SDSS 3" diameter spectroscopic 
fiber samples the central $\sim$1.6-6.7\,kpc of the galaxy.
To separate the AGN-hosts from the inactive galaxies, we require a line S/N ratio $>$3 in all BPT diagnostic lines (namely H$\alpha$, H$\beta$, [OIII]\,$\lambda$5007 and [NII]\,$\lambda$6584) and the satisfaction of a \citet{ka03} BPT prescription. If the line S/N ratio of one of the BPT lines is $<$3, or the galaxy does not satisfy the BPT prescription, it is classified as inactive. This selection yields a parent sample of 240,567 inactive galaxies and  67,753 AGN-hosts. We refer to this sample as the main sample. The samples' distributions in redshift, stellar mass, SFR and $\Sigma_{\text{SFR}}$ are shown in Figure \ref{fig:distributions}.

For starburst galaxies with inclinations less than $i\sim$60$^{\circ}$, 
\citet{heckman2000} found that a high probability ($\sim$70\%) 
existed of detecting outflows in absorption. This motivates an 
additional cut to separate galaxies based on their inclination. We therefore define a subsample from our main sample, called DISK, which includes all galaxies with a measurable inclination. In Section \ref{subsec:detections} we show that $i\sim$50$^{\circ}$ is a more suitable inclination cut. We therefore further divide the DISK sample into two sub-samples, HIGH-$i$ and LOW-$i$, with inclinations \textgreater60$^{\circ}$ and \textless60$^{\circ}$, respectively. For this, we require an $r$-band fracDeV parameter of \textless\,0.8 to ensure we select disk galaxies, from which an inclination angle can be calculated from the $r$-band axis ratio, $b/a$, as 

\begin{equation}
\mbox{i = cos$^{-1}$\bigg[\bigg($\frac{(b/a)^{2} - q^{2}}{1 - q^{2}}$\bigg) - $\frac{1}{2}$\bigg]},
\end{equation}

where $q$=0.13 \citep{giovanelli94} is the assumed intrinsic axial ratio. Finally, we define a BULGE sample with a fracDeV parameter equal to 1, to select objects completely dominated by a bulge. We do not include objects with fracDeV parameters between 0.8 and 1, since these might have some disk structure from which we cannot accurately determine an inclination and therefore do not complement the DISK or BULGE samples. The size of each sub-sample is listed in Table \ref{tab:samples}.

The global galaxy properties associated with our analysis 
come from the widely-used MPA-JHU catalog\footnote{https://wwwmpa.mpa-garching.mpg.de/SDSS/DR7/}, in which the SFRs are derived using the 4000\,\AA\ break, following the method of 
\citet{brinchmann04}. We derive $\Sigma_{\text{SFR}}$s ($\Sigma_{\text{SFR}}$ =\,SFR$\cdot$cos($i$)\,/\,$\pi$r$^{2}$), where r is the physical radius of the galaxy probed by the fiber, in kpc.

\begin{table}
  \centering
  \caption{The number of galaxies in each sub-samples defined for this study.}
  \label{tab:samples}
  \begin{tabular}{lccccc}
      \hline
        Sample & Inactive & AGN \\
        \hline
    	main & 240,567 & 67,753 \\
        DISK & 165,571 & 32,728 \\
    	LOW-$i$ & 75,739 & 13,282 \\
    	HIGH-$i$ & 86,558 & 19,446 \\
        BULGE & 43,724 & 19,558 \\
        \hline
  \end{tabular}
\end{table}

\section{Analysis}
\label{sec:analysis}
\subsection{Binning and stacking optical spectra}
\label{sec:sdssstacking}
In order to achieve high S/N, we opt for a spectral stacking analysis over planes of global galaxy properties (e.g., SFR-M$_{*}$ or $i$-$\Sigma_{\text{SFR}}$). Bins are constructed via an adaptive approach, where the edges are defined such that the resulting bin is larger than the mean uncertainty of the relevant property, and the stacked spectrum has a continuum S/N$\geqslant$100. For bins of stellar mass and SFR we require bins larger than 0.2 dex and 0.5 dex, respectively. The spectra in each bin are first sorted by parameter of interest before being corrected for galactic extinction using the Schlegel dust maps and a \citet{odonnell} Milky Way extinction curve. 

To create the stack, each galaxy spectrum is converted to air wavelengths and shifted to the rest-frame, before being interpolated over a common wavelength array. The spectrum is then normalized to the median flux between 5440\,$\mbox{\AA}$ and 5550\,$\mbox{\AA}$, where it is uncontaminated by emission or absorption lines - this normalization ensures no preferential weighting is given to the lowest redshift galaxies in our sample. The normalized spectrum is then weighted by a mask array (with values of 0 for bad pixels identified in the SDSS spectrum array, and 1 for everything else) and added to the stack. The final stack is then simply the mean over $N$ galaxies with a normalized spectrum, over each wavelength element. The flux uncertainties associated with the composite spectrum are derived by adding in quadrature the mean flux uncertainties calculated from the SDSS error arrays and the sampling error, which we estimate via a bootstrapping method with replacement.

\subsection{Fitting the stellar continuum}
\label{sec:contfitting}
\begin{figure*}
 \includegraphics[width=\textwidth]{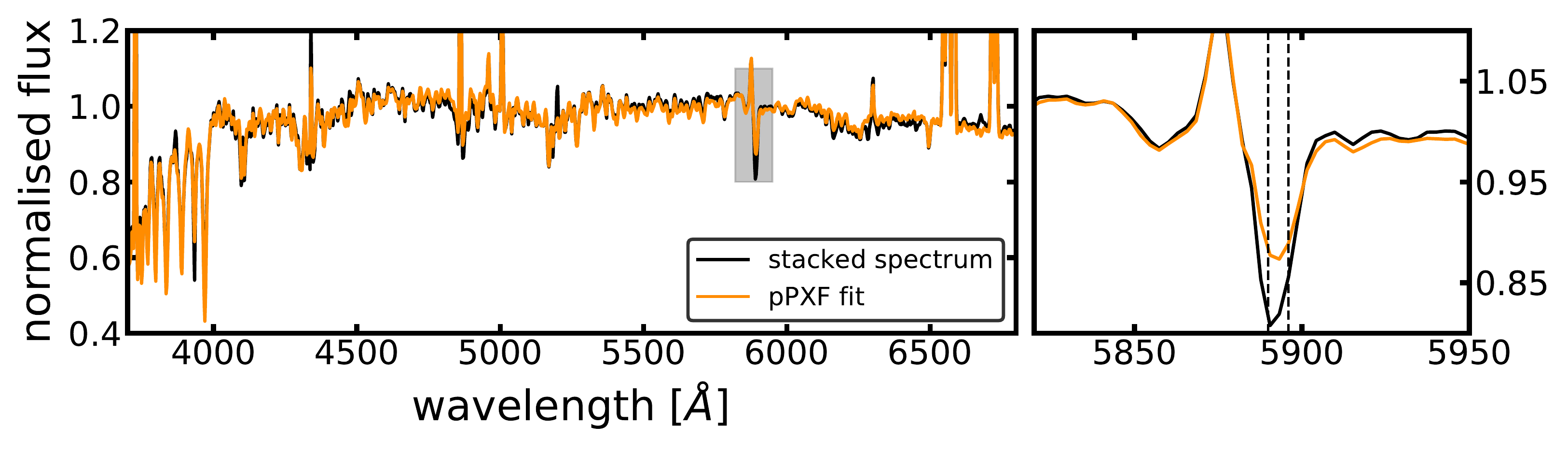}
 \caption{An example stacked spectrum (black) from the inactive main sample, 
 with its best-fit pPXF continuum (orange). The full spectrum is shown on the left, whilst the plot on the right is a zoomed-in portion of the gray-shaded region, highlighting the fit to the 
 He\,I line and the \nad\ absorption. The dashed lines represent the central blueshifted and redshifted wavelengths of the \nad\ doublet at 5889\,$\mbox{\AA}$ and 5895\,$\mbox{\AA}$, respectively.}
 \label{fig:stackfit}
\end{figure*}

The Sodium doublet is a predominantly photospheric transition and is particularly strong in the spectra of cool stars, with peak strengths for stars of types K3-M0 \citep{jacoby84}. The prevalence of bulge K-giants in nuclear regions of galaxies means they are likely to make an important contribution to the spectra of this selected sample, since the SDSS fiber probes the central regions of the galaxies at our selected redshifts. Throughout this study, however, we assume the \nad\ doublet is a feature with additional contributions from the interstellar medium (ISM) and that any signature of in/outflows will be found in this component. As such, the careful removal of any stellar contribution is imperative.
To do this, we model the stellar continuum with the Penalized Pixel-Fitting (pPXF; \citealt{cappellari17}) code to fit a non-negative linear combination of Simple Stellar Population (SSP) templates, which make use of the MILES \citep{miles} empirical stellar library, with BaSTI isochrones and [$\alpha$/Fe]-enhanced models where available. A \citet{battisti16} extinction law is assumed. We carefully mask the Ca\,II (K and H) and \nad\ transitions, since these are all in our spectral fitting range and we assume they are the result of a stellar+ISM contribution, which the models cannot account for. We also mask the red half of the He\,I emission line at 5875.67\,\AA, which is close enough to the \nad\ line that it could affect the residual profile. Furthermore, we allow for non-Gaussian line-of-sight velocity distributions (LOSVD), since there is a very small possibility this could influence our results if unaccounted for. This is most likely a very minor effect (if present), since our stacking method should blend out such cases and pPXF by definition penalises non-Gaussian LOSVDs. Additionally, the pPXF software also fits the optical nebular emission lines present in the spectrum. To remove the stellar contribution, the stacked spectrum is divided by the continuum fit. An example stacked spectrum and its best fit continuum model are shown in Figure \ref{fig:stackfit}.

To ensure a level of robustness in our continuum fits, we look at the Mg\,$I$\,$\lambda\lambda\lambda$\,5167, 5173, 5184 (Mg\,$b$) triplet. Since Mg\,$b$ has a similar ionizing potential as \nad\ and is produced in similar nuclear processes of hot stars, several studies of \nad\ outflows in ULIRGs (e.g., \citealt{martin05,rupke05a,rupke05b}) estimated the stellar contribution of \nad\ from Mg\,$b$. We instead look at the equivalent-width (EW) of the Mg\,$b$ residuals left over from our continuum fitting: assuming the stellar continuum is well modeled, the residual should be very small and can be used as a proxy for goodness-of-fit. We find that the distribution of residual EWs is roughly bimodal, with one mode containing the majority of (small) residuals and the other mode a smaller population of larger residuals, and can be well described by two Gaussian functions. We interpret the larger EWs as a result of poor continuum fitting and define a range of acceptable residuals with a lower limit EW$_{\text{Mg}}$$_{b}$$_{\text{,low}}$=0\,\AA\ and an upper limit EW$_{\text{Mg}}$$_{b}$$_{\text{,upp}}$=0.112\,\AA\ given by the 1$\sigma$ width of the main Gaussian containing the small residuals.

\subsection{Interpretation of \texorpdfstring{\nad}{TEXT} Doppler Shifts}
\label{subsec:nadinterp}
Prior to modeling the \nad\ residual, it is important to consider which types of Doppler shifts we consider to be signatures of outflows and inflows. In the Introduction we described how blueshifted or redshifted absorption can be interpreted as foreground gas moving along the line of sight, and therefore as unambiguous signatures of outflows and inflows. Being a resonance transition, \nad\ also re-emits all absorbed photons isotropically and as such, blueshifted or redshifted resonant \textit{emission} becomes an important consideration. Due to the isotropic nature of the re-emission, on average one cannot have more emission than absorption from forefront gas (the observer sees each absorption signature from the continuum but many re-emitted photos may follow a different sightline), and it is therefore reasonable to assume that absorption dominates the signatures of foreground gas. Following this logic, for a clump of gas on the backside of the galaxy, absorption signatures of the continuum are not visible but photons that are absorbed and then re-emitted by the clump can fall back along the line of sight towards the observer. If the gas is moving away from the observer, the re-emitted photons are redshifted and therefore signatures of outflowing gas, whilst if they are moving towards the observer then they are blueshifted and signatures of inflowing gas. Several studies (e.g., \citealt{chen10,rupke15}) have demonstrated a correlation between the visibility of redshifted emission and the dust content of the galaxy, suggesting the redshifted emission comes from a backside receding outflow seen through a dust-poor, face on disk. In some cases redshifted emission is also accompanied by blueshifted absorption \citep{phillips93} in the form of a P-Cygni profile. Additionally, \citet{prochaska11} showed via radiative transfer models of cold gas winds that redshifted emission was in fact a prominent and important feature to consider in outflow studies. Our interpretation of Doppler shifted \nad\ is consistent with this picture. 

To summarise, the sources of inflows and outflows from Doppler shifts (based on geometry) are: blueshifted absorption (outflow), redshifted absorption (inflow), blueshifted emission (inflow) and redshifted emission (outflow). All of these are highlighted in Figure \ref{fig:naddoppler}. However, we note that the line (amplitude) S/N ratio of absorption and emission are significantly different; absorption signatures generally have S/N ratios larger than 10, whilst emission signatures have ratios less than 10. This is important because it means emission signatures are more sensitive to noise and errors in continuum fitting, as well as residuals from fits to the He\,I line immediately blueward of \nad\ . For these reasons, we consider only blueshifted absorption, redshifted absorption, and redshifted emission as signatures of flows, since blueshifted emission is highly sensitive to a larger number of residuals and noise, and therefore much less reliable.

\subsection{Bayesian inference and \texorpdfstring{\nad}{TEXT} profile fitting}
\label{sec:nadfitting}
Prior to modeling the residual \nad\ profile, we fit a first-order polynomial to the flux immediately blueward and redward of the profile and divide the residual by this, to account for any systematic continuum-fitting errors that could give rise to artificial residuals. After this, we are free to fit the ISM residual of \nad\ with an analytical expression.
Because multiple components may contribute to the \nad\ signal, many degeneracies exist in the profile fitting.
For this reason, we employ a Bayesian inference 
approach using PyMultinest \citep{buchner14}, a Python wrapper for the popular nested sampling code, Multinest \citep{feroz09}. 
We make the assumption that our posteriors 
follow a Gaussian distribution and that our data points are 
uncorrelated. 
For our \nad\ modeling, we use the 
analytical function described by \citet{rupke05a}. The model follows the form

\begin{equation}
\label{eq:int}
I(\lambda) = 1 - C_{f} + C_{f} \times e^{-\tau_{\text{B}}(\lambda)-\tau_{\text{R}}(\lambda)} ,
\end{equation}

where C$_{f}$ is the velocity-independent covering factor, and $\tau_{\text{B}}(\lambda)$ and $\tau_{\text{R}}(\lambda)$ 
are the optical depths of the Na\,I\,$\lambda$\,5891 and Na\,I\,$\lambda$\,5897 lines, respectively. The optical depth of the line, $\tau(\lambda)$, can be expressed as

\begin{equation}
\label{eq:optic}
\tau(\lambda) = \tau_{0} \times e^{-(\lambda - \lambda_{0} + \Delta\lambda_{\text{offset}})^{2}/((\lambda_{0} + \Delta\lambda_{\text{offset}}) b_{\text{D}}/c)^{2}} ,
\end{equation}

where $\tau_{0}$ and $\lambda_{0}$ are the central optical 
depth and central wavelength of each line component, $b_{\text{D}}$ is the Doppler line width, and $c$ is the speed 
of light. The wavelength offset is converted from a velocity offset, given $\Delta\lambda_{\text{offset}}$=$\Delta$v$\cdot\lambda_{0}$/$c$.
For \nad\, $\tau_{0,\text{\text{B}}}/\tau_{0,\text{\text{R}}}=2$ \citep{morton91}, meaning the Na\,I\,$\lambda$\,5891 line has twice the depth of the Na\,I\,$\lambda$\,5897 line. The optical depth parameter can be derived from the column density of Sodium, which is described as

\begin{equation}
\label{eq:nad}
N(\text{Na\,I}) = \frac{\tau_{0}\,b}{1.497\times10^{-15}\lambda_{0}f} ,\quad [\text{cm}^{-2}],
\end{equation}

where $\lambda_{0}$ and $f$ are the rest frame wavelength (vacuum) and oscillator strength. 
Throughout this study we assume $\lambda_{0}$=5897.55\,\AA\ and $f$=0.318 \citep{morton91}.
To determine if an outflow is present, we first fit the \nad\ line with the model given in Eq. \ref{eq:nad}: once assuming $\Delta\lambda_{\text{offset}}$=0, the second time leaving it as a free parameter. If the model with $\Delta\lambda_{\text{offfset}}\neq$0, is preferred, then a flow is present. To determine this, we use a Bayesian Information Criterion (BIC), which makes use of the likelihood for each model but penalizes for additional free parameters. The BIC is defined as 

\begin{equation}
\label{eq:BIC}
\text{BIC} = -2\mathcal{L} + k\cdot\text{log}(N),
\end{equation}

where $\mathcal{L}$ is the log-likelihood, $k$ is the number of free parameters and N is the number of data points that get fit. In addition, a minimum velocity offset $\Delta$v is required to confirm a flow detection. We discuss the minimum required BIC$_{\text{fixed}}$/BIC$_{\text{offset}}$ ratio (K) and $\Delta$v in Section \ref{subsec:complete}. If a flow is detected, we characterize the total profile of the line with a two-component model. In the case of an outflow detection, the profile is also fit with a three-component model consisting of a systemic component (in absorption), a blueshifted absorption component offset by a minimum of -20 km\,s$^{-1}$ and a redshifted emission component with 20<$\Delta$v$_{\text{offset}}$<200 km\,s$^{-1}$, in order to be consistent with the findings of \citet{prochaska11}. All three profiles are allowed a maximum linewidth of 200 km\,s$^{-1}$. These priors are chosen so as to restrict the redshifted emission to near-systemic velocities (e.g., \citealt{prochaska11}) and prevent unrealistically large absorption and emission profiles which overfit the data and try to cancel each other out (e.g., \citealt{veilleux13}). A BIC ratio determines the preferred model out of the two, and the final fit is selected accordingly.

The allowed ranges for the parameters in Eq. \ref{eq:nad} are separated in two categories, "detection" and "characterization", with the former being slightly more restrictive in linewidth and velocity offset compared to the latter. These are presented in Table \ref{tab:priors}. The "detection" ranges apply to single-component fits and the "characterization" ranges to multiple-component fits. The reasons for this are a) we want to limit the amount of degenerate and unrealistic fits that are allowed in the determination of flow detections: e.g., a flow detection could be determined by unrealistically large linewidths and/or velocity offsets that attempt to fit noise or baseline residuals, and b) once a robust detection is found we wish to sample a large parameter range to ensure both the systemic and flow components are well described.

The above procedure works well for profiles of \nad\ excess in absorption and emission. In Section \ref{subsec:ismprofiles} we show that we also find excess in the form of P-Cygni profiles, which are an unambiguous detection of outflows. We are unable to accurately model a systemic galaxy component for such profiles, so we fit the profile with single blueshifted and redshifted double-Gaussian components (e.g., \citealt{veilleux13}). This limitation means we gain no physical information about the state of the gas apart from the velocity shift, which itself can be underestimated as the central wavelength of the Doppler shift will be restricted by the influence of its blueshifted/redshifted counterpart. We therefore assume such outflow velocities to be lower limits and exclude them from our analyses which require information on the physical state of the gas.

\begin{table}
  \centering
  \caption{The priors applied to our model when used for detection and characterization purposes. Note that for emission profiles the covering factor prior changes to -1 $\leqslant$ C$_{f}$ $\leqslant$ 0. The free parameters are: |C$_{f}$|, absolute covering fraction; $b_{\text{D}}$, Doppler linewidth in km/s; log N(Na\,I), column density in cm$^{-1}$; |$\Delta$v$_{\text{offset}}$|, absolute velocity offset in km/s.}
  \label{tab:priors}
  \begin{tabular}{llll}
      \hline
        Parameter & Priors & \multicolumn{2}{c}{Priors} \\
         & (detection) & \multicolumn{2}{c}{(characterization)} \\
        \hline
         &  & \textbf{Systemic} & \textbf{Flow} \\
        \hline
        |C$_{f}$| & 0-1 & 0-1 & 0-1 \\
        $b_{\text{D}}$ [km/s] & 20-300 & 50-450 & 50-450 \\
        log N(Na\,I) [cm$^{-1}$] & 9-15.3 & 9-15.3 & 9-15.3 \\
        |$\Delta$v$_{\text{offset}}$| [km/s] & 0-200 & -- & 0-500 \\
        \hline \\
  \end{tabular}
\end{table}

\subsection{Model Completeness and Reliability}
\label{subsec:complete}
\begin{figure*}
 \includegraphics[width=0.8\textwidth]{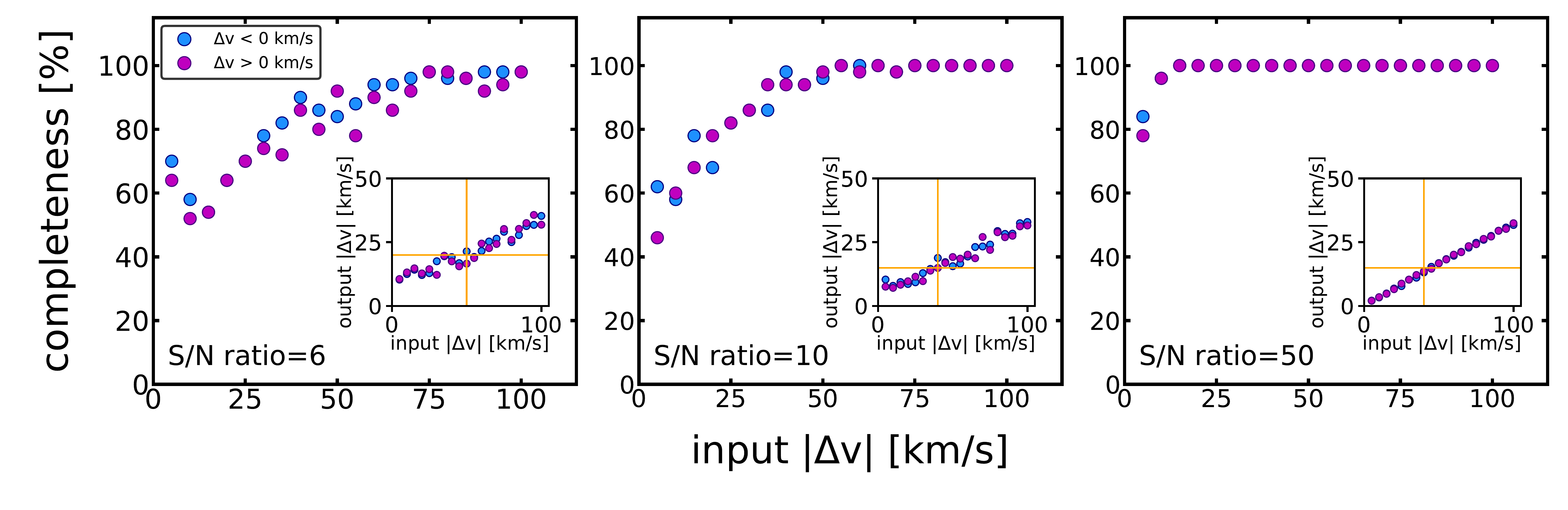}
 \caption{Plots of completeness versus input velocity for our detection procedure. The procedure is calculated for line S/N ratios of 6 (left), 10 (middle) and 50 (right), for both negative offset velocities (blue circles) characteristic of outflows and positive offset velocities (magenta circles) characteristic of inflows.
Each inset plot shows the linear evolution of |$\Delta$v|$_{\text{output}}$ vs |$\Delta$v|$_{\text{input}}$ in each completeness plot.}
 \label{fig:completeness}
\end{figure*}

It is fundamental that the limitations of our fitting models and procedures be understood, and their completeness and reliability quantified.
For completeness, we generate synthetic spectra consisting of systemic and offset components. The velocities of the offset components range from -100 to 100 km\,s$^{-1}$ in 5 km\,s$^{-1}$ intervals. Each profile is convolved to the FWHM resolution of SDSS with a Gaussian function, before adding random Gaussian noise. This is repeated for three different S/N ratios of 6, 10 and 50 at each velocity offset. The spectrum is then fitted according to our detection technique described in Section \ref{sec:nadfitting} and the measured blue/redshifted velocity recorded. To ensure the result is not dependent on the random noise added to the spectrum, we repeat this sequence 50 times for each S/N ratio, each with different random Gaussian noise. The completeness is defined as the fraction of recovered non-zero velocity offsets as a function of input $\Delta$v, for each S/N ratio. This is shown in Figure \ref{fig:completeness}. In the inset plots of Figure \ref{fig:completeness} we show the $measured$ $\Delta$v as a function of input $\Delta$v for each completeness plot. Based on these results, we adopt a |$\Delta$v|$_{\text{input}}$ threshold of 40 km\,s$^{-1}$ for line S/N ratios greater than 10 and 50 km\,s$^{-1}$ for line S/N ratios less or equal to 10, which corresponds to >90\,\% and $\sim$85\% completeness, respectively. This translates to |$\Delta$v|$_{\text{output}}$=15 km\,s$^{-1}$ and |$\Delta$v|$_{\text{output}}$=20 km\,s$^{-1}$, respectively, using the linear |$\Delta$v| evolution shown in the inset plots of Figure \ref{fig:completeness}.

Arguably the most important test, however, is the reliability of our detections, since there are a number of factors that could mimic a Doppler shift: the main culprit of this would likely be ISM residuals or artifacts created from bad continuum-fitting. For this test, we use our HIGH-$i$ sample (for both inactive galaxies and AGN hosts), which we assume will not display outflows due to unfavorable inclinations. We define bins of SFR-M$_{*}$ and create 50 stacked bootstrap samples for each bin, in the same fashion as described in Section \ref{sec:sdssstacking}. Each stacked spectrum is then fitted with pPXF and the \nad\ residual put through our detection procedure and all $measured$ (output) $\Delta$vs are recorded. Since these stacks are meant to represent spectra with no outflow signatures, all detections are considered false positives. The reliability of each bin is defined as the difference between a perfect case of no false-positives (100\% reliability) and the percentual number of false positives detected out of the 50 stacks, allowed by a set of selection criteria.
Our selection criteria should rely on a combination of thresholds given by a K ratio, a minimum measured $\Delta$v, and a quantity to guard against residuals left from bad continuum fitting. For this latter consideration, we look at the Mg\,$b$ absorption residuals, since they are stellar in origin. For the minimum measured velocity we use |$\Delta$v|$_{\text{output}}$>15 km\,s$^{-1}$ and |$\Delta$v|$_{\text{output}}$>20 km\,s$^{-1}$, as derived from our completeness tests and K>1. The reliability for our samples based on these criteria remains above 85\% over the whole plane.
Although our criteria do a good job of guarding against false positive outflow detections, we note that these tests cannot be performed in the same manner for inflowing gas since we have no \textit{a priori} information on their angle of incidence. However, the selected thresholds should also limit the number of false positive inflow detections, since the main culprit for these would be bad continuum fitting, which we account for.

\section{Stacking results}
\label{sec:results}
\subsection{\texorpdfstring{\nad}{TEXT} profiles across the \texorpdfstring{SFR-M$_{*}$}{TEXT} plane}
\label{subsec:profiles}
\begin{figure}
 \includegraphics[width=1.\columnwidth]{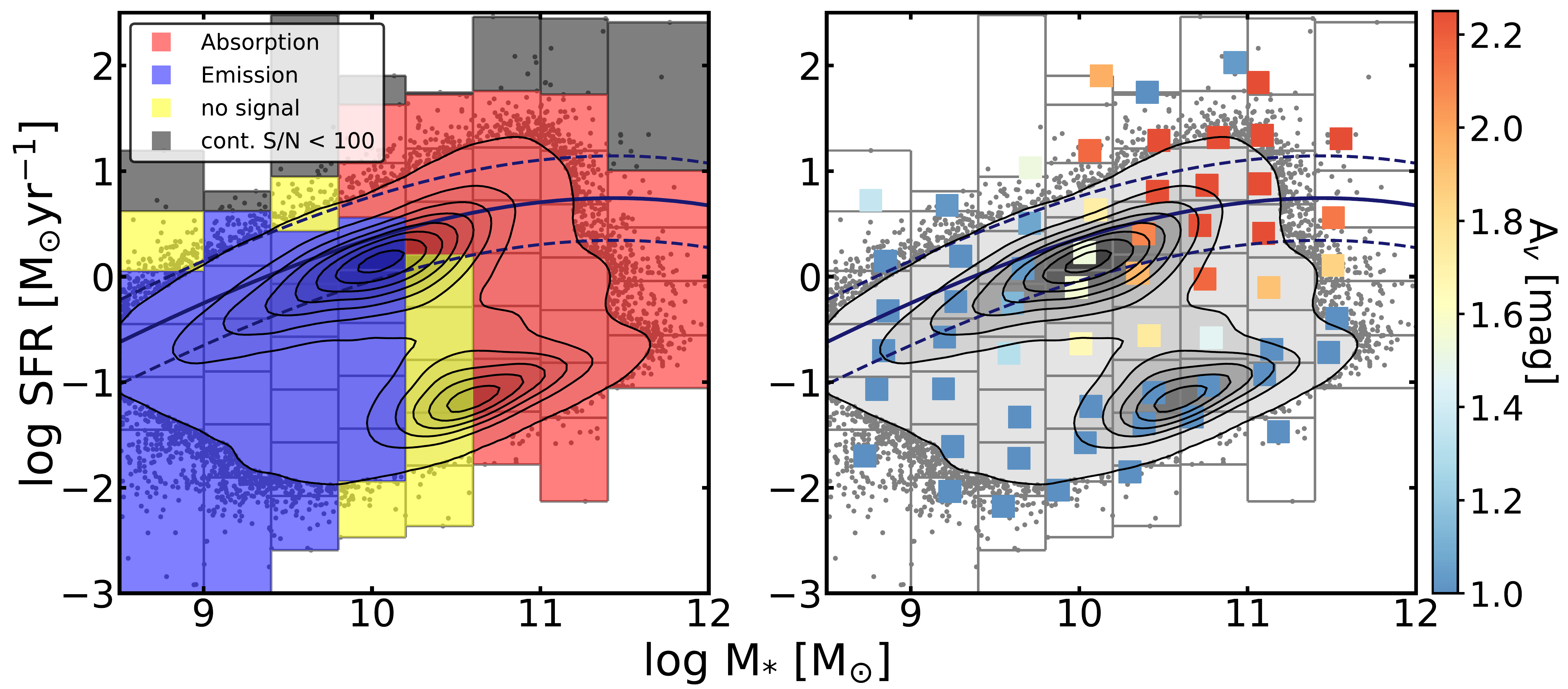}
 \caption{Left:The \nad\ residual profiles for inactive galaxies from our stacked spectra over the main sample SFR-M$_{*}$ plane, as a result of the division of the best fit continuum given by pPXF. Absorption and emission profiles dominate the high- and low-mass galaxies, respectively, with a separation at log\,M$_{*}$/M$_{\odot}\sim$10-10.5 characterised by low line S/N ratios and P-Cygni profiles. The solid and dashed lines mark the main sequence relation defined by \citet{saintonge16}, with a +0.35 dex offset in log SFR to account for the different median redshifts of our and their sample. Right: The same plots as the left but with the mean dust A$_{v}$ values for each stack, highlighting correlation between the dust content and the \nad\ residual profile.}
 \label{fig:ismprofiles}
\end{figure}

The profiles (absorption, emission, P-Cygni or unknown) of the \nad\ residual in each stack are identified via visual examination, and reveal a stark bimodality in type occurring between low mass (log M$_{*}$/M$_{\odot}$ < 10) and high mass (log M$_{*}$/M$_{\odot}$ > 10) galaxies, with the former showing average profiles in emission and the latter in absorption. A few profiles at log M$_{*}$/M$_{\odot}\sim$10-10.5 have near zero line amplitude or show a P-Cygni profile. This distribution is shown in Figure \ref{fig:ismprofiles} and is similar to the distribution of Sodium excess found by \citet{concas17}. 
The change in \nad\ profile type with stellar mass is most likely attributed to the nebular dust attenuation in each stack. It is well known that \nad\ has a low ionizing potential (5.14 eV) and therefore requires dust shielding and high gas filling factors in the ISM for its survival. At high mass, galaxies have sufficient amounts of dust to allow \nad\ to survive and therefore absorb incident photons. Inclination can also play an important effect, since highly inclined galaxies are viewed along the plane of the disk, with an increased quantity of intervening dust. The exception to these rules are red sequence galaxies below the main sequence, which have low dust contents and filling factors yet still show profiles in absorption. We find that in such cases the EW of the \nad\ residual correlates with the Mg\,$b$ residual, and is therefore attributed to template mismatch - we do not consider flow detections in these galaxies.

At low masses (log M$_{*}$/M$_{\odot}\lesssim$10-10.5) the \nad\ profile is seen in emission. The reasons for this are not fully understood. As discussed in \citet{chen10}, this could be due to a template mismatch in the continuum fitting. Whilst we cannot completely rule out this possibility, we greatly reduce such a risk by constructing very high S/N stacked spectra and by checking the quality of our continuum fits through the Mg\,$b$ residual. Another possibility is that the emission excess is caused by our choice of SSP models and continuum fitting code. In Appendix \ref{sec:SSPcompare}, however, we demonstrate that the strong log M$_{*}$/M$_{\odot}$ dependence on the ISM profile is reproduced using several different codes and SSP models. Finally, due to the fact it is possible to observe \nad\ in emission, we must also consider that these profiles are real: a decrease in continuum intensity with stellar mass results in a reduced absorption line profile, since absorbed photons become scarce. However, a non-negligible fraction of the re-emitted photons still make their way to the observer to "fill in" \citep{martin12} the near-absent absorption. The resulting net profile can therefore be in emission if enough photons are re-emitted along the line of sight. Since absorption occurs along many other sightlines from the galaxy, re-emission from these can also be scattered toward the observer and contribute to the line profile. Such an effect is only seen in cases of scarce absorption at low mass (due to a weaker continuum). This is also observed with other resonant absorption lines (e.g., \citealt{martin12}). Over all SFR-M$_{*}$ stacks for inactive and AGN galaxies we find absorption profiles in $\sim$50\% of all high continuum S/N bins, emission profiles in $\sim$35\%, P-Cygni profiles in $\sim$0.3\%, and $\sim$14\% of profiles are classified as 'unknown'.


\subsection{Flow detection rates and inclination dependence}
\label{subsec:detections}
\subsubsection{$i$-$\Sigma_{\text{SFR}}$ plane}
\label{subsubsec:ivsfrd}
\begin{figure}
 \includegraphics[width=\columnwidth]{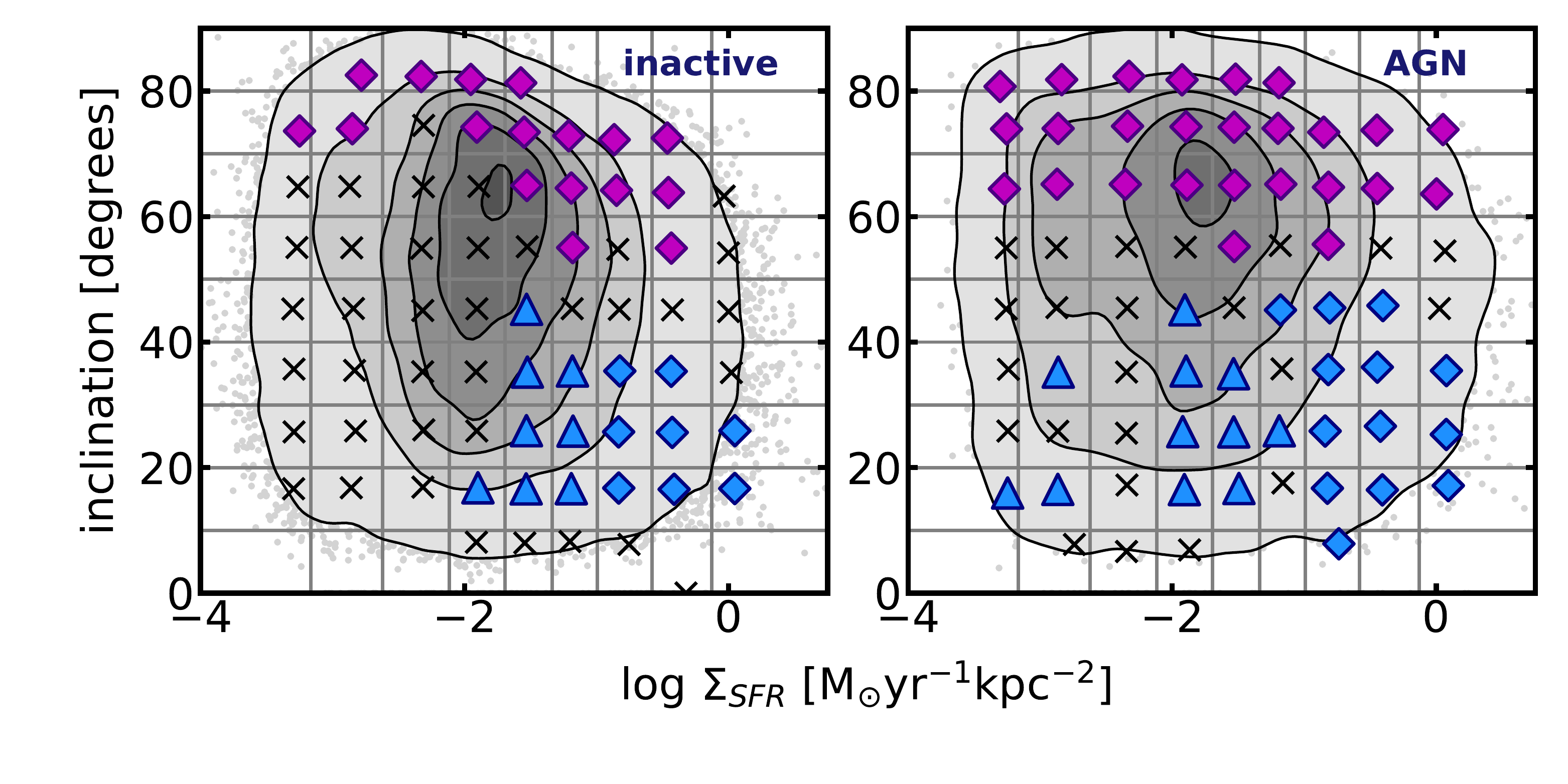}
 \caption{The inclination and $\Sigma_{\text{SFR}}$ dependence of inflows and outflows for our DISK sample. The left and right panels shows the results for the inactive and AGN objects, respectively. The symbol and color convention follow those of Figure \ref{fig:incdep}.}
 \label{fig:incvsfrd}
\end{figure}

Many studies have found strong dependencies of outflow detection rates on $\Sigma_{\text{SFR}}$ \citep{heckman2000} and inclination (e.g., \citealt{chen10, concas17}). To test the prevalence of inflows and outflows in our sample, we therefore begin by analyzing the \nad\ ISM component in bins of $i$-$\Sigma_{\text{SFR}}$ for the DISK sample. The results of the stacks are shown in Figure \ref{fig:incvsfrd} for inactive galaxies and AGN-hosts.
We also observe a clear dependence of outflow detections on $\Sigma_{\text{SFR}}$ and inclination: outflows are found most prominently in face-on systems that are characterised by low inclinations ($i$<50$^{\circ}$) and high $\Sigma_{\text{SFR}}$s. \citet{heckman2000} found outflows to be ubiquitous above a threshold of $\Sigma_{\text{SFR}}$> 0.1\,M$_{\odot}$yr$^{-1}$kpc$^{-2}$ and with low ($i$<60$^{\circ}$) inclinations. Our results decrease the former threshold by an order of magnitude (to 0.01\,M$_{\odot}$yr$^{-1}$kpc$^{-2}$) and reduce the latter to $i$<50$^{\circ}$ (in agreement with results found by \citealt{concas17}). We measure the detection rate of outflows as the number of bins with detections divided by the total number of bins in a sample or set of thresholds. The detection rate over these thresholds is 74\% (inactive and AGN). All detections with $\Sigma_{\text{SFR}}$>0.1\,M$_{\odot}$yr$^{-1}$kpc$^{-2}$ are characterised by profiles in absorption, whilst those with 0.01<$\Sigma_{\text{SFR}}$<0.1\,M$_{\odot}$yr$^{-1}$kpc$^{-2}$ are found in emission or via P-Cygni profiles, highlighting the necessity to consider all sources of Doppler shifted gas. 



We also find a large number of inflow detections in regions of high inclinations ($i$>50$^{\circ}$) and a large range of $\Sigma_{\text{SFR}}$s, with a detection rate that mildly increases with higher $\Sigma_{\text{SFR}}$s. Such a clear inclination dependence for inflowing gas has not been seen before, with several studies claiming contrasting results: e.g., \citet{rubin12} found that out of a sample of six disk-like galaxies, five displayed inflow at signatures at high inclinations ($i$>55$^{\circ}$), yet \citet{martin12} reported that out of four galaxies reporting inflows, only one had a similarly high inclination ($i\sim$61$^{\circ}$) and the remaining three had low inclinations ($i$<55$^{\circ}$; \citealt{kornei12}).

The properties of the $i$-$\Sigma_{\text{SFR}}$ detections are discussed throughout the rest of Section \ref{sec:results}, although due to the slightly uncertain nature of the detections in emission, we focus only on detections in absorption, where the nature of the residual is better understood..

\subsubsection{SFR-M$_{*}$ plane}
\label{subsec:MSplane}
\begin{figure*}
 \includegraphics[width=\textwidth]{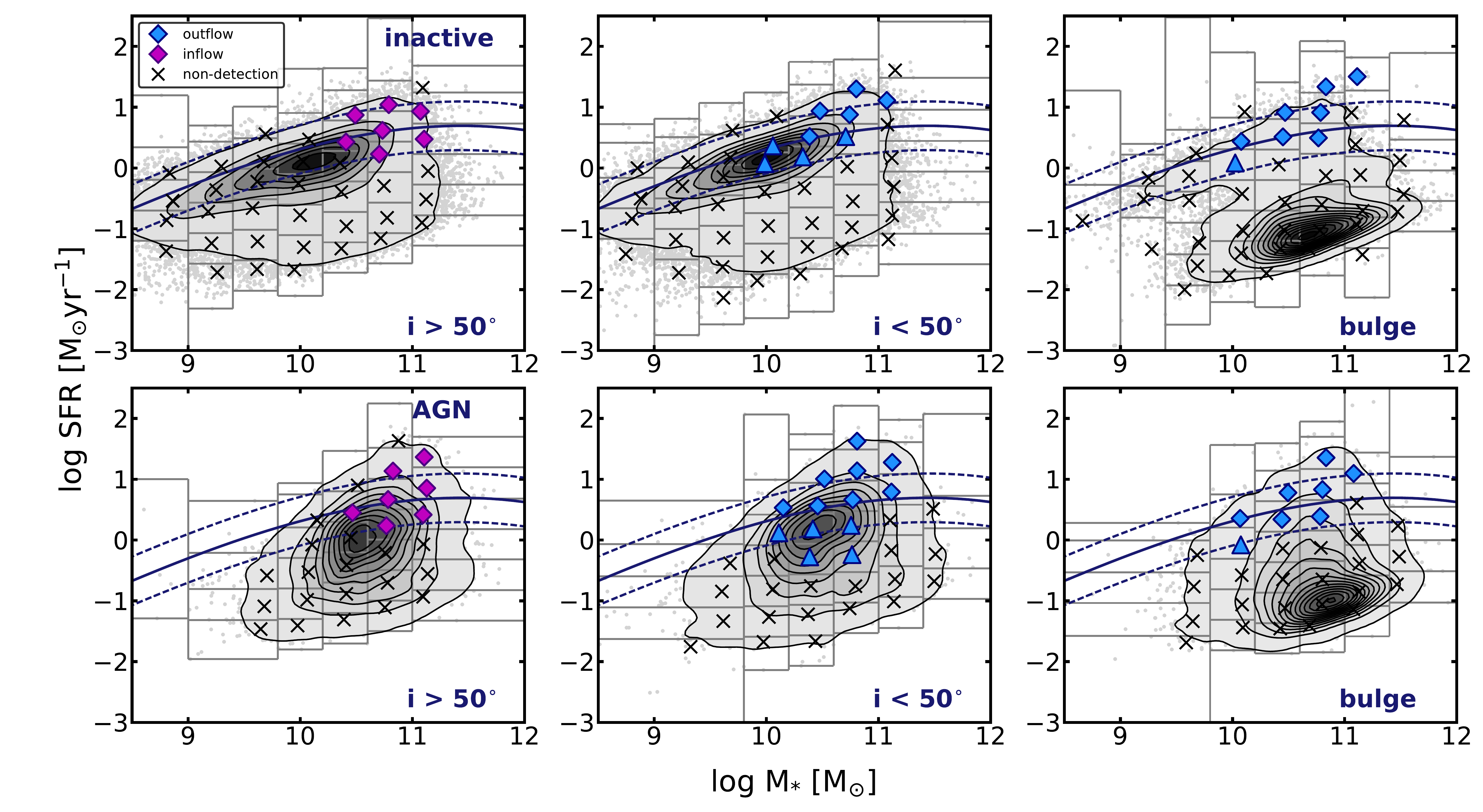}
 \caption{The detections of inflows and outflows across the SFR-M$_{*}$ plane for disk galaxies (left and middle columns) separated by inclination, and bulge galaxies for which inclinations cannot be accurately established (right column). The top row shows shows our sample of inactive galaxies whilst the bottom row is for AGN hosts.}
 \label{fig:incdep}
\end{figure*}

From the above results, we can now repeat our analysis over the SFR-M$_{*}$ plane for our samples of disk galaxies with inclinations less or greater than $i$=50$^{\circ}$, and bulge-dominated objects. The results for these are shown in Figure \ref{fig:incdep}. Similarly to our findings over the $i$-$\Sigma_{\text{SFR}}$ plane, we find a high number of outflow detections in star-forming regions (log SFR$\gtrsim$0\,M$_{\odot}$yr$^{-1}$) of high mass (log M$_{*}$/M$_{\odot}\gtrsim$10) galaxies with low inclinations. Detections are found in absorption, emission and in P-Cygni profiles. No outflow detections are found in low mass (log M$_{*}$/M$_{\odot}\lesssim$10) galaxies or galaxies with high inclinations. This applies to both inactive galaxies and AGN hosts. The detection rates and median galaxy-host properties of our detections are shown in Table \ref{tab:mainseqtab}, whilst the properties of the gas flows are presented in Table \ref{tab:flowpropstab} and Table \ref{tab:flowpropstab2}. For bins with log SFR>-0.5\,M$_{\odot}$yr$^{-1}$ - which roughly coincides with the lower limit of the star-forming main sequence at low mass - over our LOW-$i$ and BULGE samples, we find an outflow detection rate of 53.5\%.

We find detections of inflows in star-forming galaxies with high inclinations. No inflow detections are found in low inclination galaxies or bulge-dominated galaxies. If we apply the same SFR lower limit as above to the HIGH-$i$ sample, we find an inflow detection rate of 43.7\% for inactive galaxies and AGN.

\begin{table}
  \caption{The detection rate of inflows and outflows in the SFR-M$_{*}$ plane across all bins with continuum S/N>100, and median properties of their galaxy hosts.}
  \label{tab:mainseqtab}
  \begin{tabular}{lccccc}
      \hline
     & \multicolumn{2}{c}{\bf{Inactive}} & \multicolumn{2}{c}{\bf{AGN}} \\
        Sample & Inflows & Outflows & Inflows & Outflows \\
        \hline
    	HIGH-$i$ & 18\% & 0\% & 26\% & 0\% \\
        LOW-$i$ & 0\% & 20\% & 0\% & 39\% \\
        BULGE & 0\% & 19\% & 0\% & 26\% \\
        \hline
        \multicolumn{5}{c}{Median Properties of host Galaxy}
    \\[0.15ex]
    \hline
    \\[-2.75ex]
    SFR (M$_{\odot}$yr$^{-1}$) & 4.15 & 3.29 & 4.64 & 3.63 \\
    $\Sigma_{\text{SFR}}$ (M$_{\odot}$yr$^{-1}$kpc$^{-2}$) & 0.09 & 0.21 & 0.12 & 0.11 \\
    M$_{*}$ (log M$_{\odot}$) & 10.73 & 10.47 & 10.82 & 10.75 \\
    nebular A$_{\text{v}}$ (mag) & 2.94 & 2.20 & 3.22 & 2.35 \\
    D$_{\text{n}}$4000 & 1.42 & 1.30 & 1.45 & 1.35 \\
    Concentration Index & 2.45 & 2.53 & 2.54 & 2.43 \\
    \hline
  \end{tabular}
\end{table}

\subsection{Covering fractions}
The covering fraction of the flow, C$_{f}$, is a measure of the local clumpiness of the gas along the line of sight.
In Table \ref{tab:flowpropstab} and Table \ref{tab:flowpropstab2} we report the covering fractions determined by our analysis for inactive galaxies and AGN-hosts, respectively. For each of our samples, the covering fractions span the full range of allowed values and there appears to be no difference between the covering fractions of outflows and inflows. We note, however, that in many cases we also find flows characterized by very low covering fractions, |C$_{f}|\lesssim$0.25. Unlike for point sources at high redshift where the gas completely covers the background source, for low redshift sources where the background source subtends a large angle on the sky, a covering fraction less than unity is not unexpected. However, such low covering fractions are likely not a result of geometry alone. Very low fractions have also been observed by \citet{chen10} who stack over similar samples of galaxies. One explanation to describe such low values is that we are observing small amounts of neutral \nad\ gas with low dust shielding in very dense clouds within the outflow, where ionizing radiation and shocks no longer dominate.

\subsection{Equivalent-widths}
\label{subsec:ismprofiles}
\begin{figure}
 \centering
 \includegraphics[width=\columnwidth]{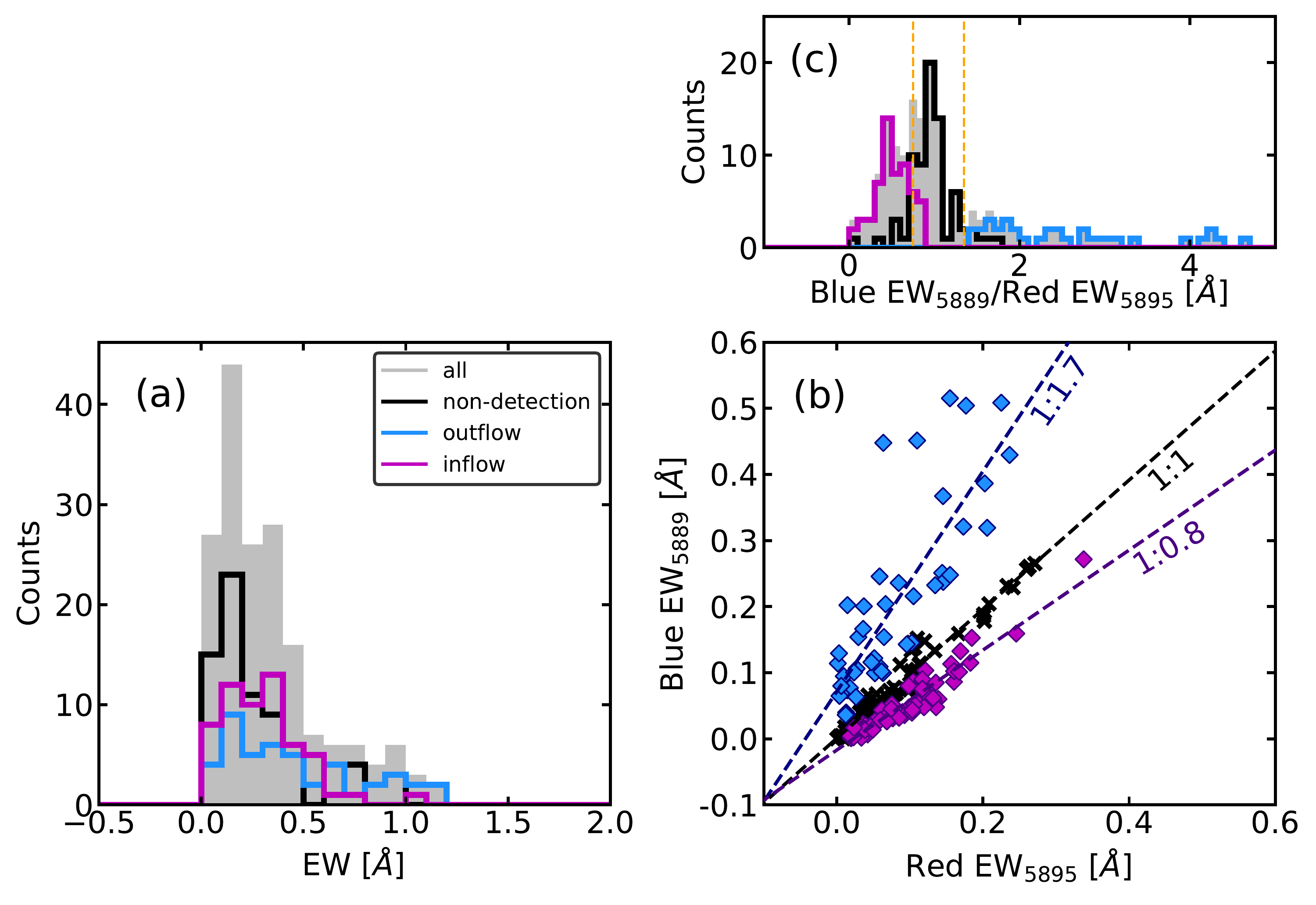}
 \caption{(a): The distribution of total EWs for inactive galaxies and AGN. The gray bars represent the full distribution, whilst the blue bars represent outflow detections and inflows are marked by the magenta bars. Black bars are non-detections. (b): A comparison of the EWs measured blueward and redward of the 5889\,$\mbox{\AA}$ and 5895\,$\mbox{\AA}$ Na\,I lines, where outflows (blue) and inflows (magenta) are expected to be seen, respectively. The dotted black line is a straight line fit to the systemic components (black x's) whilst the dashed blue and magenta lines are best fit linear functions to outflow and inflow detections. (c) histogram distributions of the EW$_{\text{blue}}$/EW$_{\text{red}}$ ratio for inflows, outflows and non-detections. The orange vertical lines represent our suggested cuts to isolate each distribution.}
 \label{fig:EW}
\end{figure}

The equivalent width (EW) of a line is a measure of its strength, and therefore can provide information about the strength of an outflow or inflow. Figure \ref{fig:EW} plots the distributions of EWs measured from fits to our \nad\ absorption profiles over the \textit{i}-$\Sigma_{\text{SFR}}$ and SFR-M$_{*}$ planes. The measurements are also presented in Table \ref{tab:flowpropstab} and Table \ref{tab:flowpropstab2}. Figure \ref{fig:EW}a plots the distribution of the \textit{total} line EW. We report a narrow range of outflow EWs (0<EW$_{\text{\nad}}$<1.2) for the combined inactive and AGN samples, with a median 0.24\,\AA\ and a standard deviation 0.26\,\AA. These values are similar to those found by \citet{chen10} but have a median which is an order of magnitude smaller than that found by \citet{rupke05b}, who report a median of 3.3\,\AA\ and a maximum of value 9.1\,\AA\ for \nad\ in (U)LIRGs. The higher values found by \citet{rupke05b} are most likely attributed to the increased column densities found in their samples, whilst \citet{chen10} study galaxies more closely matched to this sample. We note that our AGN sample have a slightly higher median value of 0.29\,\AA\ compared to 0.25\,\AA\ for inactive galaxies, and a higher maximum value of 1.2\,\AA\ (AGN) compared to 0.93\,\AA\ (inactive). A Kolmogorov-Smirnov (K-S) test between the two distributions, however, reveals a low coefficient of 0.16 suggesting the two distributions are still very similar.

The difference in reported values between \citealt{rupke05b} and the distributions in Figure \ref{fig:EW}a clearly illustrate the difference in outflow strength between normal galaxies and more extreme objects. By preselecting systems with large \nad\ residuals, it is likely that a large number of weaker outflow signatures would be overlooked.
In Figure \ref{fig:EW}b we show an alternative measurement of EW, where we plot the EW of the flux blueward of the Na\,I\,5889\,\AA\ line vs the EW of the flux redward of the Na\,I\,5895\,\AA\ line. A clear separation of outflow detections, inflow detections, and non-detections becomes evident, which is not apparent from measurements of the \textit{total} EW of the \nad\ doublet. A histogram of the EW$_{\text{blue}}$/EW$_{\text{red}}$ ratio is shown in Figure \ref{fig:EW}c and three distinct distributions appear. By applying a cut of EW$_{\text{blue}}$/EW$_{\text{red}}$>1.35 for outflows and a cut of EW$_{\text{blue}}$/EW$_{\text{red}}$<0.75 for inflows, one selects 100\% of outflow detections and 86\% of inflow detections, with only $\sim$10\% contamination from the non-detections (subject to uncertainties in the EW measurements). Using "edge-EWs" instead of the total EW of \nad\ provides a more complete and unbiased way to select potential outflow candidates.

\subsection{Flow velocities}
\label{subsec:flowvel}
\begin{figure*}
 \includegraphics[width=\textwidth]{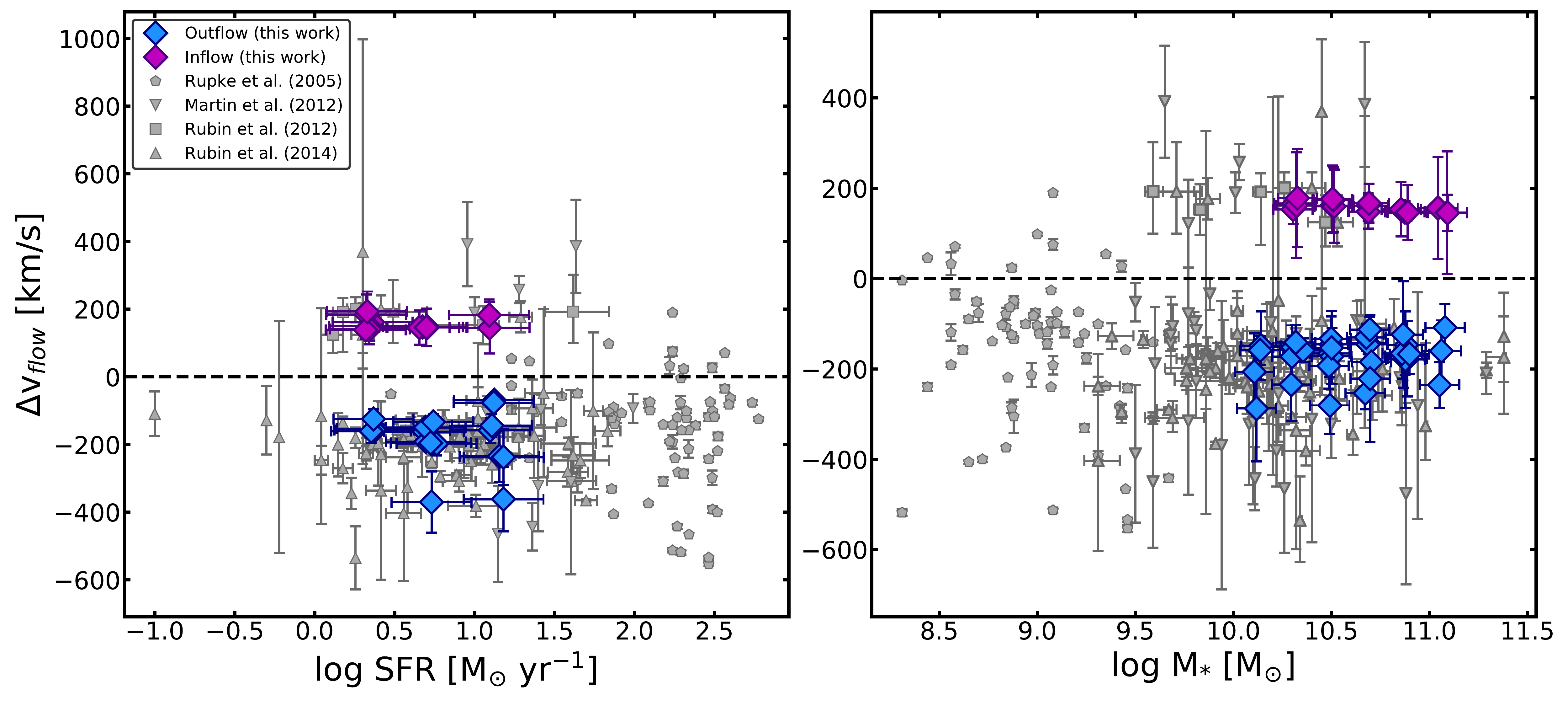}
 \caption{Left: The central inflow and outflow velocities (uncorrected for inclination) as a function of SFR. Right: The same as the left panel but as a function of stellar mass. Results from \citet{rupke05a,rupke05b}, \citet{martin12}, \citet{rubin12} and \citet{rubin14} are overplotted for comparison, where available. The symbol and color convention follow those of Figure \ref{fig:incdep}.}
 \label{fig:dvs}
\end{figure*}

The central velocity of a flow is a measure of the velocity at which the bulk of the material is traveling. In Figure \ref{fig:dvs} we plot the central velocity measurements of inflow and outflow detections in absorption as a function of global galaxy properties, and compare them to results in the literature which study samples of outflows in galaxies at $z$<1. The stacks shown in these plots are created from a sample of high mass (log M$_{*}$/M$_{\odot}$>10) and high $\Sigma_{\text{SFR}}$ ($\Sigma_{\text{SFR}}$s > 0.1 M$_{\odot}$yr$^{-1}$kpc$^{-2}$) DISK galaxies, since in Sections \ref{subsec:detections} and \ref{subsec:MSplane} we have shown these thresholds to be important in finding outflows.
The left panel of Figure \ref{fig:dvs} shows stacks binned by \textit{i}-log\,SFR, whilst the points in the right panel are binned by \textit{i}-log\,M$_{*}$.
   
We report absolute outflow velocities in the range 69-370 km/s with a median of 160 km/s, consistent with results for samples of normal star-forming galaxies (e.g., \citealt{chen10,martin12,rubin12,rubin14}). Our reported values are not characteristic of particularly high outflow velocities compared to some cases of extreme starburst or AGN hosts, which are able to launch $\sim$1000-2000 km/s outflows with different gas phases (e.g., \citealt{tremonti07,chung11,cicone14,carniani15}). We find no significant difference between outflow velocities from the inactive sample compared to the AGN hosts: we report medians of 155 km/s (inactive) and 167 km/s (AGN), with maximum central velocities of 234 km/s (inactive) and 370 km/s (AGN). This suggests that whilst the presence of an optically-selected AGN might slightly enhance an outflow's velocity, it does not do so by a significant amount.

In the left panel, we see there appears to be little to no correlation of outflow velocity with total SFR (unlike in eg., \citealt{heckman15}) within our sample, although the scatter appears to increase with SFR. We also note that a correlation may not appear present due to the small range of SFRs probed by our stacks, which also appears to be the case in \citet{chen10} for a near identical sample and SFR range. In the right panel we also find little to no correlation between outflow velocity and increasing stellar mass.
Inflow velocities are also consistent with the results from the studies mentioned above, spanning a range 139-193 km/s with a median central velocity of 151 km/s. Only a $\sim$6 km/s difference exists between the median inactive and AGN inflow velocities. Furthermore, we find no correlations of velocity with SFR or stellar mass.

It is important to note that none of these velocities have been corrected for inclination, and as such they may be (and are likely to be) underestimated (we observe a difference of $\sim$20-30 km\,s$^{-1}$ between the inclination-corrected and uncorrected median outflow velocities in Figure \ref{fig:dvs}). Since the velocity offset is used in several calculations (e.g., the mass outflow rates in Section \ref{subsec:dMsec}), this underestimation is propagated throughout the analysis and therefore such outflow quantities serve as lower limits. We present the inclination-corrected velocities in Tables \ref{tab:flowpropstab} and \ref{tab:flowpropstab2}, however do not use these in our plots for the sole purpose of facilitating comparison with other results in the literature, who also use uncorrected velocities.

\subsection{Mass outflow rates and mass-loading factor}
\label{subsec:dMsec}
Two of the most important quantities one can derive in studies of galactic-scale flows are the mass outflow rate ($\dot{M}_{\text{out}}$) and mass-loading factor ($\eta$), which describes the rate of mass ejected from the galaxy per unit of SFR. These measurements help quantify the rate at which galaxies are expelling mass and the extent to which they are able to quench the star formation. Before deriving these rates, however, there are several important assumptions to consider. For a spherically symmetric, mass conserving wind that travels at velocity $v$, the average mass flow rate across a radius $r$ can be expressed as the following:

\begin{equation}
\label{eq:massfloin}
\dot{M}_{\text{out}} = \Omega\,\mu\,m_{\text{H}}\,N(\text{H})\,v\,r\,,
\end{equation}

where $\Omega$ is the solid angle subtended by the wind at its origin (i.e., the global covering factor of the wind), m$_{\text{H}}$ is the mean atomic weight (with a $\mu$=1.4 correction for relative He abundance), N(H) is the column density of Hydrogen along the line of sight, $v$ is the central velocity of the wind, and r is the distance from the galaxy.

Here we make the same assumptions made by \citet{rupke05b} and refer the reader to their paper for details. In short, we assume a solid angle less than 4$\pi$, a radius of 5 kpc, and that the column density of Hydrogen can be expressed as 

\begin{equation}
\label{eq:Nh_column}
 N(\text{H}) = \frac{N(\text{Na\,I})}{\chi(\text{Na\,I})\,d(\text{Na\,I})\,Z(\text{Na\,I})} ,
\end{equation}
\\
where N(Na\,I) is the Sodium column density, $\chi$(Na\,I)=N(Na\,I)/N(Na) is the assumed ionization fraction, $d$(Na\,I) is the depletion onto dust, and Z(Na\,I) is the Na abundance. We assume a 90\% ionization fraction ($\chi$(Na\,I)=0.1), a Galactic value \citep{savage96} for the depletion onto dust (log\,$d$(Na\,I)=-0.95), and solar metallicity (Z(Na\,I)=log[N(Na)/N(H)]$_{\odot}$=-5.69). 
We report a wide distribution of total outflow column densities for our $i$-$\Sigma_{\text{SFR}}$ and SFR-M$_{*}$ stacks of 17.85 < log $N$(H)/cm$^{-2}$ < 21.98, with a median of 19.77 cm$^{-2}$. We observe little difference between the medians of the inactive objects (19.46 cm$^{-2}$) and AGN-hosts (19.89 cm$^{-2}$). These values are similar (albeit slightly lower) to those observed for (U)LIRGs at low-$z$ \citep{rupke05b,cazzoli16}. The distribution of column densities for the inflows is somewhat narrower and shifted towards lower values, with a range 18.94 < log $N$(H)/cm$^{-2}$ < 20.28 and median 19.60 cm$^{-2}$.

From the above assumptions, Equation \ref{eq:massfloin} can be expressed as 

\begin{equation}
\label{eq:finaldM}
\begin{split}
\dot{M}_{\text{out}} = 11.5\sum\,\bigg(\frac{C_{\Omega}}{0.4}\,C_{f}\bigg)\,&\bigg(\frac{r}{10\,\text{kpc}}\bigg) \\
&\times\bigg(\frac{N(\text{H})}{10^{21}\,\text{cm}^{-2}}\bigg)\bigg(\frac{|\Delta\,v|}{200\,\text{km\,s}^{-1}}\bigg) \,\text{M}_{\odot}\,\text{yr}^{-1},
\end{split}
\end{equation}

\begin{figure}
 \includegraphics[width=\columnwidth]{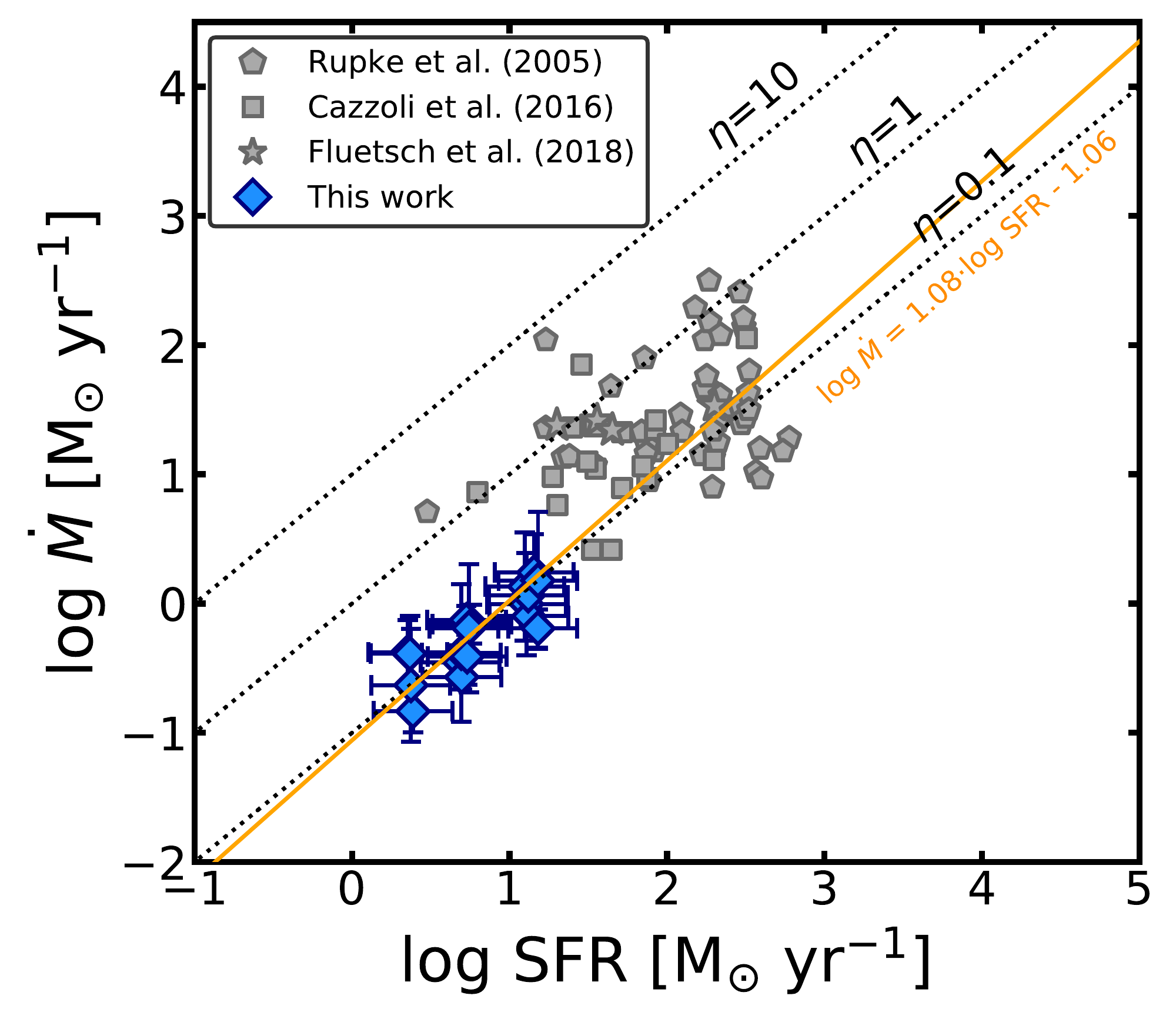}
 \caption{The mass outflow rates for the LOW-$i$ samples of inactive and AGN galaxies. A first-order polynomial fit to our data is shown in orange. Overplotted are the results from \citet{rupke05b} (gray pentagons) and \citet{fluetsch18} (blue stars), where available.}
 \label{fig:dmdt}
\end{figure}

Figure \ref{fig:dmdt} shows the derived mass outflow rates versus SFR for the \textit{i}-SFR stacks, and we compare these to the (U)LIRGs of \citet{rupke05b} and \citet{cazzoli16}, as well as the H\,II galaxies of \citet{fluetsch18}, who all use the same tracer and similar assumptions. All uncertainties associated with our calculated values incorporate those from the fit free parameters. We note that the main drivers of the mass outflow uncertainties are the covering factor and the assumed radius of the wind.
We report mass outflows rates of -0.83 $\lesssim$ log $\dot{M}_{\text{out}}$/M$_{\odot}$yr$^{-1}$ $\lesssim$ 0.24 for a SFR range of -0.16 $\lesssim$ log SFR/M$_{\odot}$yr$^{-1}$ $\lesssim$ 1.23. These rates are similar to those found by \citet{rubin12} for galaxies of similar stellar mass but slightly higher redshift ($z\sim$0.5). Our detections extend to lower SFRs. We find that mass loss rates are generally $\sim$10\% of the SFR and that a positive (linear) correlation between the two quantities exists with a near constant mass-loading factor $\eta\approx$0.1. The relation has a measured Pearson coefficient of r$_{\text{p}}$=0.85 using our data only and a slight decreased coefficient  r$_{\text{p}}$=0.77 when also using the results from \citet{rupke05b}, \citet{cazzoli16} and \citet{fluetsch18}. A first-order polynomial fit to our data returns 

\begin{equation}
\label{eq:massload}
\text{log}\,\dot{M}_{\text{out}} = (1.08\pm0.17)\cdot\text{log\,SFR} - (1.05\pm0.14).
\end{equation}

The near constancy of $\eta$ is perhaps surprising, however such a value is consistent with other low factors observed in studies of similar objects (e.g., \citealt{veilleux05,martin12,rubin14}), suggestive of low-$z$ starbursts and Milky Way-type galaxies being unable to drive strong winds (defined by high mass-loading factors). We also note a mean difference of 0.11\,M$_{\cdot}$\,yr$^{-1}$ between inclination-corrected outflow rates and the uncorrected rates presented above.

\subsection{Comparison to other gas phases}
In the previous section we derived mass outflow rates using the \nad\ tracer of neutral gas. However, this is only one gas phase and does not account for the molecular and ionized gas phases, which likely contribute non-negligible amounts of ejected gas. A direct and comprehensive comparison is challenging due to the lack of uniform datasets, however some studies have made notable attempts. Recently, \citet{fluetsch18} looked at molecular outflows with ALMA CO data and cross-matched their sample with optical data (where available). They found that for star-forming galaxies the ratio of molecular mass outflow rates ($\dot{M_{H_{2}}}$) to ionized ($\dot{M_{\text{ion}}}$) mass outflow rates was close to unity, whilst AGN-hosts displayed much stronger molecular mass outflow rates (correlating with AGN luminosity). Of particular interest to this study is the comparison of $\dot{M_{H_{2}}}$ to the atomic mass outflow rates ($\dot{M_{H\,I}}$): for their sample of AGN-hosts, they find $\dot{M_{H_{2}}}$ is generally $\sim$1 order of magnitude higher than $\dot{M_{H\,I}}$ using \nad. However, for their star-forming sample large scatter dominates and prevents a clear conclusion. To work around this, an alternative tracer (C+) is used and the ratio $\dot{M_{H_{2}}}$/$\dot{M_{H\,I}}$ is found to be roughly equal for AGN. They tentatively conclude that for starburst-driven galaxies, the ionized, atomic and molecular phases contribute in roughly equal quantities to the total mass outflow rate. As such, it is likely our mass outflow rates are only lower limits and a multiwavelength estimation of such rates would lead to more complete and slightly higher values, given the added mass from the other gas phases.

\subsection{Upper limits on mass inflow rates}
\label{subsec:inflowdisc}
In Figure \ref{fig:incvsfrd} and Figure \ref{fig:incdep} we find detections of inflowing gas among disk galaxies. The infalling gas could come from cosmological filaments, from galactic fountains, minor mergers, or from gas cooling from the CGM. Due to the uncertain source of the inflows, the assumptions made for Equation \ref{eq:finaldM} may not hold. In particular, assumptions about the metal content, ionization fraction and depletion onto dust become highly uncertain when converting to a column density of Hydrogen. Nonetheless, we can assume these as upper limits to convert to mass inflow rates, since it is likely metallicity and abundances would decrease outside of the galaxy disk. With this in mind, we report upper limit inflow rates of 0.08-0.38 M$_{\odot}$yr$^{-1}$. No significant trend is found with the SFRs or stellar mass of the galaxies.

In Figure \ref{fig:incdep} we see that inflows have a strong inclination dependence, and are only seen at high ($i$>50$^{\circ}$) inclinations. This suggests that we are seeing the gas accreting along the plane of the disk. \citet{ho17} used Mg\,II absorption and quasar sightlines to probe the CGM of a sample of 15 highly-inclined, local star-forming galaxies with known rotation curves. They showed that much of the Doppler shifted Mg\,II gas was consistent with the rotational motion of the host galaxies and the implication for this was radial infall of gas into the disk plane. It is possible that our results suggest a similar scenario, where inflowing gas (from a variety of sources) falls radially before becoming dominated by the circular motions of the galaxy disk.

\section{Discussion}
\label{sec:discussion}
\subsection{The prevalence of outflows and inflows}
\label{subsec:ubiquity}
Several studies have claimed a ubiquity of outflows over the star-forming main sequence (e.g., \citealt{weiner09, rubin14}). Our results are partially consistent with this picture in that outflows appear prevalent in star-forming systems with SFR>1\,M$_{\odot}$yr$^{-1}$ or $\Sigma_{\text{SFR}}$$\geqslant$0.01\,M$_{\odot}$yr$^{-1}$kpc$^{-2}$ and stellar masses log M$_{*}$/M$_{\odot}$>10. We don't, however, find outflows in low mass galaxies. Reasons for this could be due to lower $\Sigma_{\text{SFR}}$s, or limitations of \nad\ as a tracer (e.g., in the absence of dust). We therefore cannot claim ubiquity over the whole of the main sequence. Additionally, we find that outflows are also found in bulge-dominated objects with sufficiently high SFRs, and therefore are not dependent on morphology. We find this to be true for both AGN and inactive galaxies.

\subsection{Comparison to simulations}
\label{subsec:simucompare}
In this section, we compare the flow properties derived in this study to results from simulations, namely those of \citet{muratov15} and \citet{aa17} using the Feedback in Realistic Environments (FIRE) simulations at $z<$0.5, as well as those from \citet{oppenheimer10}.
\begin{enumerate}
\item \textit{The prevalence of inflows and outflows in star-forming galaxies.}

The prevalence of outflows in our samples of star-forming galaxies appears only partially consistent with results from simulations. Both \citet{muratov15} and \citet{oppenheimer10} find that high-mass galaxies have stable discs and a more continuous, quiescent mode of star formation at $z$<1 that can no longer drive very \textit{strong} outflows into the halo. Dwarf galaxies instead maintain a bursty state of star formation which allows them to produce outflows \citep{muratov15}. Our results both agree and contrast with these simulations in that we find low-velocity outflows to be common among star-forming galaxies with high stellar mass but no detections in low-mass (log M$_{*}$/M$_{\odot}$\,<\,10) galaxies, whose $\Sigma_{\text{SFR}}$s are significantly lower. If outflows are indeed present at low-mass, it is possible that we are unable to detect them due to a) the outflows being too weak for our code to detect, or b) a resolution issue where the velocities are blended by the SDSS spectral resolution, or c) the unreliability of \nad\ as a tracer in low A$_{\text{v}}$ environments.

The above simulations also predict non-negligible amounts of accreting gas onto star-forming galaxies. Our study agrees with this, as we find inflow detections in star-forming, high mass disk galaxies. The source of the inflowing gas is impossible to ascertain from our data, however it is likely a combination of material coming from pristine cold gas, gas from nearby companions, minor mergers, and/or recycled gas ("galactic fountains"). \\

\item \textit{Outflow central velocities and mass loading factors.}

By using the M$_{*}$-M$_{h}$ relation described in \citet{behroozi13} and Equation 1 in \citet{mo02}, we are able to compare the central velocities of our outflow detections to those reported in simulations, as a function of halo circular velocity, $v_{\text{c}}$. We find our central velocities are within the broad range of median velocities (20$\lesssim\Delta$v$\lesssim$4000 km/s) reported by \citet{muratov15} and lie right on the power law relation calibrated for their medium-$z$ (0.5 < $z$ < 2.0) and high-$z$ (2.0 < $z$ < 4.0) samples. However, our velocities appear more than an order of magnitude larger than the upper limits of their L* progenitors at $z<$0.5, which have velocities less than 100 km/s.

We also compare our mass loading factors to those found in simulations and find them to be in agreement with the upper limits for the low-$z$ L* progenitors of \citet{muratov15}. \citet{muratov15} make an approximate comparison between their mass loading factors and those derived in the Illustris project \citep{vogelsberger13}, and find the Illustris results to be systematically higher than theirs ($\eta\approx$7 for a Milky Way-mass galaxy at $z$=0, compared to $\eta\ll$1). Although we caution a direct comparison due to the differences by which the mass loading factors are measured in each study, such high mass loading factors are in contrast with our results and suggest some prescriptions may be adopting abnormally strong outflows than what are typically seen in the local Universe.
\end{enumerate}

\subsection{Star formation vs AGN}
\label{subsec:SFvsAGN}
Several recent studies have discussed the implications of AGN on the baryon cycle and their influence in the launching of outflows. In these studies, the \nad\ tracer was used to detect outflows in samples of AGN and star-forming galaxies and determine which energy source was the primary driver of the outflows. For example, \citet{sarzi16} used SDSS spectra of 456 local star-forming galaxies from the mJIVE-20 survey to determine whether these hosted both an optical outflow and showed radio emission as part of the Very Large Array's (VLA) FIRST survey. Not a single object showed an outflow detection together with radio emission and therefore the authors concluded outflows were regulated by star formation, not AGN feedback. \citet{nedelchev17} also compared the effects of AGN feedback in a sample of $\sim$9,900 SDSS Seyfert 2 galaxies and a control sample of $\sim$44,000 inactive galaxies. Only 0.5\% of their Seyfert 2 sample displayed potential outflows compared to 0.8\% for the control sample, suggesting AGN activity did not enhance outflow activity.
Figure \ref{fig:incdep} from our study extends these results to the regime of normal star-forming galaxies and modest AGN. As reported in Section \ref{subsec:flowvel} and Section \ref{subsec:dMsec}, there is a mild increase in outflow velocity and mass outflow rates with the presence of an AGN, although the differences between the median inactive and AGN values are only $\sim$12 km\,s$^{-1}$ and $\sim$0.34 M$_{\odot}$\,yr$^{-1}$. Such small values suggest these AGN do not \textit{significantly} enhance outflow activity or strength. We can therefore conclude that the presence of an optically-selected AGN does not significantly enhance outflows in \textit{normal} galaxies of the local Universe, and that such winds are unlikely to be able to quench a galaxy via "ejective" feedback, where gas is removed from the galaxy via the outflow.

\begin{figure}
 \includegraphics[width=0.9\columnwidth]{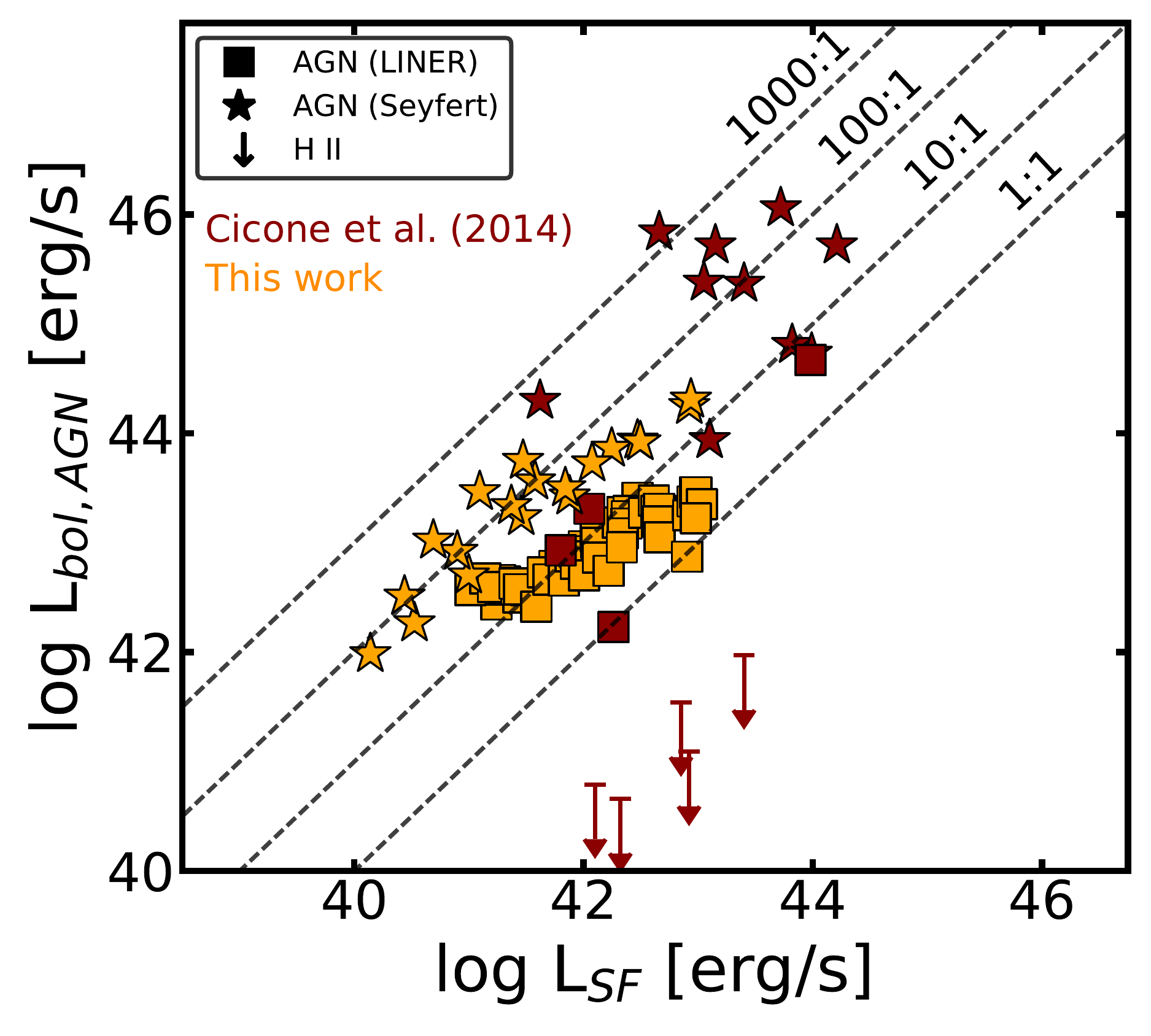}
 \caption{A comparison of the energy output from the AGN (L$_{\text{AGN}}$) and from star formation (ie., supernovae, L$_{\text{SF}}$) for our AGN stacked sample (orange) and objects from \citet{cicone14} (red). Additionally, we extract and stack the Seyfert objects in our AGN sample and plot the data points for comparison. The dashed lines denote L$_{\text{bol,AGN}}$/L$_{\text{SF}}$=1000, 100, 10 and 1.}
 \label{fig:AGNenergy}
\end{figure}

This may appear somewhat at odds with recent observations of strong AGN feedback in both the local and high-$z$ Universe (e.g., \citealt{feruglio10, alatalo11, maiolino12, cicone14}), however there are several plausible reasons for this. The first is that we may not be observing the same types of AGN: our BPT cut and binning procedure ensure we are selecting and mixing weak AGN which could be drowning out much of the signal produced by rarer and much stronger AGN (e.g., Seyferts). This is highlighted in Figure \ref{fig:AGNenergy}, where we compare the energy output from the AGN versus the energy output of supernovae.

The AGN luminosity is calculated using Equation 3 of 
\citet{netzer09}, which makes use of the [OIII] and [OI] luminosities, whilst the energy output of star formation (i.e., supernovae) is derived using the relation presented by \citet{veilleux05}:

\begin{equation}
L_{\text{SF}} = 7\times10^{41} \text{SFR}(M_{\odot}yr^{-1})\,\,\,[\text{erg s}^{\text{-1}}].
\end{equation}

For comparison, we plot the quantities (where available) for the sample of \citet{cicone14} and also stacked spectra of the Seyferts within our AGN sample (selected with an additional BPT cut of log [OIII]/H$\beta$ > 0.5).

As evident from the plot, the AGN feedback found by the aforementioned studies are observed in extreme objects which host very strong AGN, not typical of the samples of galaxies that we probe. We find a median luminosity (uncorrected for dust) log L$_{\text{AGN, bol}}$=42.8 erg/s over the DISK AGN sample. For comparison, the median log L$_{\text{AGN}}$ of \citet{cicone14}'s extended sample is log L$_{\text{AGN}}$=44.76 erg/s - about two orders of magnitude higher. This highlights the comparative weakness of optically-selected AGN in normal galaxies. Additionally, it is important to note that SF can significantly contribute to the [OIII] flux and therefore deducing an accurate L$_{\text{AGN}}$ value from this method is challenging. These values are, in essence, upper limits of the true AGN energy contribution. Nevertheless, a comparison of AGN in normal star-forming objects - not just in 
extreme objects - remains useful towards constraining the extent to which an active nucleus may impact the prevalence and properties of galactic winds.

A second, less likely, reason is to do with the dynamical timescales of AGN activity and outflows: it is possible that we are also observing a) objects with AGN that are in the process of turning off due to reduced rates of gas accretion and/or b) outflows which are relics of the strong feedback found in more extreme objects or at high-$z$. All of these scenarios are consistent with our observations and findings and our study does not rule out strong feedback by more extreme AGN.

\subsection{The fate of outflows}
\label{subsec:fate}
\begin{figure}
 \includegraphics[width=1.\columnwidth]{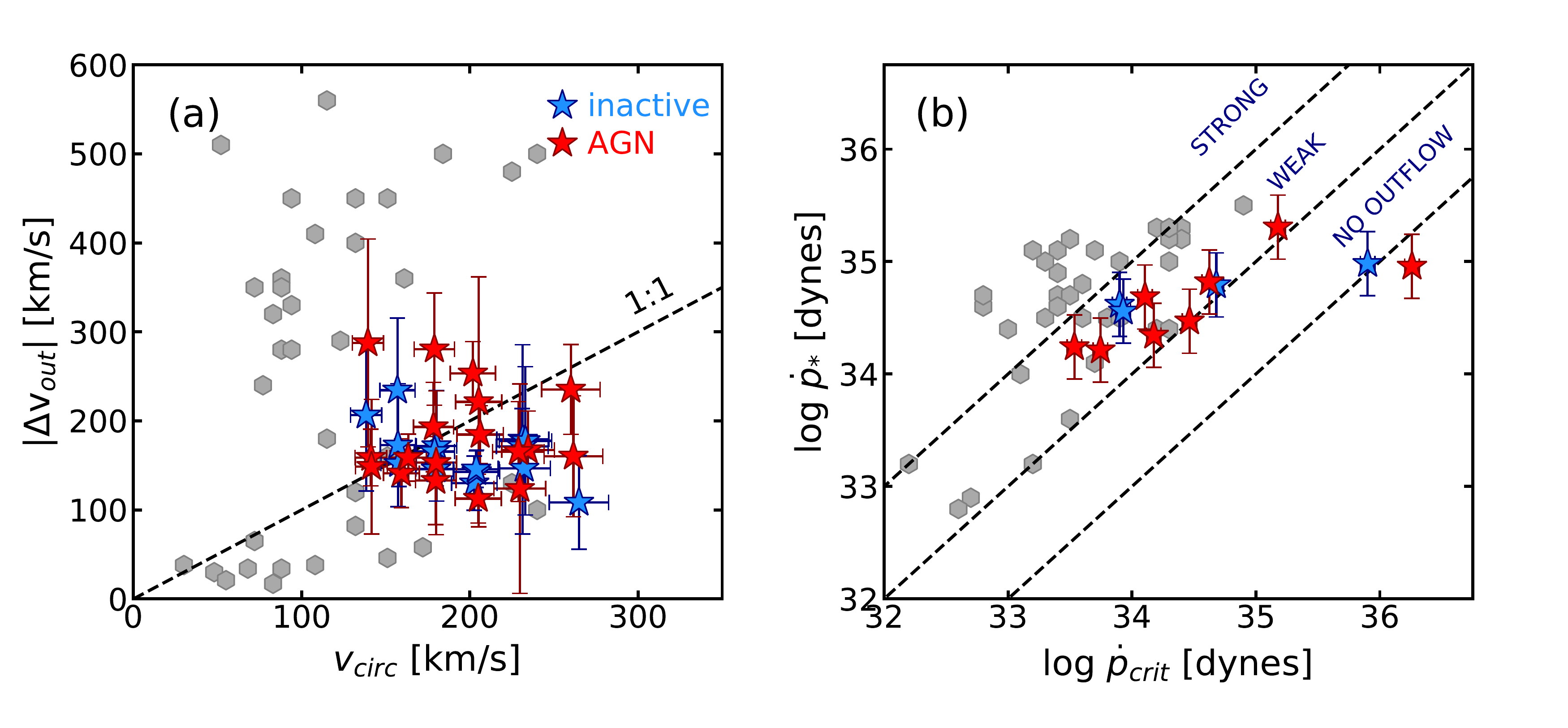}
 \caption{\textit{Left}: The outflow velocity as a function of the galaxy's circular velocity, compared to the results of \citet{heckman15}. Blue stars are the inactive galaxies and the red stars are the AGN of the $i$-log\,M$_{*}$ sample. \textit{Right}: A comparison of the momentum flux (or force) provided by the galaxy starburst versus the critical momentum flux necessary for the net force acting on a cloud to be outwards. The dashed diagonal lines denote $\dot{p}_{*}$/$\dot{p}_{crit}$=10, 1 and 0.1 as well as three outflow regimes: "no outflow", "weak outflow", and "strong outflow".}
 \label{fig:p_h15}
\end{figure}

Several useful quantities exist to obtain an approximation of an outflow's energy relative to the gravitational well of the host galaxy. In Figure \ref{fig:p_h15}a we plot the outflow velocity versus the circular velocity ($v_{\text{circ}}$) of the host galaxy for stacks over the $i$-log\,M$_{*}$ plane, and add the results of \citet{heckman15} for local star-forming galaxies for comparison. This provides us with an idea of whether an outflow is traveling at speeds close to the escape velocity of the galaxy or not. The circular velocity is derived from the stellar mass of the host galaxy: $v_{\text{circ}}$=$\sqrt2S$, where $S$ is the kinematic parameter \citep{weiner06, kassin07} found to have a good fit with stellar mass for low-$z$ star-forming galaxies, log\,$S$ = 0.29\,log\,M$_{*}$ - 0.93 \citep{simons15, heckman15}. We see that most of our detections (23/33) lie below the 1:1 line, suggesting the outflow velocity does not exceed the circular velocity of the host. However, we also notice there are some detections (10/33) which have outflow velocities greater than the circular velocity of the galaxy. These all occur in the lower mass systems, suggesting that outflowing gas may become unbound from the galaxy's gravitational potential.

Another useful comparison is of the force provided by the host galaxy's starburst (caused by stellar winds, supernovae and radiation pressure) to the critical force needed to have a net force acting outward on the outflow. Assuming a momentum-driven outflow consisting of a population of filamentary clouds (e.g., \citealt{chevalier85}) dense enough to produce the observed absorption line profile (e.g., the outflow in M82), the momentum flux (or force) provided by the starburst is $\dot{p}_{*}$ = 4.8$\times$10$^{33}$\,SFR and the critical momentum flux acting on a cloud needed for the net force acting on it to be outward is $\dot{p}_{\text{crit}}$=4$\pi  r_{*}N(H)\,m_{\text{H}}\,v_{\text{circ}}^{2}$ (for more details, see Section 4 of \citealt{heckman15}). In Figure \ref{fig:p_h15}b we plot these two quantities for the LOW-$i$ SFR-M$_{*}$ stacks and compare them to the results of \citet{heckman15}. We find that 10/12 detections fall under the "weak outflow" regime defined by \citet{heckman15}, where the starburst provides $\dot{p}_{*}\sim$1-10 $\dot{p}_{\text{crit}}$, and 2/12 detections fall under the "no-outflow" regime where $\dot{p}_{*}$<$\dot{p}_{\text{crit}}$ and the starburst cannot match or exceed the force needed to overcome gravity. None of our detections fall in the "strong outflow" regime where $\dot{p}_{*}$>10$\dot{p}_{\text{crit}}$ and the outflow exceeds the escape velocity of the galaxy.

These basic results provide rough approximations of the force provided to the outflows and suggest the vast majority of our detections are unable to escape the host galaxy's gravitational hold. In fact, such arguments are based on ballistic models which do not account for the presence of a surrounding gaseous corona, while in reality hydrodynamical processes should play a crucial role in slowing down the outflow, making it even more difficult to escape the galaxy system. This is likely to play an even more important role in the most massive systems, since they reside in denser environments \citep{oppenheimer08}.
Given the low velocities of our inflow/outflow detections, the inclination dependence and the relatively low median SFRs of our stacks, it is likely we are viewing aspects of a galactic fountain scenario, where the gas is expelled from the galaxy disk into the surrounding medium, before it mixes and cools with potential pristine gas to fall back down into the disk as an inflow. The low velocities are unlikely to be enough to escape the host system and it is therefore not unreasonable to assume these outflows could be fueling (in part) the extra-planar gas observed in external galaxies (e.g., \citealt{fraternali02, oosterloo07, rossa03}) and the Milky Way \citep{marasco11}. The simultaneous detections of outflows \textit{and} inflows in virtually the same regions of parameter space - separated only by inclination effects - are most easily explained by the scenario of a galactic fountain \citep{fraternali06}.

\section{Summary and Conclusions}
\label{sec:conclusions}
In this study we conduct a stacking analysis of 240,567 inactive galaxies and 67,753 AGN-hosts from the SDSS DR7 survey. We stack spectra over bins of global galaxy properties and place constraints on the detection rates and properties of inflows and outflows in the local Universe. Our main conclusions can be summarized as the following:

\begin{itemize}
\item Signatures of outflowing gas are detected along the main sequence of star-forming galaxies for a large range of stellar masses (10$\lesssim$log M$_{*}$/M$_{\odot}\lesssim$11.5). We also find detections of inflows in star-forming, disk galaxies over a similar range of stellar mass (10$\lesssim$log M$_{*}$/M$_{\odot}\lesssim$11). These results hold for both inactive galaxies and AGN-hosts.
\\
\item We find a strong inclination dependence for the detection rates of both outflows and inflows in disk galaxies, with outflows prevalent at low inclinations ($i\lesssim$50$^{\circ}$) and inflows at high inclinations ($i\gtrsim$60$^{\circ}$). This is suggestive of outflowing gas perpendicular to the galaxy disk and accretion along the plane of the disk. Galaxy morphology does not appear to play a major role in the detection rates of outflows.
\\
\item We report low ($\sim$0.14-1.74\,M$_{\odot}$yr$^{-1}$) mass outflow rates and compare these to other results in the literature. These comparisons reveal a strong linear relationship between the mass outflow rate and the SFR of the host galaxy, and a prescription is provided. The mass loading factor, given by the ratio of these two quantities, is calculated to be near-constant ($\eta$ $\approx$0.1) for local, normal star-forming objects.
\\
\item We find only minor differences in outflow detection rates and properties of inactive and AGN galaxies, suggesting that the presence of a weak AGN does not significantly enhance either. Neither galaxy type appears able to launch winds strong enough to quench a galaxy.
\end{itemize}

Galaxy-scale outflows are an integral element of galaxy evolution models. They play a key role in shaping the environments and mass build up of galaxies across cosmic time. Here we have studied outflows in stacks of large samples of local galaxies over a range of stellar mass and SFRs and found them to be common among star-forming galaxies. However, none of the outflows are powerful enough to quench their hosts via ejective feedback, but may nonetheless be able to significantly influence the surrounding environments of the galaxies. To verify this, more work is required to link the properties of outflows to the gas content and distribution in the CGM. 
To obtain a better understanding and a more comprehensive picture of outflows, large dedicated surveys (UV, optical and submillimetre) and IFU observations of neutral, ionized and molecular gas in normal star-forming objects are required in order to constrain and link the multiwavelength nature of outflows. Such observations would also allow more concrete constraints on the geometries of outflows, which have until now relied on crude and unconstrained assumptions. Finally, still required are detailed analyses of inflows and their interplay with outflows and the host galaxies. In combination with simulations that track the accretion of pristine, merged, and recycled gas, such observations would greatly complement and enhance our knowledge of the conditions necessary to fuel star formation across cosmic time.

\section*{Acknowledgements}
This research was supported by grants from the Royal Society.

We thank the anonymous referee for a thoughtful report which strengthened this study. GRB would like to thank Yan-Mei Chen, Filippo Fraternali, Isabella Lamperti, Giocchino Accurso, Tim Heckman, Kate Rubin and Sylvain Veilleux for helpful conversations that improved this work. GRB would also like to thank Claudia Maraston for providing SSP models that were used in this study.

Funding for the Sloan Digital Sky Survey (SDSS) has been provided
by the Alfred P. Sloan Foundation, the Participating Institutions,
the National Aeronautics and Space Administration, the
National Science Foundation, the U.S. Department of Energy, the
Japanese Monbukagakusho and the Max Planck Society. The SDSS
web site is http://www.sdss.org/.

The SDSS is managed by the Astrophysical Research Consortium
(ARC) for the Participating Institutions. The Participating
Institutions are The University of Chicago, Fermilab, the Institute
for Advanced Study, the Japan Participation Group, The Johns
Hopkins University, Los Alamos National Laboratory, the Max-Planck-Institute
for Astronomy (MPIA), the Max-Planck-Institute
for Astrophysics (MPA), New Mexico State University, University
of Pittsburgh, Princeton University, the United States Naval Observatory, and the University of Washington.




\appendix
\section{Comparison of SSP models}
\label{sec:SSPcompare}
To ensure the distribution of line residuals we discuss in Section \ref{subsec:profiles} are not dependent (to within reasonable margins of uncertainty) on the codes or SSP models used in our study, we repeat our continuum fitting on the main sample for inactive galaxies using a custom made Python continuum-fitting code in combination with the codes from \citet{bc03} (henceforth, BC03) and \citet{maraston11} (henceforth, M11). We makes use of both the MILES and STELIB libraries for each code and compare these to the results derived in pPXF. The results are presented in Figure X. The distribution of residual \nad\ ISM profiles remains constant throughout all four cases, demonstrating that this result is independent of the choice of SSP models or codes.

\begin{figure*}
 \centering
 \includegraphics[width=0.7\columnwidth]{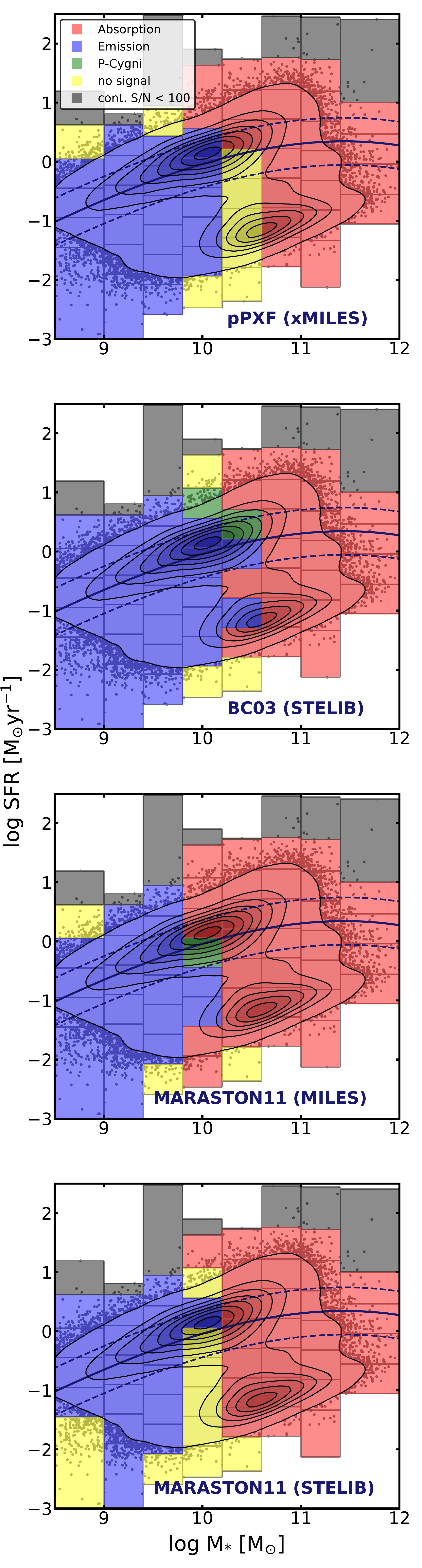}
 \caption{The distribution of \nad\ ISM profiles across the SFR-M$_{*}$ plane derived by dividing the stacked spectrum in each bin by the best fit continuum obtained using the code and SSP models stated on the bottom right of each plot.}
 \label{fig:modelscomp}
\end{figure*}

\section{\texorpdfstring{\nad}{TEXT} fitting properties}

\begin{table*}
    \caption{The properties of the flow parameters measured over the DISK, HIGH-$i$, LOW-$i$, BULGE, $i$-log\,SFR and $i$-log\,M$_{*}$ samples for inactive galaxies. For profiles with blueshifted absorption and redshifted emission, a $b$ postscript indicates the blueshifted component, whilst the $r$ postscript refers to the redshifted component of the profile.}
    \label{tab:flowpropstab}
    \begin{tabular}{cccccccccccc}
        \hline
        Sample & SFR & log M$_{*}$/M$_{\odot}$ & EW & $\Delta$v & $\Delta$v/cos($i$) & b$_{\text{D}}$ & C$_{f}$ & N(Na I) & N(H) & Profile & Type \\
        \hline
DISK & 0.13 & 10.59 & 0.07 & 147$\pm$49 & 521 & 118$\pm$43 & 0.75$\pm$0.26 & 11.94 & 19.61 & Absorption & Inflow \\
.............. & 0.18 & 10.09 & 0.06 & 133$\pm$20 & 484 & 136$\pm$19 & 0.84$\pm$0.26 & 11.86 & 19.54 & Absorption & Inflow \\
.............. & 0.21 & 9.63 & 0.06 & 69$\pm$132 & 535 & 97$\pm$104 & 0.71$\pm$0.28 & 11.90 & 19.58 & Absorption & Inflow \\
.............. & 0.36 & 9.89 & 0.07 & 67$\pm$19 & 503 & 111$\pm$20 & 0.87$\pm$0.25 & 11.88 & 19.56 & Absorption & Inflow \\
.............. & 1.08 & 9.94 & 0.01$^{b}$ & -260$\pm$87$^{b}$ & -271$^{b}$ & 61$\pm$42$^{b}$ & 0.26$\pm$0.27$^{b}$ & 11.58$^{b}$ & 19.25$^{b}$ & Absorption & Outflow \\
.............. &  &  & -0.24$^{r}$ & 21$\pm$5$^{r}$ & 22$^{r}$ & 199$\pm$10$^{r}$ & -0.06$\pm$0.01$^{r}$ & 13.84$^{r}$ & 21.52$^{r}$ & Emission & Outflow \\
.............. & 0.67 & 10.06 & 0.05 & 77$\pm$7 & 284 & 117$\pm$8 & 0.83$\pm$0.22 & 11.79 & 19.47 & Absorption & Inflow \\
.............. & 0.70 & 10.18 & 0.04 & 173$\pm$15 & 1226 & 71$\pm$13 & 0.50$\pm$0.25 & 11.82 & 19.50 & Absorption & Inflow \\
.............. & 1.48 & 10.04 & 0.02$^{b}$ & -260$\pm$64$^{b}$ & -271$^{b}$ & 56$\pm$35$^{b}$ & 0.48$\pm$0.26$^{b}$ & 11.36$^{b}$ & 19.04$^{b}$ & Absorption & Outflow \\
.............. &  &  & -0.20$^{r}$ & 25$\pm$10$^{r}$ & 26$^{r}$ & 200$\pm$10$^{r}$ & -0.04$\pm$0.00$^{r}$ & 13.98$^{r}$ & 21.66$^{r}$ & Emission & Outflow \\
.............. & 1.50 & 10.00 & 0.01$^{b}$ & -235$\pm$25$^{b}$ & -262$^{b}$ & 70$\pm$15$^{b}$ & 0.90$\pm$0.23$^{b}$ & 11.04$^{b}$ & 18.72$^{b}$ & Absorption & Outflow \\
.............. &  &  & -0.22$^{r}$ & 22$\pm$7$^{r}$ & 25$^{r}$ & 200$\pm$3$^{r}$ & -0.05$\pm$0.00$^{r}$ & 13.86$^{r}$ & 21.54$^{r}$ & Emission & Outflow \\
.............. & 1.43 & 10.02 & 0.01$^{b}$ & -210$\pm$20$^{b}$ & -256$^{b}$ & 77$\pm$11$^{b}$ & 0.49$\pm$0.23$^{b}$ & 11.30$^{b}$ & 18.97$^{b}$ & Absorption & Outflow \\
.............. &  &  & -0.22$^{r}$ & 21$\pm$3$^{r}$ & 26$^{r}$ & 198$\pm$1$^{r}$ & -0.05$\pm$0.00$^{r}$ & 13.91$^{r}$ & 21.59$^{r}$ & Emission & Outflow \\
.............. & 1.39 & 10.04 & 0.01$^{b}$ & -282$\pm$13$^{b}$ & -401$^{b}$ & 51$\pm$6$^{b}$ & 0.40$\pm$0.24$^{b}$ & 11.27$^{b}$ & 18.95$^{b}$ & Absorption & Outflow \\
.............. &  &  & -0.12$^{r}$ & 21$\pm$2$^{r}$ & 29$^{r}$ & 200$\pm$1$^{r}$ & -0.02$\pm$0.00$^{r}$ & 14.45$^{r}$ & 22.13$^{r}$ & Emission & Outflow \\
.............. & 1.14 & 10.12 & 0.04 & 82$\pm$7 & 193 & 104$\pm$7 & 0.89$\pm$0.21 & 11.61 & 19.29 & Absorption & Inflow \\
.............. & 1.10 & 10.24 & 0.04 & 180$\pm$7 & 631 & 50$\pm$5 & 0.46$\pm$0.23 & 11.76 & 19.43 & Absorption & Inflow \\
.............. & 1.09 & 10.27 & 0.04 & 170$\pm$125 & 1128 & 80$\pm$75 & 0.84$\pm$0.26 & 11.61 & 19.28 & Absorption & Inflow \\
.............. & 1.79 & 10.08 & -0.05 & 245$\pm$45 & 256 & 175$\pm$33 & -0.63$\pm$0.25 & 11.90 & 19.57 & Emission & Outflow \\
.............. & 1.95 & 10.11 & -0.06 & 267$\pm$30 & 297 & 178$\pm$29 & -0.94$\pm$0.24 & 11.76 & 19.44 & Emission & Outflow \\
.............. & 1.84 & 10.10 & -0.04 & 286$\pm$19 & 351 & 170$\pm$17 & -0.84$\pm$0.24 & 11.67 & 19.35 & Emission & Outflow \\
.............. & 1.76 & 10.17 & 0.01 & 195$\pm$7 & 341 & 50$\pm$3 & 0.32$\pm$0.25 & 11.62 & 19.30 & Absorption & Inflow \\
.............. & 1.69 & 10.25 & 0.03 & 184$\pm$6 & 428 & 50$\pm$4 & 0.88$\pm$0.25 & 11.45 & 19.13 & Absorption & Inflow \\
.............. & 1.73 & 10.37 & 0.05 & 186$\pm$11 & 631 & 66$\pm$13 & 0.22$\pm$0.25 & 12.32 & 20.00 & Absorption & Inflow \\
.............. & 2.60 & 10.17 & 0.11 & -184$\pm$56$^{b}$ & -193$^{b}$ & - & - & - & - & P-Cygni & Outflow \\
.............. &  &  & -0.06 & 184$\pm$73$^{r}$ & 192$^{r}$ & - & - & - & - & P-Cygni & Outflow \\
.............. & 2.71 & 10.22 & 0.18 & -35$\pm$28$^{b}$ & -39$^{b}$ & - & - & - & - & P-Cygni & Outflow \\
.............. &  &  & -0.13 & 158$\pm$15$^{r}$ & 175$^{r}$ & - & - & - & - & P-Cygni & Outflow \\
.............. & 2.81 & 10.22 & 0.00$^{b}$ & -91$\pm$60$^{b}$ & -112$^{b}$ & 163$\pm$34$^{b}$ & 0.43$\pm$0.17$^{b}$ & 10.18$^{b}$ & 17.86$^{b}$ & Absorption & Outflow \\
.............. &  &  & -0.09$^{r}$ & 180$\pm$6$^{r}$ & 220$^{r}$ & 197$\pm$2$^{r}$ & -0.02$\pm$0.00$^{r}$ & 13.83$^{r}$ & 21.51$^{r}$ & Emission & Outflow \\
.............. & 2.79 & 10.39 & 0.05 & 187$\pm$7 & 429 & 64$\pm$9 & 0.95$\pm$0.23 & 11.67 & 19.35 & Absorption & Inflow \\
.............. & 2.76 & 10.48 & 0.08 & 185$\pm$64 & 607 & 77$\pm$44 & 0.74$\pm$0.23 & 11.98 & 19.66 & Absorption & Inflow \\
.............. & 4.13 & 10.28 & 0.13 & -191$\pm$40 & -199 & 187$\pm$25 & 0.15$\pm$0.23 & 12.96 & 20.64 & Absorption & Outflow \\
.............. & 4.50 & 10.32 & 0.03$^{b}$ & -189$\pm$21$^{b}$ & -210$^{b}$ & 56$\pm$48$^{b}$ & 0.56$\pm$0.24$^{b}$ & 11.94$^{b}$ & 19.62$^{b}$ & Absorption & Outflow \\
.............. &  &  & -0.11$^{r}$ & 191$\pm$9$^{r}$ & 212$^{r}$ & 198$\pm$6$^{r}$ & -0.14$\pm$0.25$^{r}$ & 12.89$^{r}$ & 20.56$^{r}$ & Emission & Outflow \\
.............. & 5.04 & 10.37 & 0.07$^{b}$ & -185$\pm$4$^{b}$ & -227$^{b}$ & 50$\pm$6$^{b}$ & 0.61$\pm$0.22$^{b}$ & 12.30$^{b}$ & 19.98$^{b}$ & Absorption & Outflow \\
.............. &  &  & -0.03$^{r}$ & 196$\pm$43$^{r}$ & 241$^{r}$ & 197$\pm$34$^{r}$ & -0.32$\pm$0.26$^{r}$ & 12.00$^{r}$ & 19.68$^{r}$ & Emission & Outflow \\
.............. & 5.32 & 10.46 & 0.04 & 204$\pm$14 & 355 & 57$\pm$19 & 0.32$\pm$0.26 & 12.13 & 19.80 & Absorption & Inflow \\
.............. & 5.37 & 10.53 & 0.08 & 196$\pm$13 & 444 & 69$\pm$19 & 0.75$\pm$0.24 & 11.99 & 19.67 & Absorption & Inflow \\
.............. & 5.23 & 10.61 & 0.11 & 174$\pm$57 & 581 & 116$\pm$28 & 0.68$\pm$0.24 & 12.21 & 19.89 & Absorption & Inflow \\
.............. & 4.81 & 10.14 & 0.18 & -237$\pm$72 & -247 & 299$\pm$35 & 0.54$\pm$0.24 & 12.51 & 20.19 & Absorption & Outflow \\
.............. & 8.18 & 10.41 & 0.19 & -209$\pm$64 & -232 & 244$\pm$18 & 0.28$\pm$0.22 & 12.82 & 20.50 & Absorption & Outflow \\
HIGH-$i$ & 2.68 & 10.41 & 0.02 & 189$\pm$5 & 375 & 50$\pm$2 & 0.94$\pm$0.25 & 11.34 & 19.01 & Absorption & Inflow \\
.............. & 7.42 & 10.49 & 0.02 & 196$\pm$25 & 368 & 51$\pm$27 & 0.32$\pm$0.27 & 11.88 & 19.56 & Absorption & Inflow \\
.............. & 1.69 & 10.71 & 0.04 & 187$\pm$7 & 495 & 70$\pm$10 & 0.87$\pm$0.23 & 11.66 & 19.34 & Absorption & Inflow \\
.............. & 4.16 & 10.73 & 0.03 & 199$\pm$6 & 424 & 56$\pm$7 & 0.14$\pm$0.24 & 12.44 & 20.12 & Absorption & Inflow \\
.............. & 11.02 & 10.79 & 0.05 & 210$\pm$31 & 395 & 91$\pm$34 & 0.40$\pm$0.26 & 12.06 & 19.74 & Absorption & Inflow \\
.............. & 2.99 & 11.11 & 0.08 & 180$\pm$19 & 443 & 81$\pm$24 & 0.82$\pm$0.25 & 11.96 & 19.63 & Absorption & Inflow \\
.............. & 8.55 & 11.07 & 0.03 & 189$\pm$31 & 389 & 54$\pm$33 & 0.73$\pm$0.27 & 11.48 & 19.16 & Absorption & Inflow \\
LOW-$i$ & 1.18 & 9.99 & 0.02$^{b}$ & -234$\pm$11$^{b}$ & -293$^{b}$ & 53$\pm$6$^{b}$ & 0.93$\pm$0.24$^{b}$ & 11.13$^{b}$ & 18.81$^{b}$ & Emission & Outflow \\
.............. &  &  & -0.22$^{r}$ & 20$\pm$1$^{r}$ & 25$^{r}$ & 200$\pm$0$^{r}$ & -0.04$\pm$0.00$^{r}$ & 14.34$^{r}$ & 22.02$^{r}$ & Emission & Outflow \\
.............. & 2.32 & 10.06 & 0.01$^{b}$ & -266$\pm$16$^{b}$ & -328$^{b}$ & 59$\pm$9$^{b}$ & 0.53$\pm$0.16$^{b}$ & 11.22$^{b}$ & 18.90$^{b}$ & Absorption & Outflow \\
.............. &  &  & -0.12$^{r}$ & 20$\pm$2$^{r}$ & 25$^{r}$ & 199$\pm$1$^{r}$ & -0.02$\pm$0.00$^{r}$ & 14.59$^{r}$ & 22.27$^{r}$ & Emission & Outflow \\
.............. & 1.52 & 10.33 & 0.01$^{b}$ & -201$\pm$7$^{b}$ & -254$^{b}$ & 78$\pm$10$^{b}$ & 0.33$\pm$0.24$^{b}$ & 11.65$^{b}$ & 19.33$^{b}$ & Absorption & Outflow \\
.............. &  &  & -0.27$^{r}$ & 20$\pm$1$^{r}$ & 26$^{r}$ & 200$\pm$1$^{r}$ & -0.07$\pm$0.01$^{r}$ & 13.76$^{r}$ & 21.43$^{r}$ & Emission & Outflow \\
.............. & 3.29 & 10.39 & 0.04 & -193$\pm$8$^{b}$ & -241$^{b}$ & - & - & - & - & P-Cygni & Outflow \\
.............. &  &  & -0.03 & 231$\pm$13$^{r}$ & 288$^{r}$ & - & - & - & - & P-Cygni & Outflow \\
        \hline
    \end{tabular}
\end{table*}

\begin{table*}
    \contcaption{}
    \begin{tabular}{cccccccccccc}
        \hline
        Sample & SFR & log M$_{*}$/M$_{\odot}$ & EW & $\Delta$v & $\Delta$v/cos($i$) & b$_{\text{D}}$ & C$_{f}$ & N(Na I) & N(H) & Profile & Type \\
        \hline
.............. & 8.67 & 10.48 & 0.06$^{b}$ & -188$\pm$45$^{b}$ & -237$^{b}$ & 53$\pm$28$^{b}$ & 0.71$\pm$0.23$^{b}$ & 12.14$^{b}$ & 19.82$^{b}$ & Absorption & Outflow \\
.............. &  &  & -0.04$^{r}$ & 195$\pm$54$^{r}$ & 246$^{r}$ & 196$\pm$44$^{r}$ & -0.33$\pm$0.27$^{r}$ & 12.04$^{r}$ & 19.72$^{r}$ & Emission & Outflow \\
.............. & 3.29 & 10.71 & 0.00$^{b}$ & -226$\pm$117$^{b}$ & -284$^{b}$ & 57$\pm$42$^{b}$ & 0.61$\pm$0.28$^{b}$ & 10.39$^{b}$ & 18.07$^{b}$ & Absorption & Outflow \\
.............. &  &  & -0.06$^{r}$ & 195$\pm$5$^{r}$ & 245$^{r}$ & 200$\pm$3$^{r}$ & -0.68$\pm$0.23$^{r}$ & 11.93$^{r}$ & 19.61$^{r}$ & Emission & Outflow \\
.............. & 7.54 & 10.74 & 0.04$^{b}$ & -189$\pm$2$^{b}$ & -239$^{b}$ & 51$\pm$5$^{b}$ & 0.93$\pm$0.22$^{b}$ & 11.86$^{b}$ & 19.54$^{b}$ & Absorption & Outflow \\
.............. &  &  & -0.07$^{r}$ & 189$\pm$9$^{r}$ & 239$^{r}$ & 196$\pm$5$^{r}$ & -0.02$\pm$0.29$^{r}$ & 13.58$^{r}$ & 21.25$^{r}$ & Emission & Outflow \\
.............. & 19.98 & 10.80 & 0.34 & -157$\pm$99 & -200 & 188$\pm$55 & 0.08$\pm$0.28 & 13.95 & 21.63 & Absorption & Outflow \\
.............. & 12.93 & 11.07 & 0.20 & -84$\pm$103 & -111 & 243$\pm$43 & 0.77$\pm$0.26 & 12.41 & 20.08 & Absorption & Outflow \\        
BULGE & 1.21 & 10.02 & -0.04 & 303$\pm$26 & 393 & 84$\pm$41 & -0.53$\pm$0.26 & 11.92 & 19.60 & Emission & Outflow \\
.............. & 2.73 & 10.08 & 0.10 & -401$\pm$88 & -544 & 340$\pm$58 & 0.25$\pm$0.27 & 12.62 & 20.30 & Absorption & Outflow \\
.............. & 3.25 & 10.45 & 0.11 & -163$\pm$16 & -219 & 196$\pm$24 & 0.64$\pm$0.24 & 12.25 & 19.93 & Absorption & Outflow \\
.............. & 8.20 & 10.47 & 0.25 & -180$\pm$32 & -236 & 164$\pm$24 & 0.10$\pm$0.24 & 13.62 & 21.30 & Absorption & Outflow \\
.............. & 3.13 & 10.77 & 0.03 & -177$\pm$67 & -248 & 75$\pm$65 & 0.38$\pm$0.27 & 11.99 & 19.66 & Absorption & Outflow \\
.............. & 8.20 & 10.79 & 0.29 & -111$\pm$23 & -149 & 200$\pm$17 & 0.17$\pm$0.25 & 13.33 & 21.01 & Absorption & Outflow \\
.............. & 21.67 & 10.83 & 0.67 & -145$\pm$37 & -202 & 266$\pm$26 & 0.37$\pm$0.22 & 13.36 & 21.04 & Absorption & Outflow \\
.............. & 31.86 & 11.11 & 0.47 & -113$\pm$117 & -162 & 282$\pm$61 & 0.14$\pm$0.25 & 13.72 & 21.40 & Absorption & Outflow \\

$i$-log\,SFR & 4.85 & 10.48 & 0.15 & -162$\pm$19 & -169 & 143$\pm$30 & 0.71$\pm$0.24 & 12.35 & 20.02 & Absorption & Outflow \\
.............. & 12.52 & 10.75 & 0.41 & -158$\pm$37 & -164 & 214$\pm$26 & 0.15$\pm$0.24 & 13.63 & 21.30 & Absorption & Outflow \\
.............. & 2.44 & 10.21 & 0.06$^{b}$ & -153$\pm$6$^{b}$ & -170$^{b}$ & 116$\pm$13$^{b}$ & 0.48$\pm$0.22$^{b}$ & 12.17$^{b}$ & 19.84$^{b}$ & Absorption & Outflow \\
.............. &  &  & -0.05$^{r}$ & 195$\pm$8$^{r}$ & 216$^{r}$ & 191$\pm$5$^{r}$ & -0.60$\pm$0.24$^{r}$ & 11.92$^{r}$ & 19.60$^{r}$ & Emission & Outflow \\
.............. & 4.97 & 10.46 & 0.10$^{b}$ & -192$\pm$16$^{b}$ & -213$^{b}$ & 149$\pm$13$^{b}$ & 0.90$\pm$0.21$^{b}$ & 12.06$^{b}$ & 19.73$^{b}$ & Absorption & Outflow \\
.............. &  &  & -0.15$^{r}$ & 55$\pm$23$^{r}$ & 61$^{r}$ & 117$\pm$18$^{r}$ & -0.62$\pm$0.20$^{r}$ & 12.41$^{r}$ & 20.09$^{r}$ & Emission & Outflow \\
.............. & 12.80 & 10.70 & 0.36 & -144$\pm$35 & -161 & 206$\pm$19 & 0.16$\pm$0.24 & 13.48 & 21.16 & Absorption & Outflow \\
.............. & 2.37 & 10.19 & 0.05 & -157$\pm$20 & -194 & 55$\pm$26 & 0.15$\pm$0.26 & 12.86 & 20.54 & Absorption & Outflow \\
.............. & 4.94 & 10.46 & 0.12 & -149$\pm$9 & -183 & 82$\pm$17 & 0.06$\pm$0.15 & 13.53 & 21.20 & Absorption & Outflow \\
.............. & 13.27 & 10.69 & 0.46 & -69$\pm$8 & -84 & 152$\pm$11 & 0.15$\pm$0.02 & 13.74 & 21.42 & Absorption & Outflow \\
.............. & 13.15 & 10.77 & 0.23 & -76$\pm$16 & -107 & 114$\pm$16 & 0.05$\pm$0.01 & 14.31 & 21.99 & Absorption & Outflow \\
.............. & 2.30 & 10.27 & 0.02 & 162$\pm$28 & 283 & 50$\pm$19 & 0.36$\pm$0.26 & 12.21 & 19.88 & Absorption & Inflow \\
.............. & 2.19 & 10.30 & 0.06 & 150$\pm$54 & 345 & 60$\pm$37 & 0.23$\pm$0.23 & 12.74 & 20.42 & Absorption & Inflow \\
.............. & 4.74 & 10.59 & 0.08 & 148$\pm$4 & 338 & 53$\pm$11 & 0.88$\pm$0.24 & 12.36 & 20.04 & Absorption & Inflow \\
.............. & 12.42 & 10.82 & 0.10 & 145$\pm$76 & 331 & 65$\pm$47 & 0.26$\pm$0.23 & 12.83 & 20.51 & Absorption & Inflow \\
.............. & 2.11 & 10.39 & 0.03 & 184$\pm$59 & 606 & 57$\pm$37 & 0.62$\pm$0.25 & 11.70 & 19.38 & Absorption & Inflow \\
.............. & 4.52 & 10.68 & 0.10 & 146$\pm$51 & 487 & 57$\pm$31 & 0.36$\pm$0.23 & 12.79 & 20.47 & Absorption & Inflow \\
$i$-log\,M$_{*}$ & 3.62 & 10.30 & 0.15 & -152$\pm$26 & -159 & 160$\pm$35 & 0.58$\pm$0.25 & 12.46 & 20.14 & Absorption & Outflow \\
.............. & 4.50 & 10.50 & 0.16 & -146$\pm$19 & -152 & 173$\pm$26 & 0.91$\pm$0.24 & 12.27 & 19.94 & Absorption & Outflow \\
.............. & 6.42 & 10.69 & 0.24 & -143$\pm$24 & -149 & 216$\pm$38 & 0.60$\pm$0.25 & 12.64 & 20.32 & Absorption & Outflow \\
.............. & 8.99 & 10.89 & 0.20 & -178$\pm$83 & -184 & 250$\pm$57 & 0.67$\pm$0.26 & 12.50 & 20.18 & Absorption & Outflow \\
.............. & 2.52 & 10.11 & 0.01$^{b}$ & -206$\pm$85$^{b}$ & -228$^{b}$ & 65$\pm$42$^{b}$ & 0.74$\pm$0.28$^{b}$ & 10.94$^{b}$ & 18.61$^{b}$ & Absorption & Outflow \\
.............. &  &  & -0.09$^{r}$ & 198$\pm$8$^{r}$ & 219$^{r}$ & 199$\pm$7$^{r}$ & -0.73$\pm$0.26$^{r}$ & 12.12$^{r}$ & 19.80$^{r}$ & Emission & Outflow \\
.............. & 3.71 & 10.30 & 0.02$^{b}$ & -173$\pm$69$^{b}$ & -192$^{b}$ & 71$\pm$43$^{b}$ & 0.59$\pm$0.25$^{b}$ & 11.75$^{b}$ & 19.43$^{b}$ & Absorption & Outflow \\
.............. &  &  & -0.09$^{r}$ & 199$\pm$11$^{r}$ & 222$^{r}$ & 199$\pm$5$^{r}$ & -0.20$\pm$0.26$^{r}$ & 12.70$^{r}$ & 20.37$^{r}$ & Emission & Outflow \\
.............. & 4.95 & 10.49 & 0.13 & -166$\pm$12 & -184 & 54$\pm$36 & 0.10$\pm$0.27 & 13.65 & 21.33 & Absorption & Outflow \\
.............. & 6.61 & 10.69 & 0.18 & -146$\pm$21 & -162 & 187$\pm$19 & 0.90$\pm$0.24 & 12.34 & 20.02 & Absorption & Outflow \\
.............. & 8.09 & 10.88 & 0.17 & -147$\pm$23 & -163 & 143$\pm$28 & 0.42$\pm$0.26 & 12.64 & 20.32 & Absorption & Outflow \\
.............. & 3.63 & 10.30 & 0.00$^{b}$ & -234$\pm$81$^{b}$ & -287$^{b}$ & 86$\pm$39$^{b}$ & 0.67$\pm$0.26$^{b}$ & 10.80$^{b}$ & 18.48$^{b}$ & Absorption & Outflow \\
.............. &  &  & -0.08$^{r}$ & 194$\pm$32$^{r}$ & 238$^{r}$ & 195$\pm$5$^{r}$ & -0.02$\pm$0.26$^{r}$ & 13.96$^{r}$ & 21.64$^{r}$ & Emission & Outflow \\
.............. & 4.89 & 10.50 & 0.03$^{b}$ & -172$\pm$62$^{b}$ & -211$^{b}$ & 88$\pm$35$^{b}$ & 0.34$\pm$0.21$^{b}$ & 11.96$^{b}$ & 19.64$^{b}$ & Absorption & Outflow \\
.............. &  &  & -0.08$^{r}$ & 150$\pm$21$^{r}$ & 184$^{r}$ & 192$\pm$11$^{r}$ & -0.02$\pm$0.02$^{r}$ & 14.47$^{r}$ & 22.15$^{r}$ & Emission & Outflow \\
.............. & 6.88 & 10.68 & 0.18 & -130$\pm$30 & -159 & 117$\pm$28 & 0.07$\pm$0.29 & 13.63 & 21.31 & Absorption & Outflow \\
.............. & 8.03 & 10.88 & 0.06 & -171$\pm$42 & -209 & 92$\pm$26 & 0.73$\pm$0.25 & 11.98 & 19.66 & Absorption & Outflow \\
.............. & 9.25 & 11.08 & 0.19 & -108$\pm$53 & -132 & 182$\pm$42 & 0.13$\pm$0.27 & 13.28 & 20.96 & Absorption & Outflow \\
.............. & 8.03 & 10.88 & 0.04 & -179$\pm$106 & -253 & 95$\pm$65 & 0.33$\pm$0.26 & 12.09 & 19.76 & Absorption & Outflow \\
.............. & 3.16 & 10.31 & 0.02 & 166$\pm$20 & 288 & 50$\pm$14 & 0.96$\pm$0.27 & 11.60 & 19.27 & Absorption & Inflow \\
.............. & 7.46 & 10.88 & 0.04 & 148$\pm$23 & 260 & 56$\pm$34 & 0.64$\pm$0.24 & 12.14 & 19.82 & Absorption & Inflow \\
.............. & 2.64 & 10.30 & 0.04 & 153$\pm$33 & 346 & 51$\pm$29 & 0.81$\pm$0.25 & 12.17 & 19.84 & Absorption & Inflow \\
.............. & 3.66 & 10.50 & 0.04 & 162$\pm$14 & 371 & 50$\pm$8 & 0.84$\pm$0.25 & 12.03 & 19.71 & Absorption & Inflow \\
.............. & 4.71 & 10.69 & 0.08 & 148$\pm$20 & 342 & 50$\pm$21 & 0.97$\pm$0.24 & 12.43 & 20.11 & Absorption & Inflow \\
.............. & 6.17 & 10.87 & 0.09 & 148$\pm$8 & 347 & 51$\pm$22 & 0.49$\pm$0.25 & 12.77 & 20.44 & Absorption & Inflow \\
.............. & 8.84 & 11.04 & 0.07 & 156$\pm$113 & 365 & 73$\pm$69 & 0.46$\pm$0.26 & 12.34 & 20.01 & Absorption & Inflow \\
.............. & 3.15 & 10.51 & 0.05 & 160$\pm$81 & 523 & 72$\pm$52 & 0.56$\pm$0.25 & 12.12 & 19.80 & Absorption & Inflow \\
.............. & 3.92 & 10.69 & 0.06 & 162$\pm$28 & 543 & 60$\pm$33 & 0.72$\pm$0.24 & 12.16 & 19.84 & Absorption & Inflow \\
.............. & 5.33 & 10.86 & 0.11 & 153$\pm$60 & 523 & 67$\pm$40 & 0.95$\pm$0.23 & 12.26 & 19.94 & Absorption & Inflow \\
        \hline
    \end{tabular}
\end{table*}

\begin{table*}
    \caption{The same as Table \ref{tab:flowpropstab} but for AGN.}
    \label{tab:flowpropstab2}
    \begin{tabular}{cccccccccccc}
        \hline
        Sample & SFR & log M$_{*}$/M$_{\odot}$ & EW & $\Delta$v & $\Delta$v/cos($i$) & b$_{\text{D}}$ & C$_{f}$ & N(Na I) & N(H) & Profile & Type \\
        \hline
DISK & 0.34 & 10.81 & -0.16 & 116$\pm$87 & 120 & 185$\pm$55 & -0.82$\pm$0.27 & 12.26 & 19.94 & Emission & Outflow \\
.............. & 0.26 & 10.87 & 0.02 & 174$\pm$88 & 403 & 84$\pm$67 & 0.68$\pm$0.27 & 11.47 & 19.14 & Absorption & Inflow \\
.............. & 0.25 & 10.89 & 0.13 & 98$\pm$25 & 356 & 181$\pm$21 & 0.54$\pm$0.24 & 12.37 & 20.05 & Absorption & Inflow \\
.............. & 0.25 & 10.94 & 0.22 & 132$\pm$29 & 819 & 181$\pm$28 & 0.52$\pm$0.24 & 12.61 & 20.29 & Absorption & Inflow \\
.............. & 0.27 & 10.77 & -0.03 & 180$\pm$101 & 188 & 178$\pm$59 & -0.49$\pm$0.26 & 11.79 & 19.47 & Emission & Outflow \\
.............. & 0.27 & 10.76 & -0.02 & 191$\pm$116 & 234 & 69$\pm$71 & -0.23$\pm$0.26 & 11.86 & 19.54 & Emission & Outflow \\
.............. & 0.20 & 10.75 & 0.07 & 158$\pm$22 & 376 & 165$\pm$13 & 0.21$\pm$0.26 & 12.52 & 20.20 & Absorption & Inflow \\
.............. & 0.23 & 10.79 & 0.11 & 165$\pm$31 & 600 & 144$\pm$15 & 0.94$\pm$0.25 & 12.03 & 19.71 & Absorption & Inflow \\
.............. & 0.39 & 10.85 & 0.13 & 191$\pm$31 & 1342 & 121$\pm$22 & 0.67$\pm$0.26 & 12.28 & 19.95 & Absorption & Inflow \\
.............. & 0.36 & 10.55 & 0.04 & 174$\pm$30 & 415 & 96$\pm$17 & 0.59$\pm$0.25 & 11.82 & 19.49 & Absorption & Inflow \\
.............. & 0.39 & 10.61 & 0.08 & 169$\pm$31 & 630 & 95$\pm$24 & 0.63$\pm$0.23 & 12.09 & 19.77 & Absorption & Inflow \\
.............. & 0.55 & 10.63 & 0.12 & 158$\pm$30 & 1186 & 144$\pm$21 & 0.62$\pm$0.26 & 12.26 & 19.94 & Absorption & Inflow \\
.............. & 0.94 & 10.40 & -0.04 & 250$\pm$104 & 261 & 181$\pm$50 & -0.56$\pm$0.26 & 11.87 & 19.55 & Emission & Outflow \\
.............. & 0.90 & 10.39 & -0.05 & 158$\pm$84 & 175 & 195$\pm$38 & -0.86$\pm$0.25 & 11.72 & 19.40 & Emission & Outflow \\
.............. & 0.88 & 10.42 & -0.03 & 296$\pm$121 & 363 & 177$\pm$62 & -0.24$\pm$0.25 & 12.10 & 19.78 & Emission & Outflow \\
.............. & 0.79 & 10.42 & -0.02 & 290$\pm$95 & 411 & 153$\pm$65 & -0.92$\pm$0.25 & 11.37 & 19.05 & Emission & Outflow \\
.............. & 0.63 & 10.50 & 0.05 & 174$\pm$43 & 413 & 96$\pm$26 & 0.73$\pm$0.24 & 11.77 & 19.45 & Absorption & Inflow \\
.............. & 0.71 & 10.59 & 0.07 & 179$\pm$39 & 661 & 85$\pm$27 & 0.69$\pm$0.24 & 11.97 & 19.65 & Absorption & Inflow \\
.............. & 0.96 & 10.70 & 0.12 & 174$\pm$24 & 1220 & 116$\pm$16 & 0.34$\pm$0.24 & 12.55 & 20.23 & Absorption & Inflow \\
.............. & 1.39 & 10.45 & -0.06 & 185$\pm$65 & 193 & 198$\pm$40 & -0.86$\pm$0.26 & 11.85 & 19.53 & Emission & Outflow \\
.............. & 1.53 & 10.47 & -0.06 & 170$\pm$52 & 189 & 177$\pm$36 & -0.29$\pm$0.24 & 12.30 & 19.98 & Emission & Outflow \\
.............. & 1.43 & 10.45 & -0.06 & 210$\pm$48 & 257 & 185$\pm$29 & -0.37$\pm$0.26 & 12.19 & 19.87 & Emission & Outflow \\
.............. & 1.21 & 10.50 & 0.01 & 175$\pm$144 & 306 & 66$\pm$106 & 0.48$\pm$0.28 & 11.27 & 18.94 & Absorption & Inflow \\
.............. & 1.10 & 10.55 & 0.05 & 186$\pm$15 & 441 & 86$\pm$18 & 0.20$\pm$0.25 & 12.37 & 20.04 & Absorption & Inflow \\
.............. & 1.22 & 10.63 & 0.10 & 163$\pm$15 & 599 & 112$\pm$19 & 0.55$\pm$0.19 & 12.23 & 19.90 & Absorption & Inflow \\
.............. & 1.44 & 10.72 & 0.12 & 177$\pm$42 & 1255 & 95$\pm$28 & 0.92$\pm$0.26 & 12.09 & 19.76 & Absorption & Inflow \\
.............. & 2.14 & 10.40 & -0.10 & 194$\pm$25 & 216 & 207$\pm$24 & -0.35$\pm$0.25 & 12.44 & 20.12 & Emission & Outflow \\
.............. & 2.11 & 10.50 & 0.01$^{b}$ & -194$\pm$108$^{b}$ & -275$^{b}$ & 52$\pm$43$^{b}$ & 0.67$\pm$0.26$^{b}$ & 11.45$^{b}$ & 19.13$^{b}$ & Absorption & Outflow \\
.............. &  &  & -0.05$^{r}$ & 193$\pm$31$^{r}$ & 273$^{r}$ & 193$\pm$26$^{r}$ & -0.88$\pm$0.23$^{r}$ & 11.70$^{r}$ & 19.38$^{r}$ & Emission & Outflow \\
.............. & 1.85 & 10.57 & 0.05 & 194$\pm$47 & 462 & 63$\pm$35 & 0.63$\pm$0.25 & 11.84 & 19.52 & Absorption & Inflow \\
.............. & 1.90 & 10.60 & 0.08 & 179$\pm$19 & 651 & 96$\pm$21 & 0.31$\pm$0.25 & 12.39 & 20.06 & Absorption & Inflow \\
.............. & 2.36 & 10.70 & 0.08 & 132$\pm$94 & 875 & 111$\pm$52 & 0.47$\pm$0.25 & 12.22 & 19.90 & Absorption & Inflow \\
.............. & 4.23 & 10.45 & 0.16 & -389$\pm$67 & -392 & 290$\pm$80 & 0.88$\pm$0.28 & 12.23 & 19.91 & Absorption & Outflow \\
.............. & 2.85 & 10.50 & 0.13 & -238$\pm$44 & -249 & 248$\pm$30 & 0.74$\pm$0.27 & 12.22 & 19.89 & Absorption & Outflow \\
.............. & 3.12 & 10.43 & 0.07$^{b}$ & -232$\pm$26$^{b}$ & -258$^{b}$ & 140$\pm$19$^{b}$ & 0.43$\pm$0.25$^{b}$ & 12.18$^{b}$ & 19.86$^{b}$ & Absorption & Outflow \\
.............. &  &  & -0.07$^{r}$ & 190$\pm$43$^{r}$ & 211$^{r}$ & 194$\pm$36$^{r}$ & -0.57$\pm$0.27$^{r}$ & 12.06$^{r}$ & 19.74$^{r}$ & Emission & Outflow \\
.............. & 3.53 & 10.52 & 0.08$^{b}$ & -209$\pm$35$^{b}$ & -257$^{b}$ & 156$\pm$19$^{b}$ & 0.64$\pm$0.24$^{b}$ & 12.08$^{b}$ & 19.75$^{b}$ & Absorption & Outflow \\
.............. &  &  & -0.07$^{r}$ & 197$\pm$48$^{r}$ & 243$^{r}$ & 197$\pm$39$^{r}$ & -0.23$\pm$0.28$^{r}$ & 12.45$^{r}$ & 20.13$^{r}$ & Emission & Outflow \\
.............. & 3.41 & 10.53 & 0.08 & -198$\pm$39 & -282 & 248$\pm$12 & 0.72$\pm$0.27 & 12.04 & 19.72 & Absorption & Outflow \\
.............. & 3.39 & 10.55 & 0.03 & 145$\pm$85 & 256 & 126$\pm$60 & 0.62$\pm$0.25 & 11.73 & 19.41 & Absorption & Inflow \\
.............. & 3.14 & 10.57 & 0.07 & 151$\pm$22 & 352 & 132$\pm$20 & 0.79$\pm$0.23 & 11.92 & 19.60 & Absorption & Inflow \\
.............. & 3.08 & 10.64 & 0.10 & 183$\pm$17 & 642 & 86$\pm$22 & 0.37$\pm$0.23 & 12.44 & 20.12 & Absorption & Inflow \\
.............. & 4.80 & 10.44 & 0.37 & -201$\pm$30 & -210 & 261$\pm$33 & 0.64$\pm$0.25 & 12.76 & 20.44 & Absorption & Outflow \\
.............. & 5.86 & 10.63 & 0.27 & -242$\pm$72 & -271 & 300$\pm$36 & 0.70$\pm$0.26 & 12.58 & 20.26 & Absorption & Outflow \\
.............. & 6.15 & 10.59 & 0.27 & -172$\pm$76 & -212 & 269$\pm$37 & 0.18$\pm$0.24 & 13.21 & 20.89 & Absorption & Outflow \\
.............. & 6.66 & 10.62 & 0.14 & -406$\pm$66 & -582 & 442$\pm$46 & 0.77$\pm$0.27 & 12.23 & 19.91 & Absorption & Outflow \\
.............. & 6.35 & 10.66 & 0.08 & 197$\pm$25 & 458 & 73$\pm$26 & 0.75$\pm$0.25 & 12.00 & 19.67 & Absorption & Inflow \\
.............. & 6.59 & 10.71 & 0.09 & 199$\pm$23 & 712 & 71$\pm$26 & 0.75$\pm$0.25 & 12.08 & 19.75 & Absorption & Inflow \\
.............. & 11.69 & 10.57 & 0.53 & -256$\pm$96 & -268 & 263$\pm$55 & 0.21$\pm$0.25 & 13.48 & 21.16 & Absorption & Outflow \\
.............. & 12.74 & 10.72 & 0.60 & -258$\pm$75 & -286 & 304$\pm$41 & 0.27$\pm$0.25 & 13.42 & 21.09 & Absorption & Outflow \\
.............. & 13.38 & 10.67 & 0.35 & -276$\pm$88 & -339 & 322$\pm$57 & 0.22$\pm$0.27 & 13.22 & 20.89 & Absorption & Outflow \\
.............. & 13.61 & 10.76 & 0.15 & 128$\pm$62 & 289 & 146$\pm$33 & 0.60$\pm$0.22 & 12.38 & 20.06 & Absorption & Inflow \\
.............. & 11.43 & 10.69 & 0.11 & 142$\pm$142 & 510 & 136$\pm$96 & 0.95$\pm$0.27 & 12.06 & 19.74 & Absorption & Inflow \\
HIGH-$i$ & 2.82 & 10.47 & 0.03 & 199$\pm$20 & 406 & 56$\pm$21 & 0.52$\pm$0.26 & 11.78 & 19.46 & Absorption & Inflow \\
.............. & 1.72 & 10.77 & 0.05 & 185$\pm$9 & 449 & 90$\pm$11 & 0.28$\pm$0.24 & 12.28 & 19.95 & Absorption & Inflow \\
.............. & 4.64 & 10.78 & 0.04 & 188$\pm$21 & 404 & 104$\pm$21 & 0.94$\pm$0.25 & 11.61 & 19.29 & Absorption & Inflow \\
.............. & 13.61 & 10.83 & 0.05 & 180$\pm$116 & 368 & 112$\pm$72 & 0.52$\pm$0.26 & 11.94 & 19.62 & Absorption & Inflow \\
.............. & 2.61 & 11.10 & 0.09 & 145$\pm$27 & 354 & 165$\pm$20 & 0.80$\pm$0.20 & 12.05 & 19.73 & Absorption & Inflow \\
.............. & 7.13 & 11.13 & 0.10 & 113$\pm$59 & 249 & 161$\pm$32 & 0.43$\pm$0.23 & 12.36 & 20.04 & Absorption & Inflow \\
.............. & 23.11 & 11.11 & 0.13 & 159$\pm$132 & 315 & 170$\pm$82 & 0.83$\pm$0.26 & 12.17 & 19.85 & Absorption & Inflow \\
        \hline
    \end{tabular}
\end{table*}

\begin{table*}
    \contcaption{}
    \begin{tabular}{cccccccccccc}
        \hline
        Sample & SFR & log M$_{*}$/M$_{\odot}$ & EW & $\Delta$v & $\Delta$v/cos($i$) & b$_{\text{D}}$ & C$_{f}$ & N(Na I) & N(H) & Profile & Type \\
        \hline
LOW-$i$ & 1.31 & 10.11 & -0.05 & 308$\pm$69 & 376 & 196$\pm$49 & -0.46$\pm$0.26 & 12.03 & 19.71 & Emission & Outflow \\
.............. & 3.40 & 10.15 & 0.13 & -355$\pm$109 & -442 & 383$\pm$68 & 0.78$\pm$0.25 & 12.22 & 19.89 & Absorption & Outflow \\
.............. & 0.53 & 10.39 & -0.01 & 364$\pm$127 & 462 & 55$\pm$99 & -0.71$\pm$0.27 & 11.35 & 19.03 & Emission & Outflow \\
.............. & 1.53 & 10.41 & 0.03$^{b}$ & -194$\pm$80$^{b}$ & -242$^{b}$ & 87$\pm$39$^{b}$ & 0.55$\pm$0.27$^{b}$ & 11.77$^{b}$ & 19.45$^{b}$ & Absorption & Outflow \\
.............. &  &  & -0.19$^{r}$ & 21$\pm$20$^{r}$ & 27$^{r}$ & 199$\pm$5$^{r}$ & -0.04$\pm$0.00$^{r}$ & 13.98$^{r}$ & 21.66$^{r}$ & Emission & Outflow \\
.............. & 3.63 & 10.46 & 0.05$^{b}$ & -222$\pm$31$^{b}$ & -279$^{b}$ & 184$\pm$30$^{b}$ & 0.90$\pm$0.25$^{b}$ & 11.69$^{b}$ & 19.37$^{b}$ & Absorption & Outflow \\
.............. &  &  & -0.08$^{r}$ & 198$\pm$9$^{r}$ & 249$^{r}$ & 199$\pm$6$^{r}$ & -0.16$\pm$0.25$^{r}$ & 12.68$^{r}$ & 20.35$^{r}$ & Emission & Outflow \\
.............. & 10.11 & 10.52 & 0.20 & -246$\pm$70 & -319 & 299$\pm$30 & 0.93$\pm$0.26 & 12.31 & 19.98 & Absorption & Outflow \\
.............. & 0.58 & 10.76 & -0.02 & 173$\pm$125 & 217 & 157$\pm$76 & -0.74$\pm$0.27 & 11.29 & 18.97 & Emission & Outflow \\
.............. & 1.74 & 10.76 & 0.00$^{b}$ & -250$\pm$106$^{b}$ & -313$^{b}$ & 66$\pm$32$^{b}$ & 0.46$\pm$0.16$^{b}$ & 10.75$^{b}$ & 18.43$^{b}$ & Absorption & Outflow \\
.............. &  &  & -0.12$^{r}$ & 77$\pm$8$^{r}$ & 97$^{r}$ & 200$\pm$5$^{r}$ & -0.03$\pm$0.01$^{r}$ & 13.87$^{r}$ & 21.55$^{r}$ & Emission & Outflow \\
.............. & 4.60 & 10.77 & 0.05$^{b}$ & -229$\pm$29$^{b}$ & -287$^{b}$ & 157$\pm$18$^{b}$ & 0.46$\pm$0.24$^{b}$ & 12.04$^{b}$ & 19.72$^{b}$ & Absorption & Outflow \\
.............. &  &  & -0.08$^{r}$ & 199$\pm$38$^{r}$ & 249$^{r}$ & 198$\pm$33$^{r}$ & -0.08$\pm$0.27$^{r}$ & 13.02$^{r}$ & 20.70$^{r}$ & Emission & Outflow \\
.............. & 13.74 & 10.81 & 0.23 & -336$\pm$68 & -438 & 348$\pm$48 & 0.58$\pm$0.27 & 12.58 & 20.26 & Absorption & Outflow \\
.............. & 42.22 & 10.81 & 0.66 & -258$\pm$63 & -351 & 323$\pm$47 & 0.42$\pm$0.24 & 13.22 & 20.90 & Absorption & Outflow \\
.............. & 6.16 & 11.12 & 0.07 & -426$\pm$136 & -550 & 443$\pm$66 & 0.59$\pm$0.26 & 12.04 & 19.72 & Absorption & Outflow \\
.............. & 18.91 & 11.12 & 0.44 & -212$\pm$111 & -267 & 278$\pm$52 & 0.09$\pm$0.26 & 13.92 & 21.60 & Absorption & Outflow \\
BULGE & 0.83 & 10.07 & -0.03 & 312$\pm$73 & 395 & 142$\pm$62 & -0.86$\pm$0.27 & 11.58 & 19.26 & Emission & Outflow \\
.............. & 2.27 & 10.07 & 0.15 & -349$\pm$81 & -428 & 164$\pm$86 & 0.03$\pm$0.32 & 14.61 & 22.29 & Absorption & Outflow \\
.............. & 2.18 & 10.44 & 0.10 & -195$\pm$22 & -253 & 228$\pm$38 & 0.86$\pm$0.25 & 12.08 & 19.76 & Absorption & Outflow \\
.............. & 6.01 & 10.49 & 0.27 & -255$\pm$24 & -322 & 267$\pm$27 & 0.71$\pm$0.26 & 12.61 & 20.29 & Absorption & Outflow \\
.............. & 2.44 & 10.78 & 0.03 & -488$\pm$44 & -700 & 449$\pm$13 & 0.35$\pm$0.25 & 11.89 & 19.57 & Absorption & Outflow \\
.............. & 6.76 & 10.80 & 0.19 & -196$\pm$42 & -258 & 276$\pm$24 & 0.86$\pm$0.24 & 12.38 & 20.06 & Absorption & Outflow \\
.............. & 22.61 & 10.83 & 0.61 & -227$\pm$44 & -300 & 233$\pm$32 & 0.15$\pm$0.25 & 13.91 & 21.58 & Absorption & Outflow \\
.............. & 12.44 & 11.08 & 0.21 & -180$\pm$107 & -264 & 237$\pm$55 & 0.05$\pm$0.26 & 13.95 & 21.62 & Absorption & Outflow \\
$i$-log\,SFR & 5.78 & 10.65 & 0.24 & -197$\pm$35 & -206 & 228$\pm$40 & 0.68$\pm$0.26 & 12.57 & 20.25 & Absorption & Outflow \\
.............. & 2.26 & 10.37 & 0.17 & -160$\pm$34 & -178 & 208$\pm$42 & 0.35$\pm$0.26 & 12.74 & 20.42 & Absorption & Outflow \\
.............. & 5.32 & 10.63 & 0.26 & -195$\pm$25 & -218 & 259$\pm$25 & 0.92$\pm$0.24 & 12.48 & 20.16 & Absorption & Outflow \\
.............. & 14.30 & 10.87 & 0.44 & -235$\pm$77 & -262 & 305$\pm$44 & 0.28$\pm$0.26 & 13.28 & 20.96 & Absorption & Outflow \\
.............. & 2.33 & 10.32 & 0.21 & -124$\pm$28 & -155 & 205$\pm$24 & 0.27$\pm$0.25 & 12.95 & 20.63 & Absorption & Outflow \\
.............. & 5.52 & 10.63 & 0.25 & -133$\pm$22 & -163 & 210$\pm$15 & 0.11$\pm$0.24 & 13.50 & 21.18 & Absorption & Outflow \\
.............. & 15.17 & 10.90 & 0.32 & -238$\pm$82 & -293 & 245$\pm$45 & 0.11$\pm$0.27 & 13.61 & 21.29 & Absorption & Outflow \\
.............. & 5.38 & 10.63 & 0.07 & -370$\pm$90 & -529 & 387$\pm$53 & 0.44$\pm$0.27 & 12.24 & 19.92 & Absorption & Outflow \\
.............. & 15.16 & 10.83 & 0.12 & -362$\pm$95 & -517 & 373$\pm$71 & 0.89$\pm$0.28 & 12.17 & 19.84 & Absorption & Outflow \\
.............. & 2.21 & 10.41 & 0.08 & 142$\pm$34 & 250 & 69$\pm$25 & 0.66$\pm$0.22 & 12.27 & 19.95 & Absorption & Inflow \\
.............. & 2.08 & 10.45 & 0.11 & 139$\pm$26 & 326 & 66$\pm$16 & 0.33$\pm$0.22 & 12.74 & 20.42 & Absorption & Inflow \\
.............. & 5.08 & 10.67 & 0.08 & 147$\pm$10 & 343 & 50$\pm$31 & 0.22$\pm$0.26 & 13.09 & 20.77 & Absorption & Inflow \\
.............. & 2.13 & 10.48 & 0.05 & 193$\pm$59 & 683 & 61$\pm$33 & 0.31$\pm$0.24 & 12.18 & 19.86 & Absorption & Inflow \\
.............. & 5.00 & 10.71 & 0.11 & 146$\pm$56 & 509 & 53$\pm$40 & 0.31$\pm$0.24 & 13.01 & 20.69 & Absorption & Inflow \\
.............. & 12.34 & 10.95 & 0.12 & 183$\pm$46 & 651 & 105$\pm$28 & 0.45$\pm$0.23 & 12.48 & 20.16 & Absorption & Inflow \\
$i$-log\,M$_{*}$ & 1.58 & 10.12 & 0.08 & -288$\pm$117 & -300 & 211$\pm$96 & 0.53$\pm$0.29 & 12.21 & 19.89 & Absorption & Outflow \\
.............. & 3.47 & 10.49 & 0.31 & -193$\pm$50 & -200 & 256$\pm$60 & 0.51$\pm$0.28 & 12.83 & 20.51 & Absorption & Outflow \\
.............. & 7.76 & 10.67 & 0.42 & -254$\pm$36 & -263 & 214$\pm$38 & 0.12$\pm$0.27 & 13.78 & 21.46 & Absorption & Outflow \\
.............. & 1.99 & 10.14 & 0.17 & -149$\pm$76 & -165 & 190$\pm$64 & 0.65$\pm$0.27 & 12.44 & 20.12 & Absorption & Outflow \\
.............. & 2.92 & 10.36 & 0.15 & -159$\pm$26 & -178 & 112$\pm$35 & 0.67$\pm$0.27 & 12.38 & 20.06 & Absorption & Outflow \\
.............. & 3.97 & 10.49 & 0.22 & -281$\pm$63 & -311 & 309$\pm$41 & 0.56$\pm$0.27 & 12.63 & 20.31 & Absorption & Outflow \\
.............. & 5.80 & 10.70 & 0.29 & -185$\pm$32 & -206 & 263$\pm$36 & 0.97$\pm$0.25 & 12.51 & 20.19 & Absorption & Outflow \\
.............. & 7.18 & 10.86 & 0.28 & -165$\pm$56 & -183 & 238$\pm$53 & 0.08$\pm$0.27 & 13.79 & 21.47 & Absorption & Outflow \\
.............. & 10.45 & 11.05 & 0.37 & -235$\pm$51 & -263 & 243$\pm$58 & 0.30$\pm$0.27 & 13.18 & 20.86 & Absorption & Outflow \\
.............. & 2.26 & 10.14 & 0.11 & -159$\pm$32 & -197 & 118$\pm$34 & 0.40$\pm$0.27 & 12.48 & 20.15 & Absorption & Outflow \\
.............. & 2.97 & 10.32 & 0.17 & -141$\pm$39 & -172 & 182$\pm$34 & 0.81$\pm$0.25 & 12.35 & 20.02 & Absorption & Outflow \\
.............. & 4.30 & 10.50 & 0.18 & -133$\pm$49 & -164 & 186$\pm$30 & 0.10$\pm$0.25 & 13.38 & 21.06 & Absorption & Outflow \\
.............. & 6.14 & 10.70 & 0.38 & -113$\pm$28 & -138 & 200$\pm$20 & 0.15$\pm$0.25 & 13.59 & 21.27 & Absorption & Outflow \\
.............. & 7.24 & 10.90 & 0.26 & -167$\pm$44 & -207 & 186$\pm$26 & 0.08$\pm$0.25 & 13.73 & 21.41 & Absorption & Outflow \\
.............. & 11.41 & 11.06 & 0.30 & -160$\pm$68 & -198 & 220$\pm$39 & 0.11$\pm$0.26 & 13.61 & 21.29 & Absorption & Outflow \\
.............. & 4.01 & 10.50 & 0.14 & -153$\pm$81 & -217 & 187$\pm$54 & 0.04$\pm$0.28 & 13.92 & 21.60 & Absorption & Outflow \\
.............. & 5.92 & 10.70 & 0.07 & -221$\pm$140 & -318 & 186$\pm$94 & 0.86$\pm$0.27 & 11.95 & 19.63 & Absorption & Outflow \\
.............. & 8.74 & 10.87 & 0.23 & -124$\pm$118 & -179 & 198$\pm$68 & 0.05$\pm$0.27 & 14.00 & 21.68 & Absorption & Outflow \\        
.............. & 2.54 & 10.32 & 0.03 & 163$\pm$117 & 288 & 52$\pm$93 & 0.94$\pm$0.28 & 11.84 & 19.51 & Absorption & Inflow \\
.............. & 8.06 & 11.09 & 0.06 & 146$\pm$136 & 257 & 53$\pm$90 & 0.55$\pm$0.27 & 12.48 & 20.15 & Absorption & Inflow \\
.............. & 2.04 & 10.33 & 0.03 & 178$\pm$108 & 422 & 61$\pm$80 & 0.32$\pm$0.26 & 12.06 & 19.74 & Absorption & Inflow \\
        \hline
    \end{tabular}
\end{table*}

\begin{table*}
    \contcaption{}
    \begin{tabular}{cccccccccccc}
        \hline
        Sample & SFR & log M$_{*}$/M$_{\odot}$ & EW & $\Delta$v & $\Delta$v/cos($i$) & b$_{\text{D}}$ & C$_{f}$ & N(Na I) & N(H) & Profile & Type \\
        \hline
.............. & 3.51 & 10.51 & 0.03 & 174$\pm$72 & 407 & 56$\pm$39 & 0.48$\pm$0.24 & 11.97 & 19.65 & Absorption & Inflow \\
.............. & 4.48 & 10.69 & 0.12 & 148$\pm$37 & 343 & 51$\pm$33 & 0.83$\pm$0.23 & 12.65 & 20.33 & Absorption & Inflow \\
.............. & 6.44 & 10.89 & 0.07 & 146$\pm$19 & 337 & 56$\pm$31 & 0.62$\pm$0.26 & 12.44 & 20.12 & Absorption & Inflow \\
.............. & 2.94 & 10.51 & 0.04 & 176$\pm$74 & 645 & 100$\pm$45 & 0.59$\pm$0.24 & 11.89 & 19.56 & Absorption & Inflow \\
.............. & 4.33 & 10.69 & 0.06 & 167$\pm$43 & 584 & 58$\pm$34 & 0.81$\pm$0.26 & 12.05 & 19.73 & Absorption & Inflow \\
.............. & 5.58 & 10.89 & 0.15 & 147$\pm$61 & 501 & 58$\pm$42 & 0.57$\pm$0.23 & 12.76 & 20.44 & Absorption & Inflow \\
.............. & 6.98 & 11.09 & 0.18 & 146$\pm$40 & 522 & 129$\pm$24 & 0.62$\pm$0.22 & 12.49 & 20.17 & Absorption & Inflow \\
        \hline
    \end{tabular}
\end{table*}

\section{\texorpdfstring{\nad}{TEXT} profiles}
\label{sec:profilefits}
In this section we present examples of our \nad\ fits as described in Section \ref{sec:nadfitting} for the main sample of both inactive galaxies and the AGN sample.

\begin{figure*}
 \includegraphics[width=0.85\textwidth]{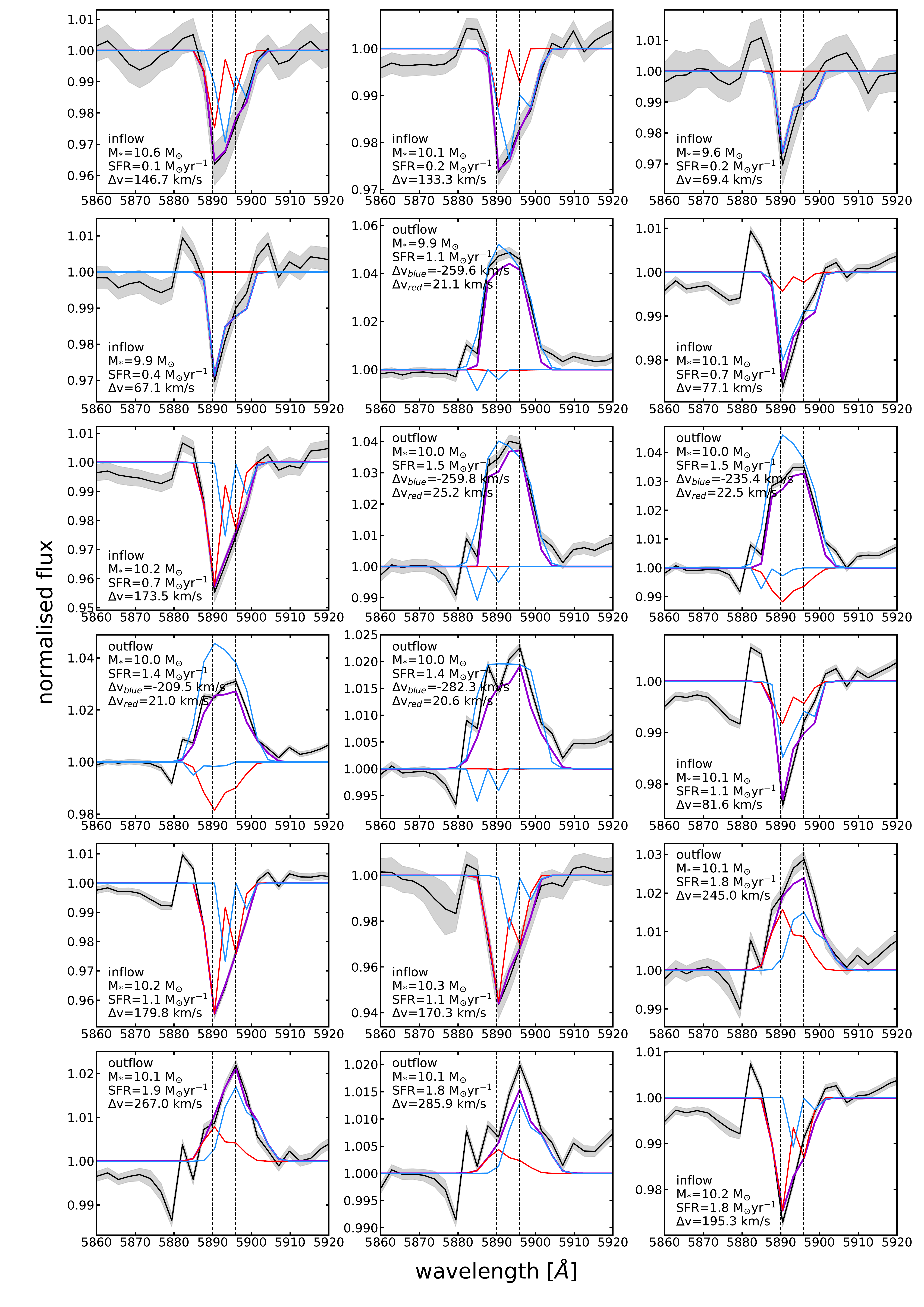}
 \caption{The normalised \nad\ ISM profiles from inflow and outflow detections in the DISK, LOW-$i$, HIGH-$i$ and BULGE samples for inactive galaxies. The black line line is the \nad\ profile and the gray shade is the flux error. The best-fit two-component models are overplotted: purple denotes the total fit, red is the systemic component, and blue represents the blueshifted or redshifted flow component.}
 \label{fig:fig_nad_inactive}
\end{figure*}

\begin{figure*}
 \includegraphics[width=0.85\textwidth]{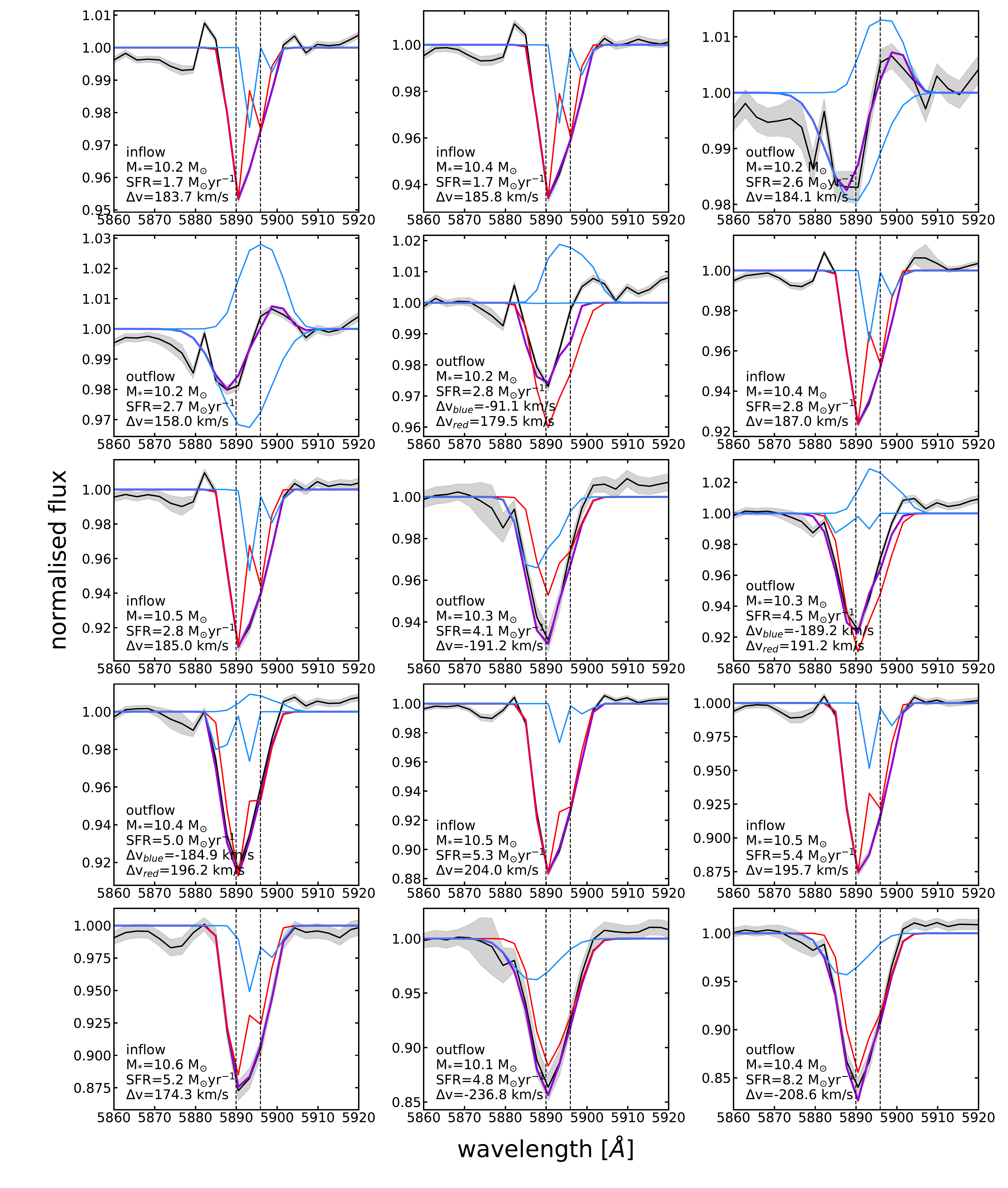}
 \contcaption{}
\end{figure*}

\begin{figure*}
 \includegraphics[width=0.85\textwidth]{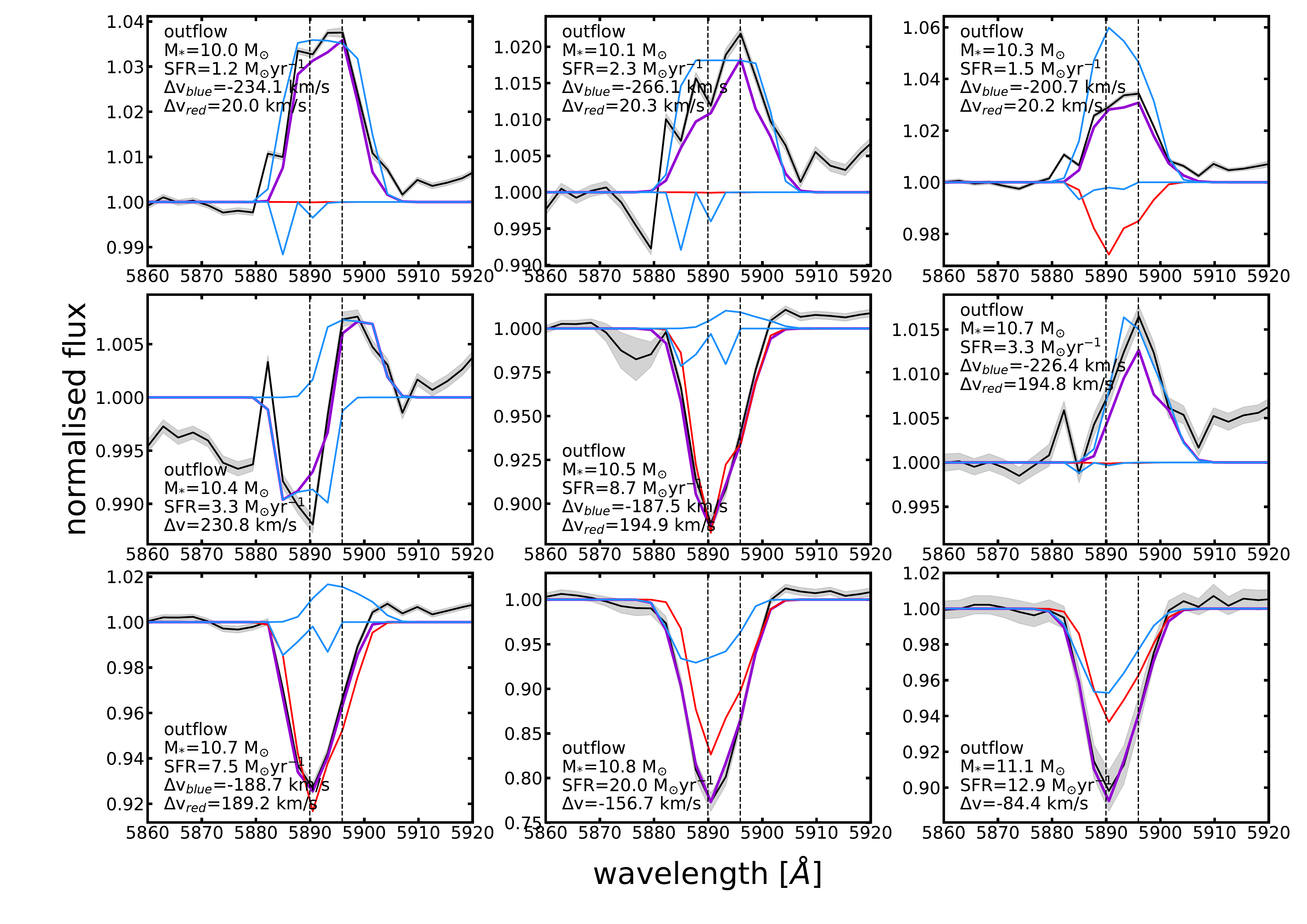}
 \contcaption{}
\end{figure*}

\begin{figure*}
 \includegraphics[width=0.85\textwidth]{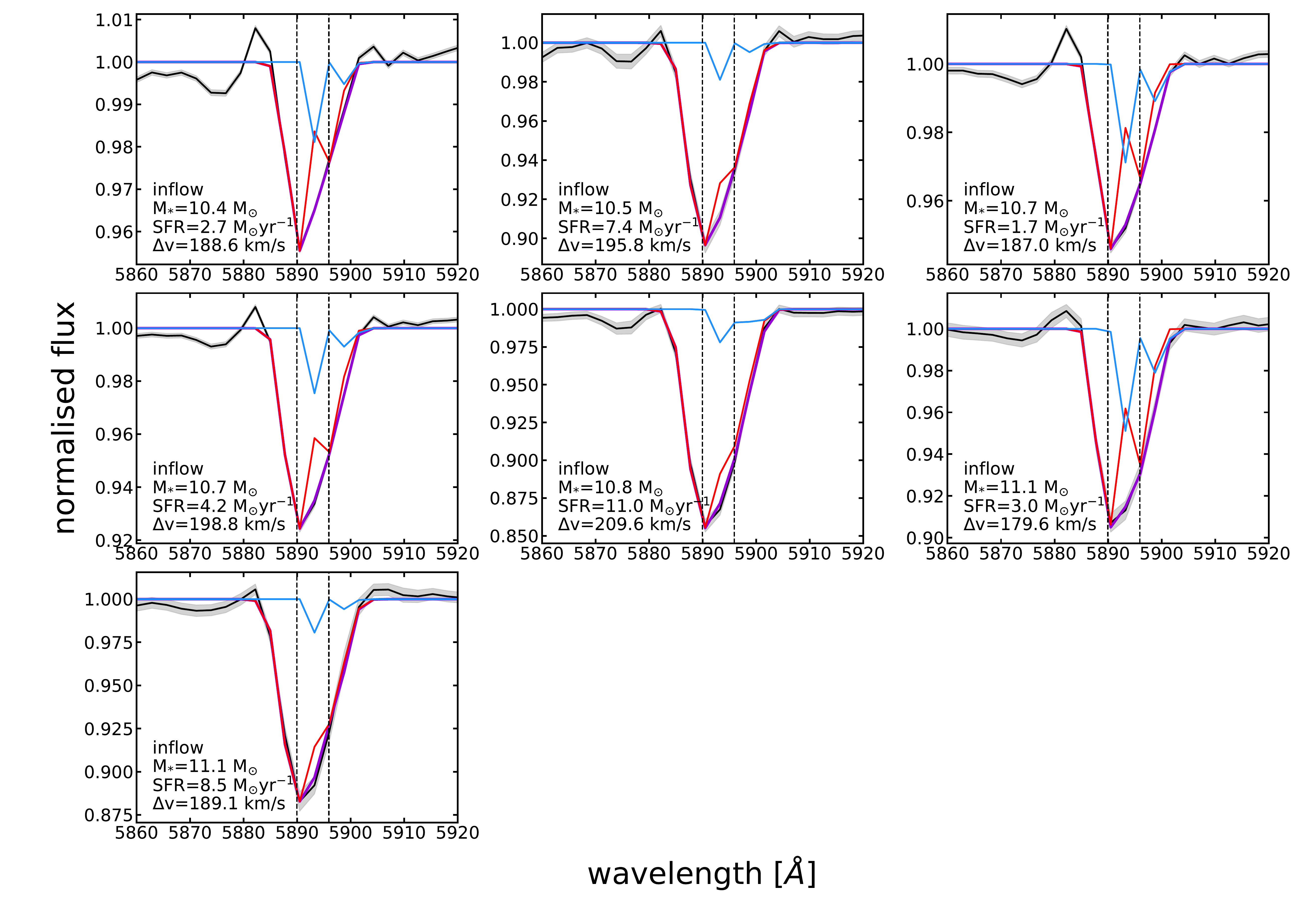}
 \contcaption{}
\end{figure*}

\begin{figure*}
 \includegraphics[width=0.85\textwidth]{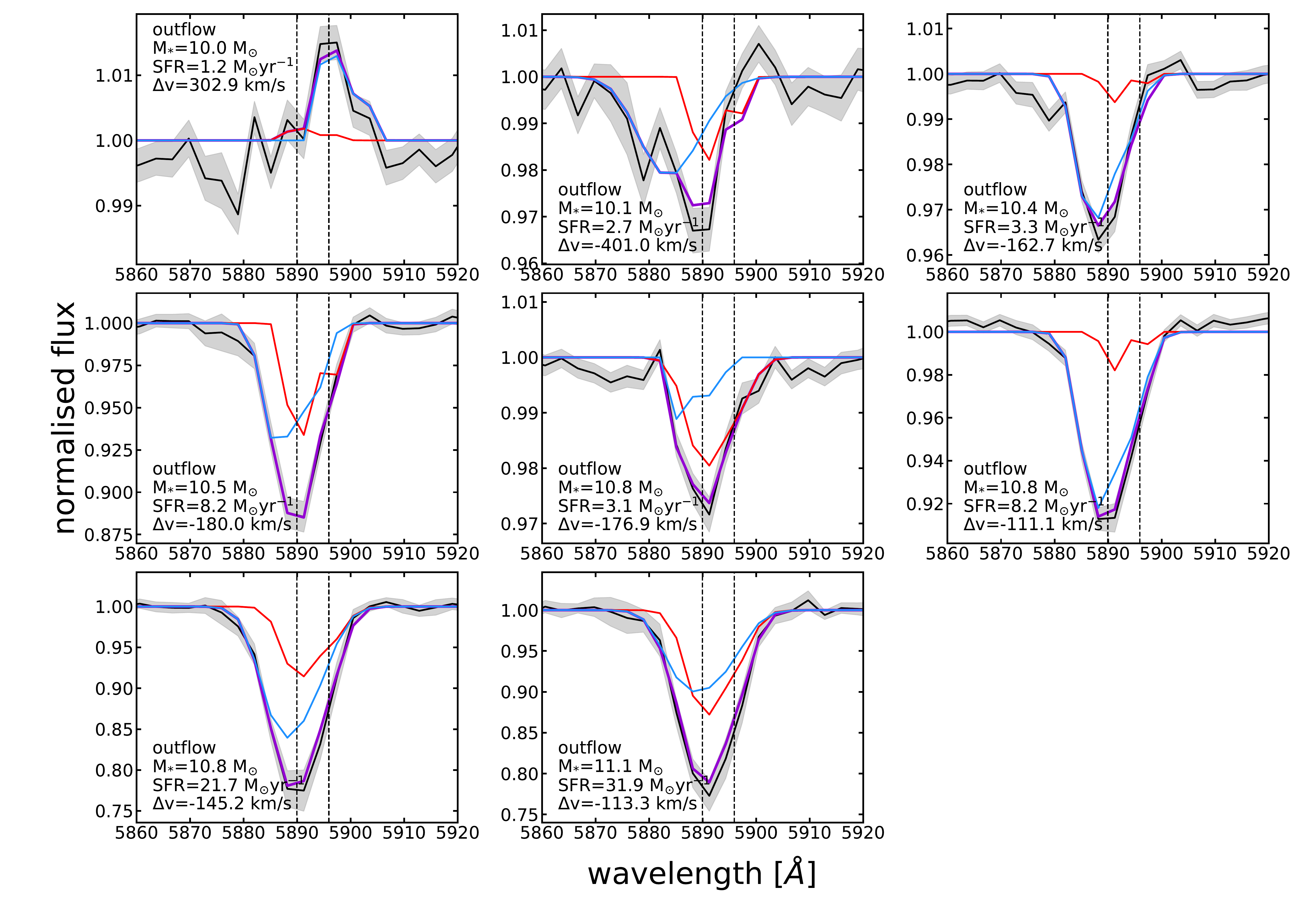}
 \contcaption{}
\end{figure*}

\begin{figure*}
 \includegraphics[width=0.85\textwidth]{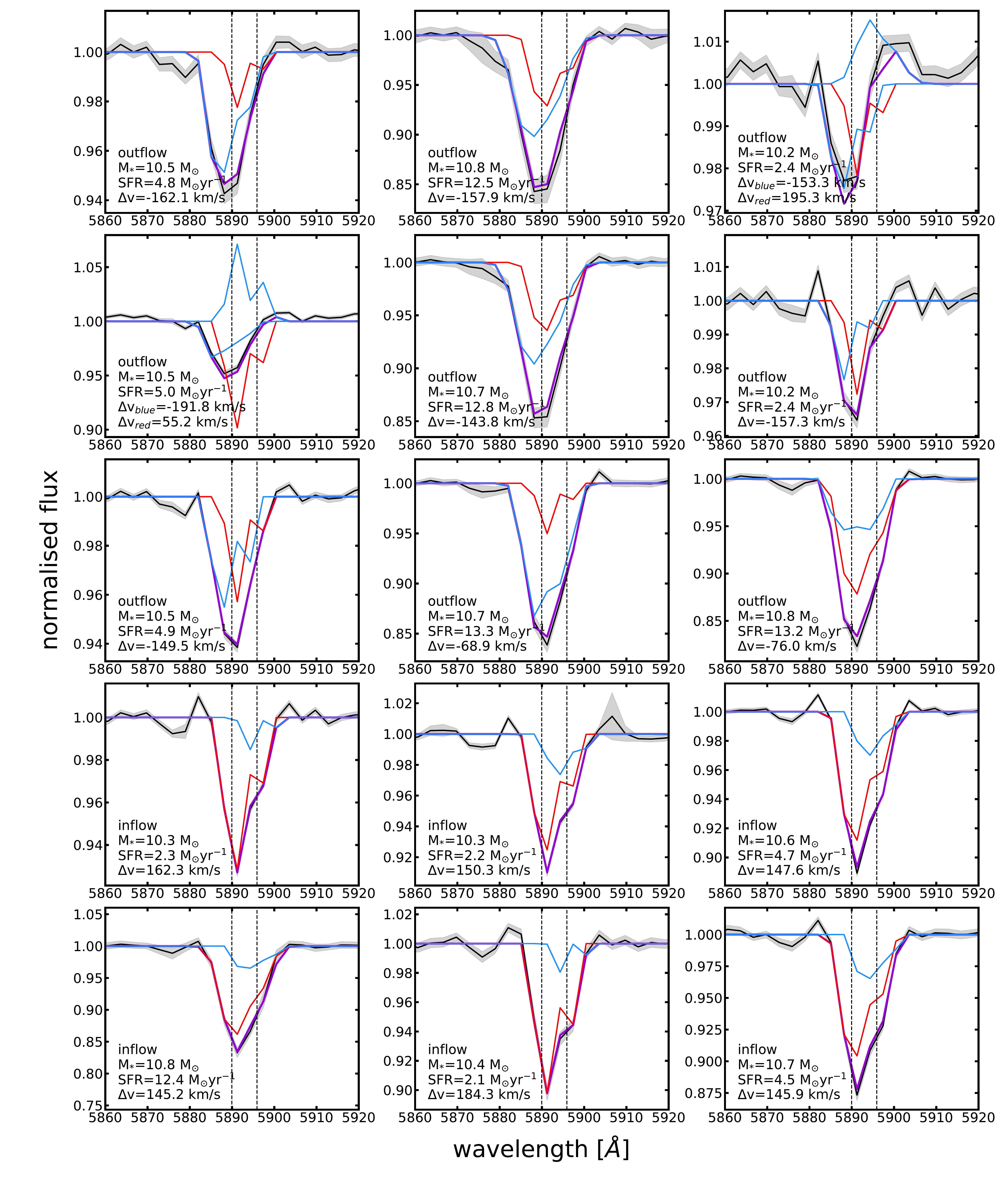}
 \contcaption{}
\end{figure*}

\begin{figure*}
 \includegraphics[width=0.85\textwidth]{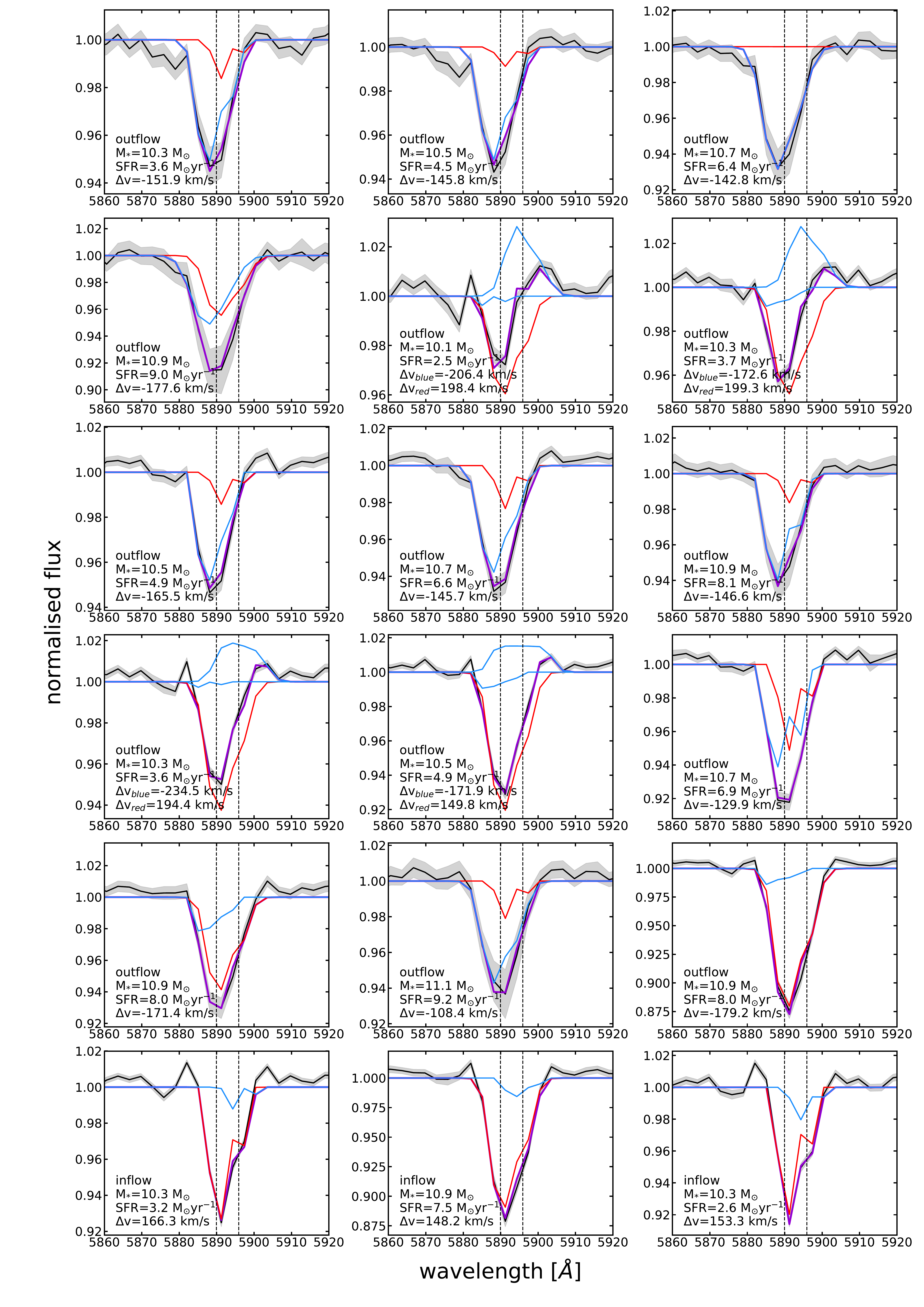}
 \contcaption{}
\end{figure*}

\begin{figure*}
 \includegraphics[width=0.85\textwidth]{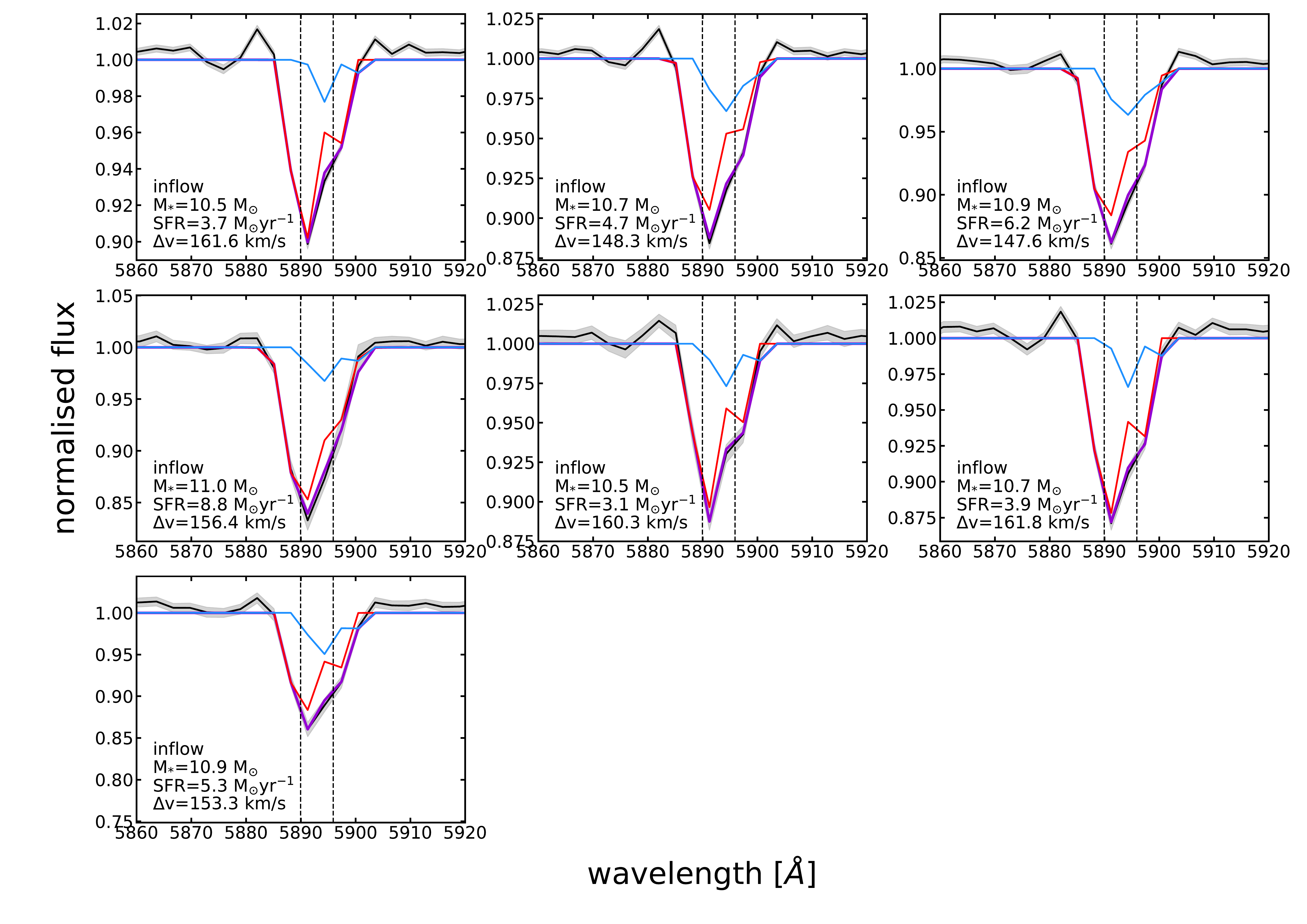}
 \contcaption{}
\end{figure*}

\begin{figure*}
 \includegraphics[width=0.85\textwidth]{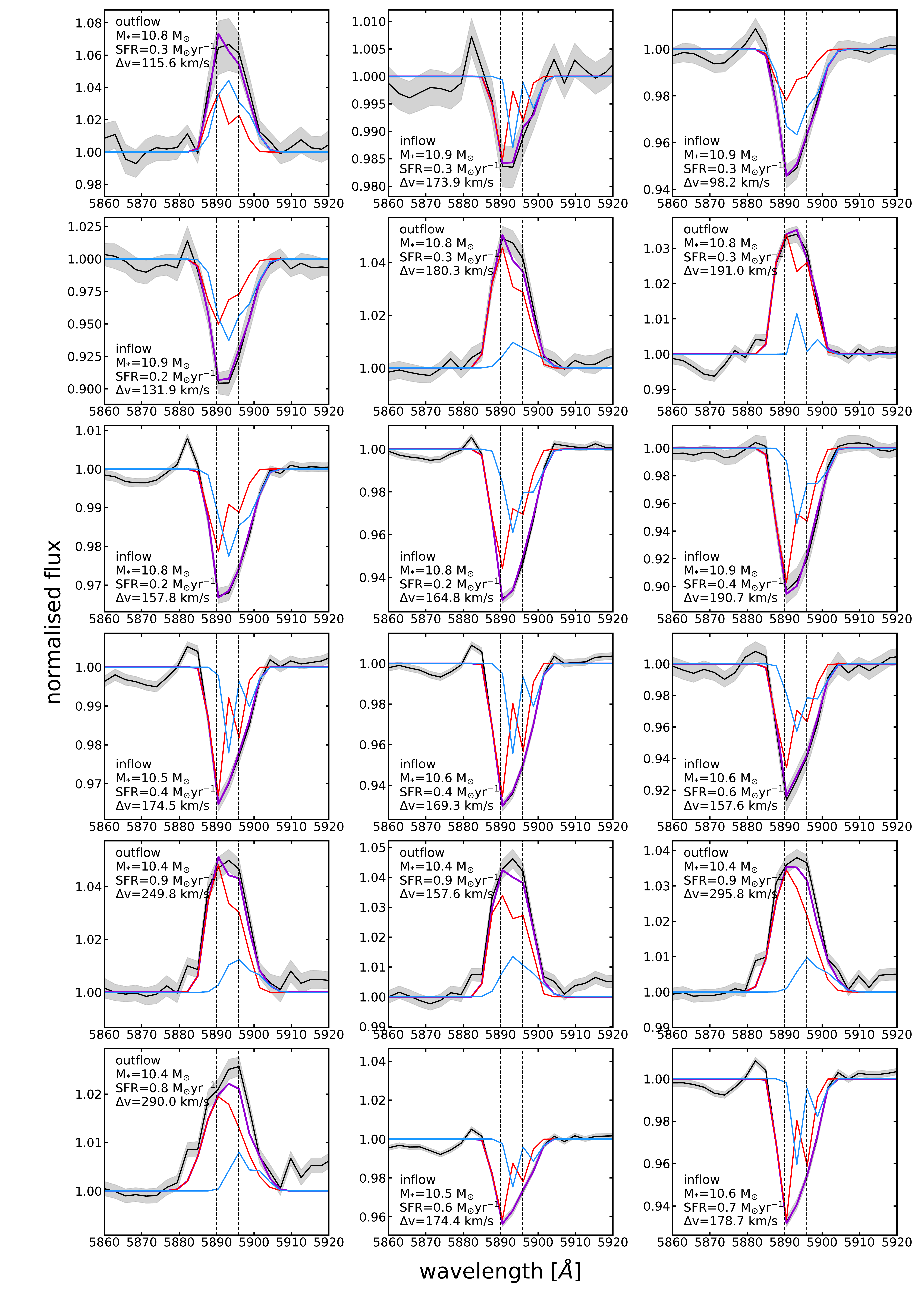}
 \caption{The same as Figure \ref{fig:fig_nad_inactive} but for the AGN sample.}
 \label{fig:fig_nad_AGN}
\end{figure*}

\begin{figure*}
 \includegraphics[width=0.85\textwidth]{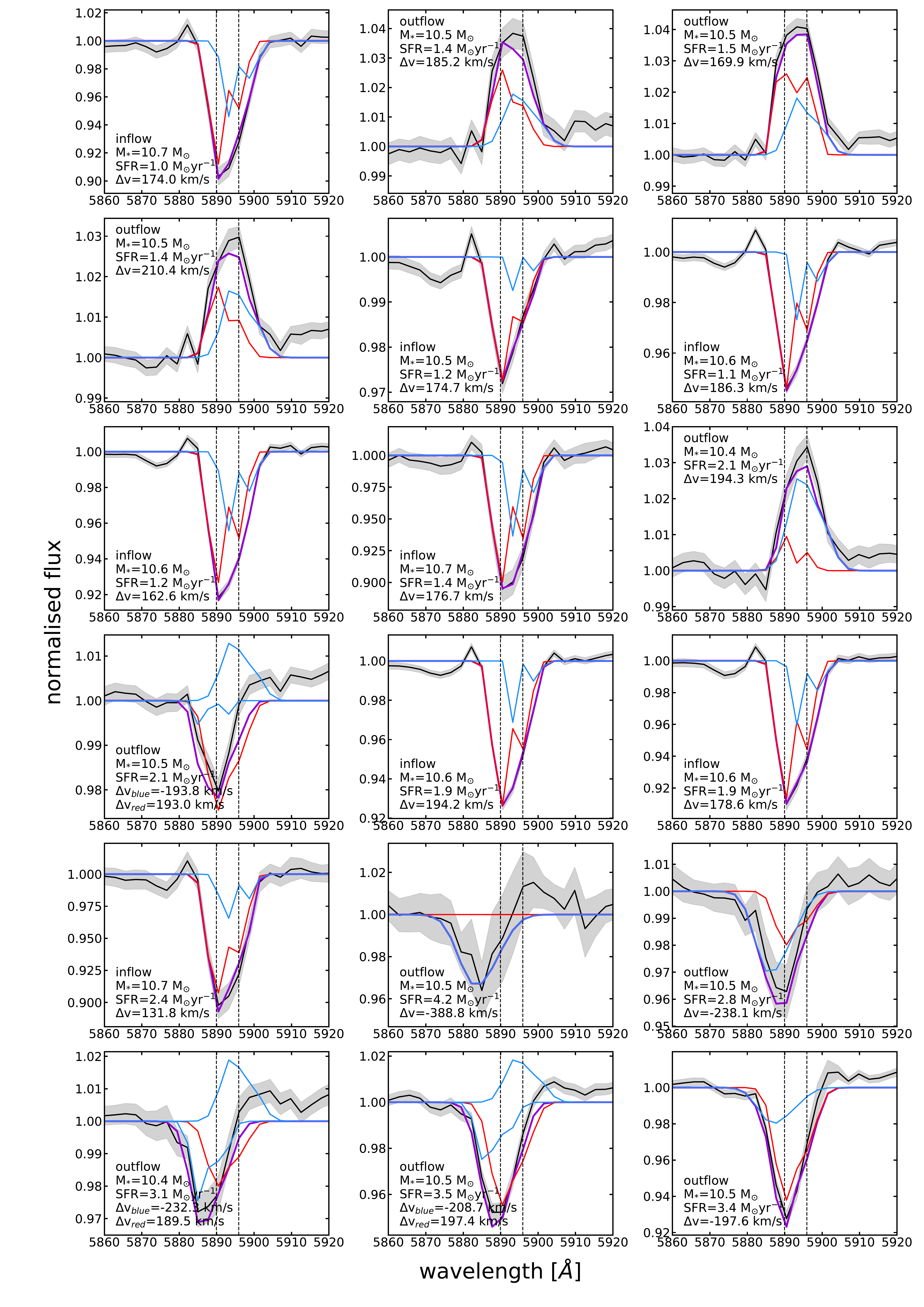}
 \contcaption{}
\end{figure*}

\begin{figure*}
 \includegraphics[width=0.85\textwidth]{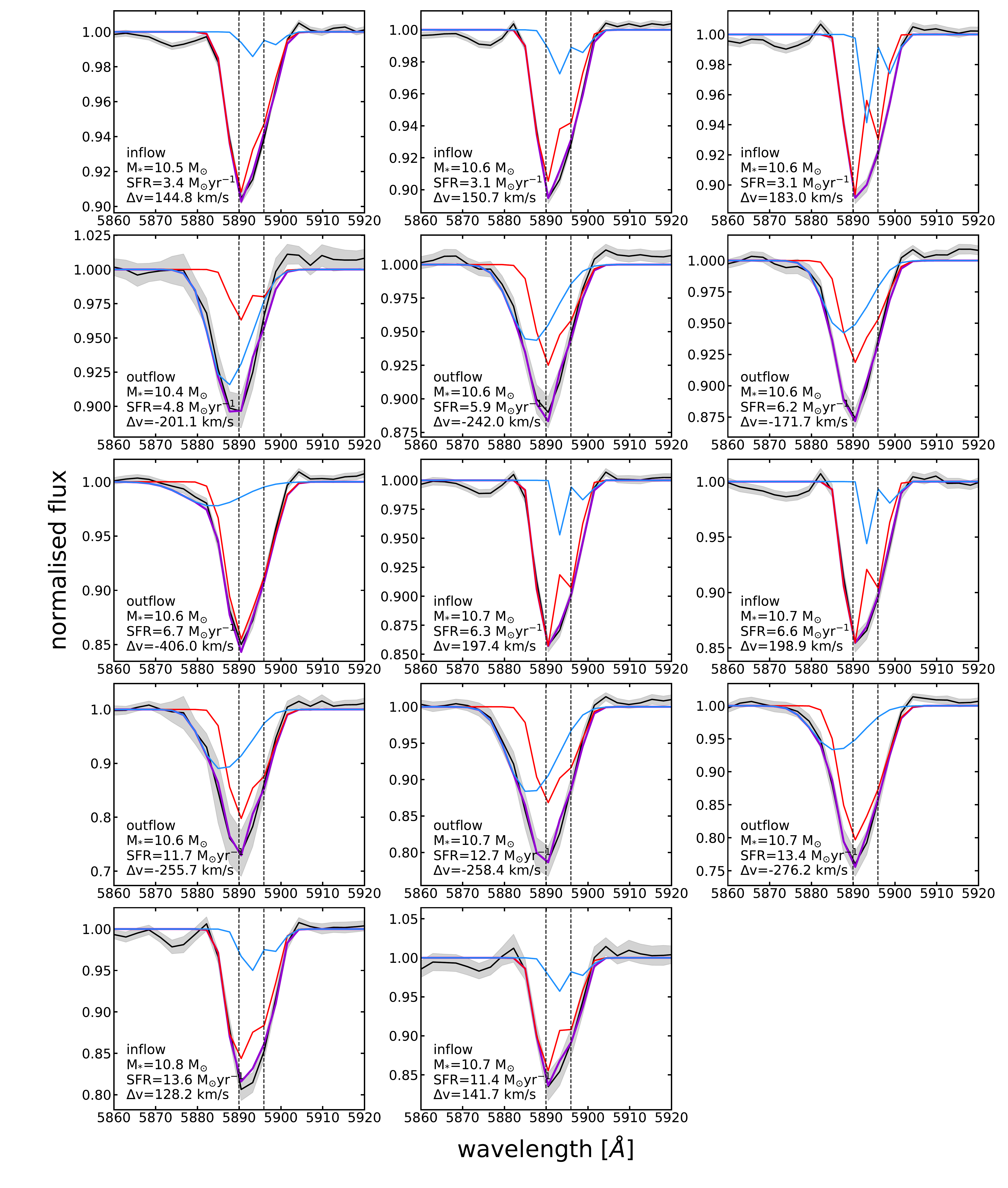}
 \contcaption{}
\end{figure*}

\begin{figure*}
 \includegraphics[width=0.85\textwidth]{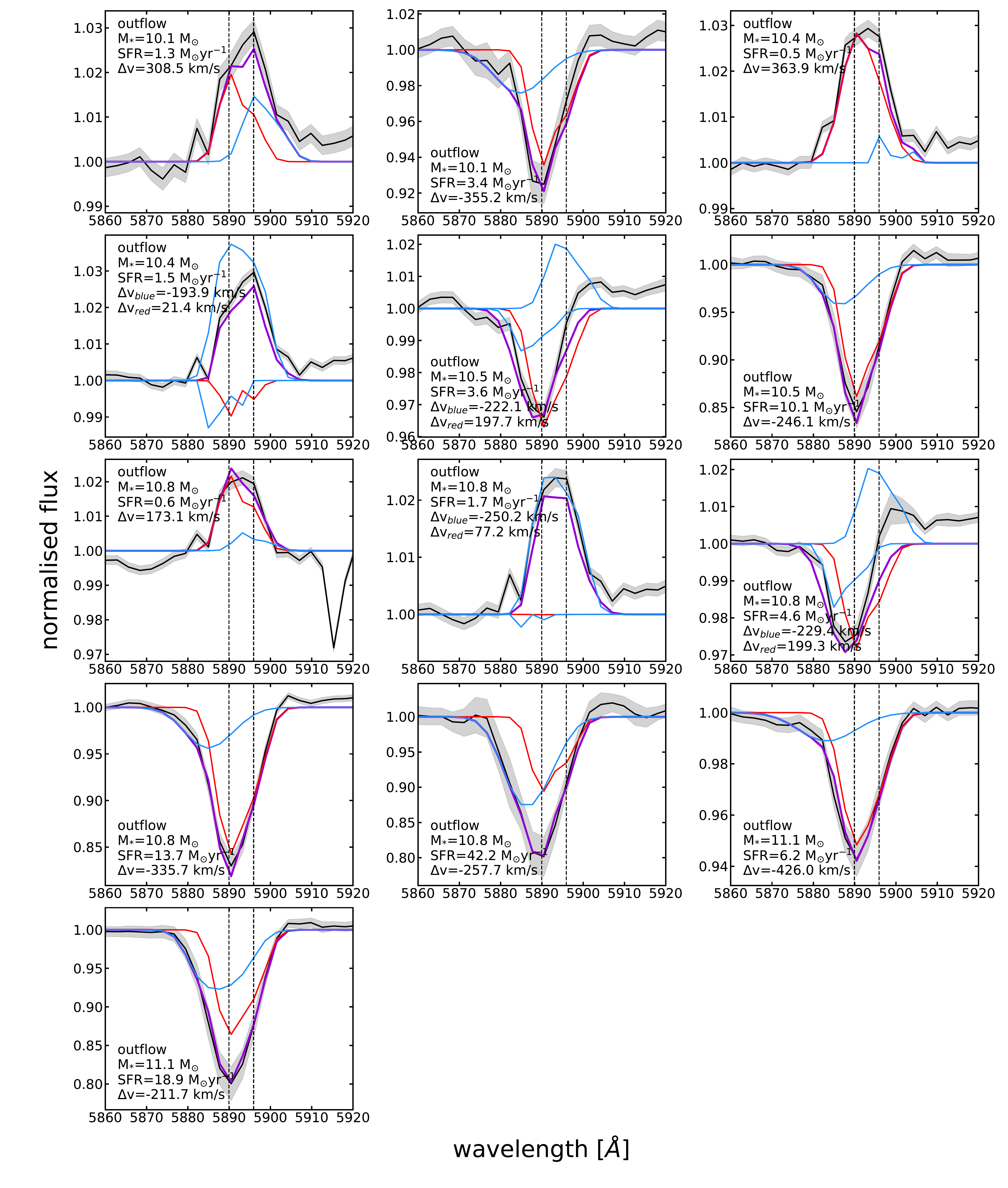}
 \contcaption{}
\end{figure*}

\begin{figure*}
 \includegraphics[width=0.85\textwidth]{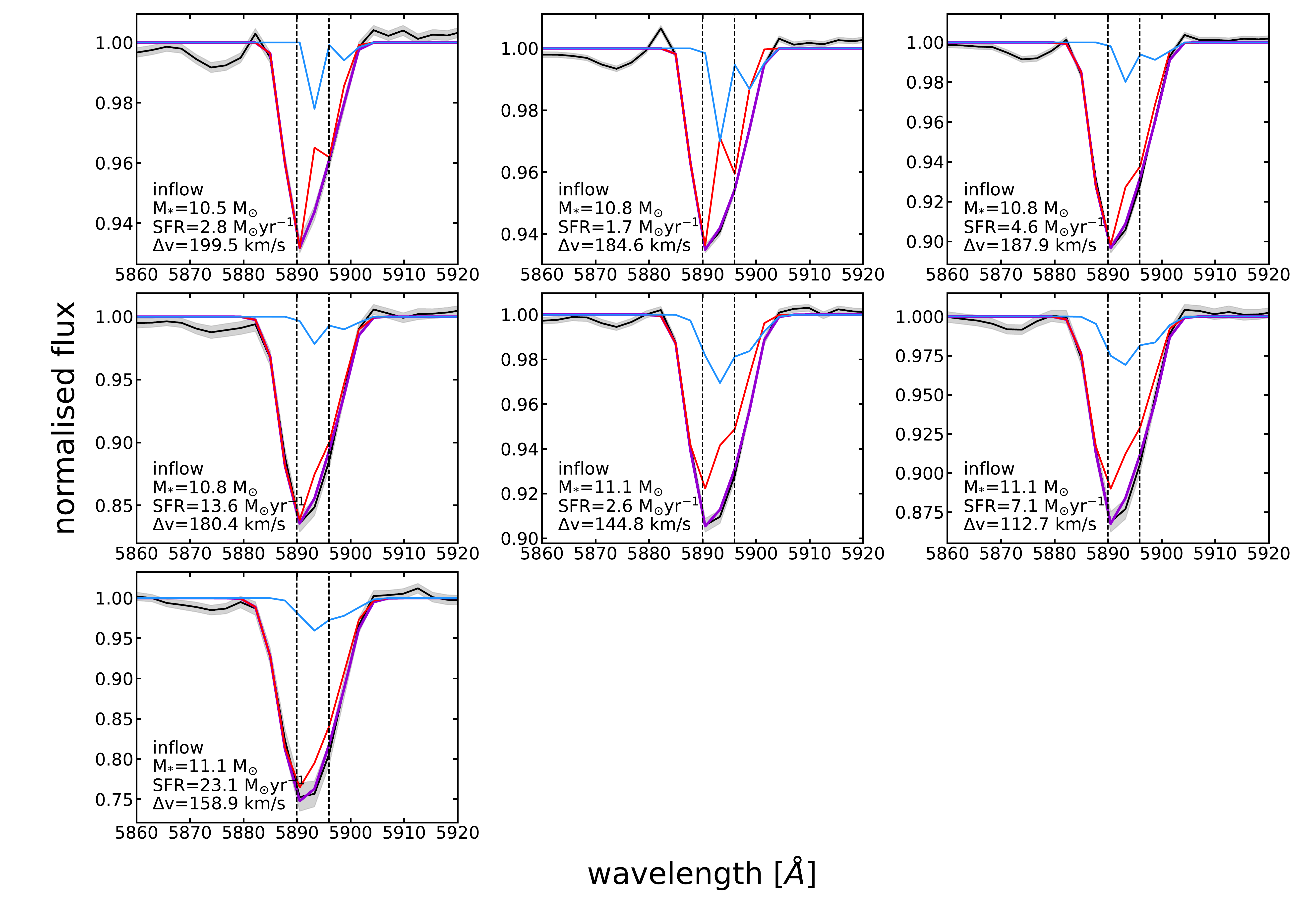}
 \contcaption{}
\end{figure*}

\begin{figure*}
 \includegraphics[width=0.85\textwidth]{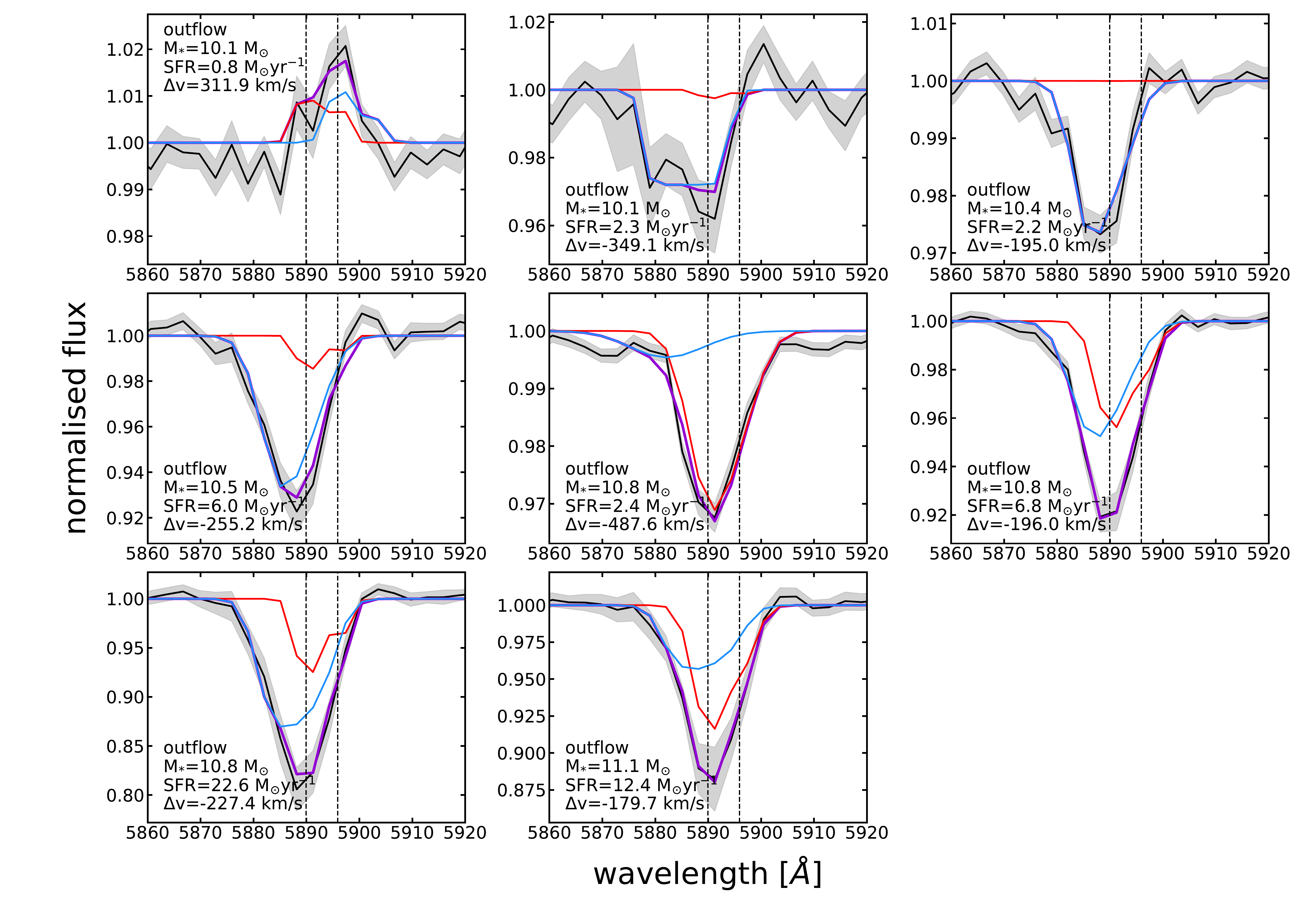}
 \contcaption{}
\end{figure*}

\begin{figure*}
 \includegraphics[width=0.85\textwidth]{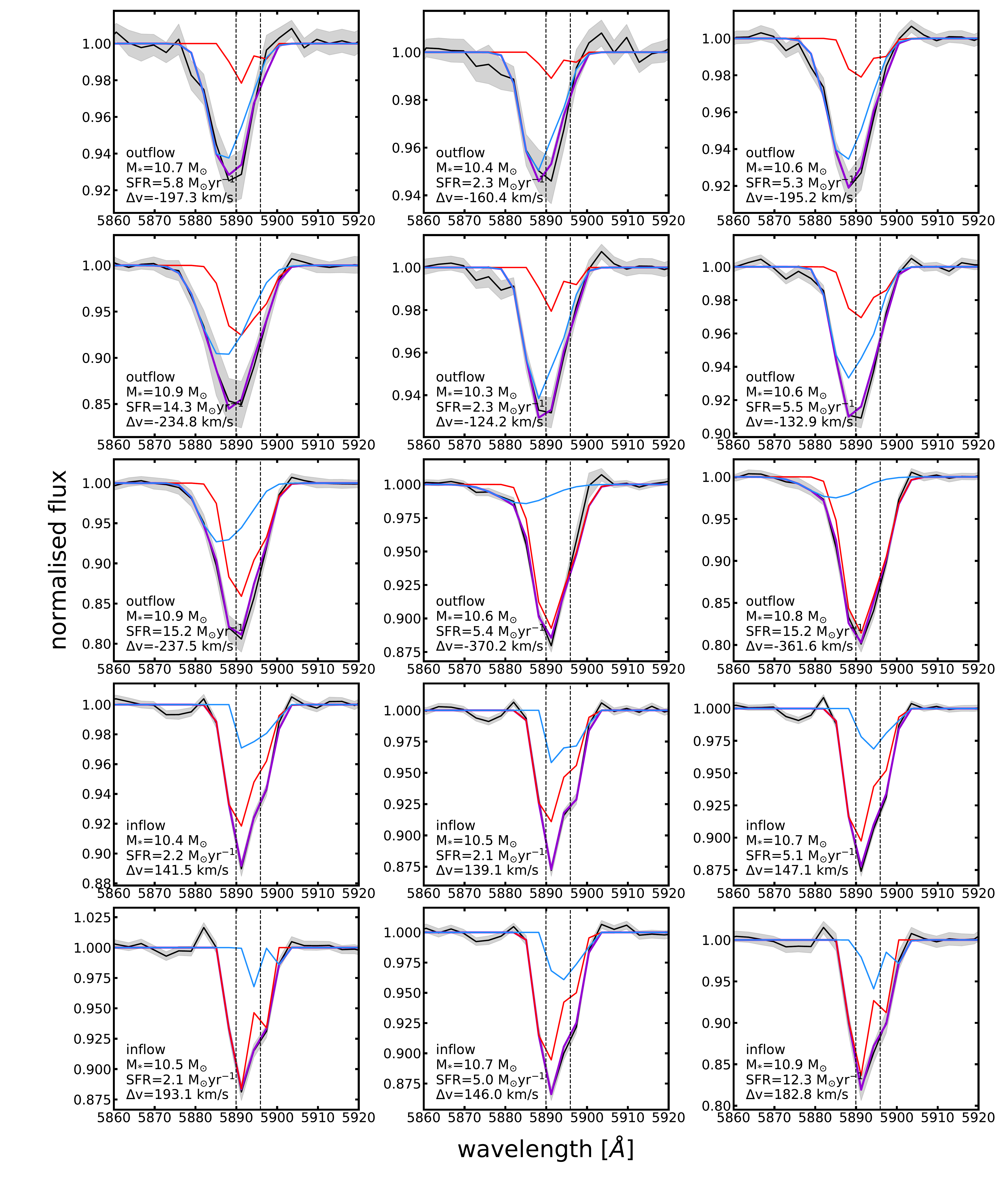}
 \contcaption{}
\end{figure*}

\begin{figure*}
 \includegraphics[width=0.85\textwidth]{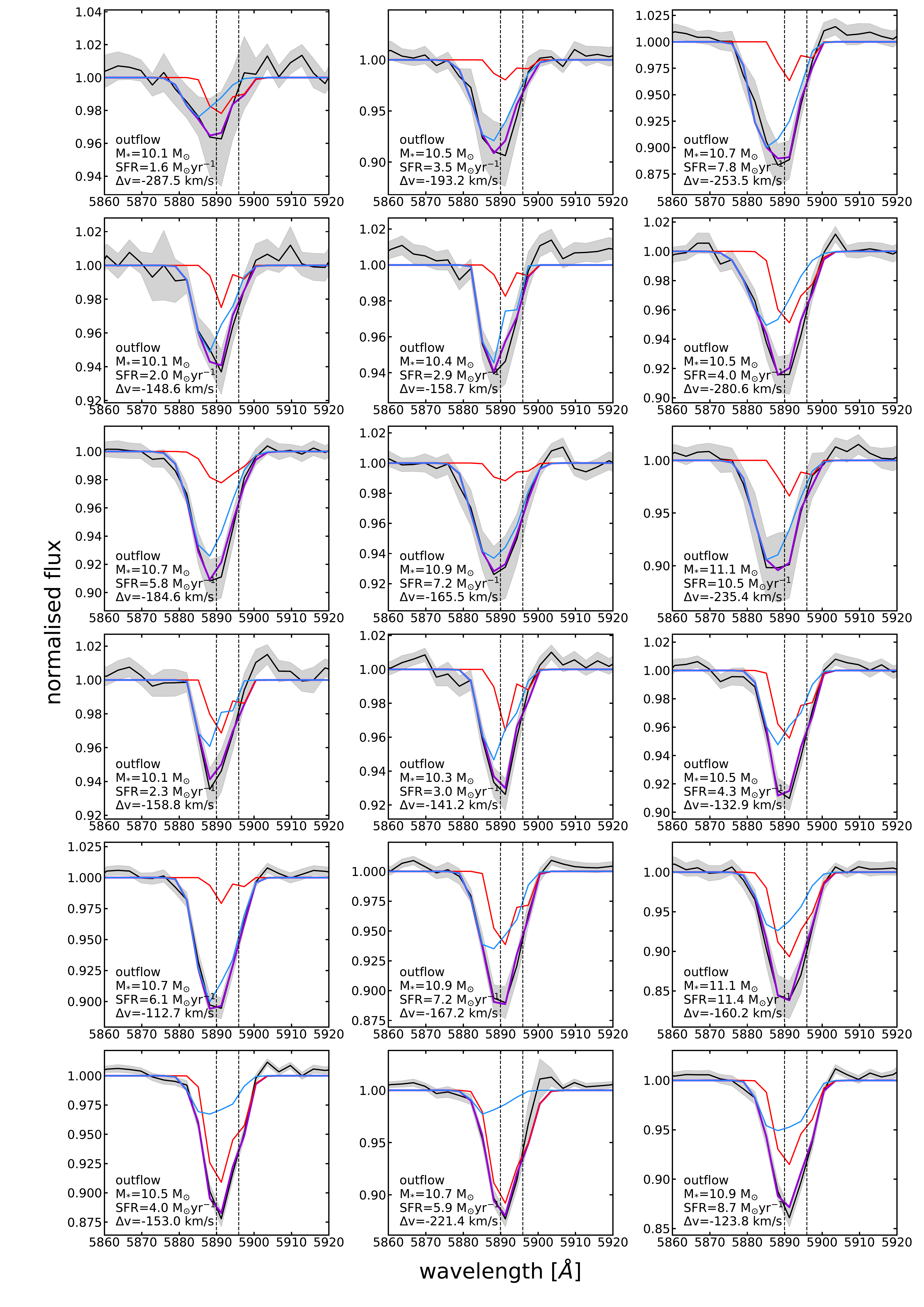}
 \contcaption{}
\end{figure*}

\begin{figure*}
 \includegraphics[width=0.85\textwidth]{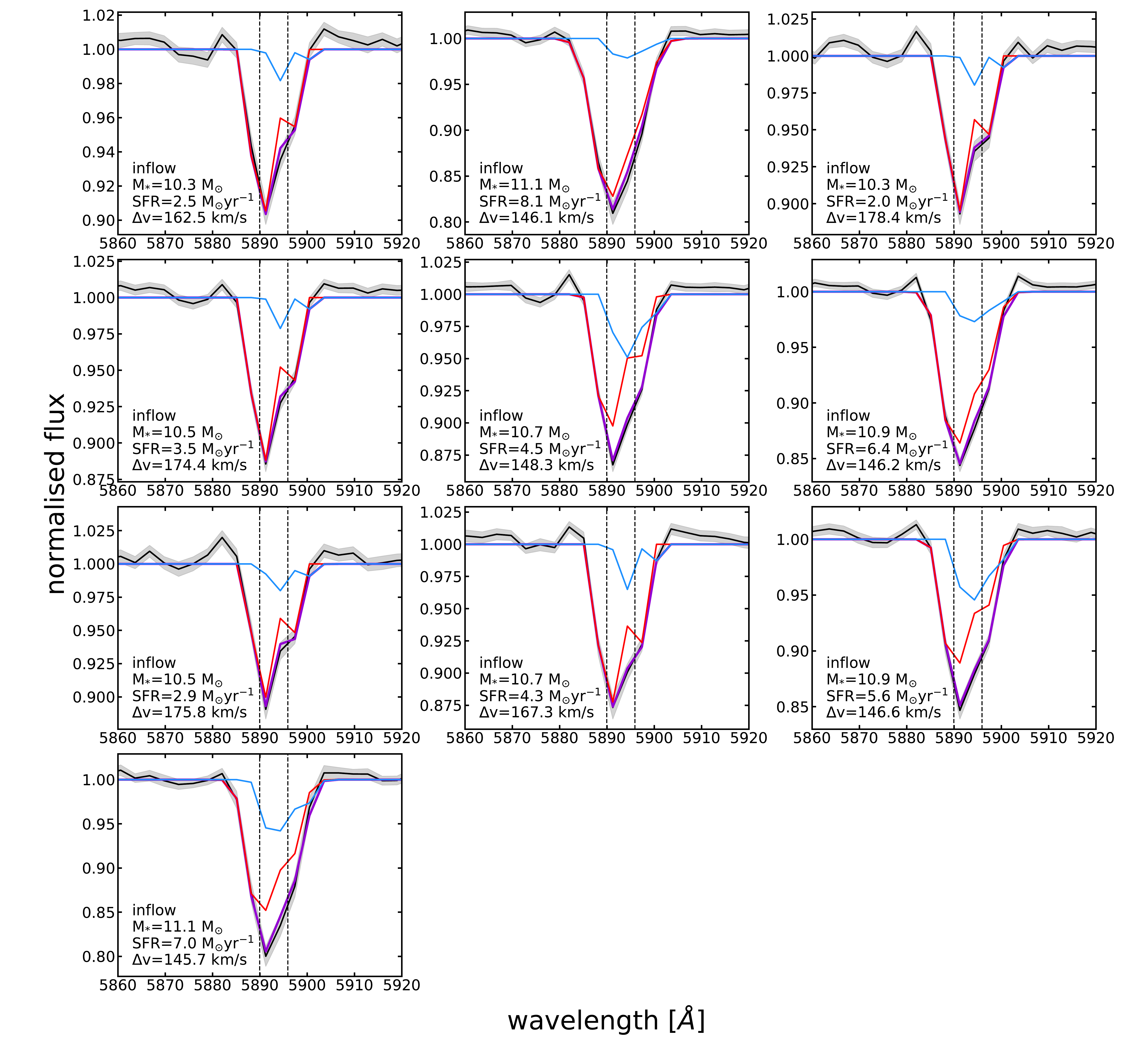}
 \contcaption{}
\end{figure*}


\bsp  
\label{lastpage}
\end{document}